





\def\marginscom{
    \voffset=0.truein
    \vsize=9.truein
    \hoffset=0.truein
    \hsize=6.5truein
    }

\voffset=0.truein
\hoffset=0.truein
\vsize=8.9truein
\hsize=6.5truein
\baselineskip=20.pt
\tolerance=10000
\pretolerance=10000
\hyphenpenalty=10000
\def\ApJstyle{\textstyle}
\def\compact{\baselineskip=12pt \marginscom \def\ApJstyle{\scriptstyle} }

\def\etal{et~al.\ }
\def\oneskip{\vskip\baselineskip}
\def\hafskip{\vskip .5\baselineskip}
\def\coverpage{
    \voffset=0.0truein
    \vsize=8.9truein
    \baselineskip=20pt
    \nopagenumbers        }


\def\makeheadline{
\vbox to 0pt{\vskip-40.1pt
  \line{\vbox to 8.5pt{}\the\headline}\vss}
  \nointerlineskip      }

\def\specialpage{\vfill\supereject\vglue .5truein\countspecial=\number\pageno
                 \counteqn=0\countfig=0}
\def\blankfootline{\hfil}
\def\blankheadline{\hfil}
\def\pagenumberfootline{\hss\tenrm\folio\hss}
\def\pagenumberheadline{\hfil\tenrm\folio}
\newcount\countspecial
\newcount\countnopagenumber
\headline={\ifnum\pageno=\countnopagenumber\blankheadline
           \else {\ifnum\pageno=\countspecial\blankheadline
                  \else\pagenumberheadline\fi}
           \fi}
\footline={\ifnum\pageno=\countnopagenumber\blankfootline
           \else {\ifnum\pageno=\countspecial\pagenumberfootline
                  \else\blankfootline\fi}
           \fi}





\newcount\counttemp
\newcount\counteqn
\counteqn=0
\def\eqnnumber#1){\number\counteqn \rm #1)}
\def\eqn#1){\global\advance\counteqn by 1   \eqnnumber \rm #1)}
\def\aeqn#1){\global\advance\counteqn by 1   \eqnnumber \rm #1)}
\def\showeqn#1){\counttemp=\counteqn \advance\counttemp by #1
                \number\counttemp)}
\def\showeqnsq#1]{\counttemp=\counteqn \advance\counttemp by #1
                \number\counttemp]}
\def\showeqnx#1,#2){\counttemp=\counteqn \advance\counttemp by #1
                \number\counttemp#2)}


\newcount\countitemc
\countitemc=0
\def\itemcnumber{\number\countitemc}
\def\itemc{\global\advance\countitemc by 1   \itemcnumber}

 
\newcount\counttable
\counttable=0
\def\tablenumber#1.{\number\counttable \rm #1.}
\def\tabler#1.{\global\advance\counttable by 1   \tablenumber \rm #1.}


\newcount\countfig
\countfig=0
\def\fignumber{\number\countfig}
\def\fig{\global\advance\countfig by 1  \fignumber}
\def\showfig#1){\counttemp=\countfig \advance\counttemp by #1
                \number\counttemp)}


\newcount\countsection
\countsection=0
\def\sectionnumber{\number\countsection}
\def\section{\global\advance\countsection by 1  \sectionnumber}
\def\showsection#1){\counttemp=\countsection \advance\counttemp by #1
                \number\counttemp)}


\def\lapprox{\hbox{\lower .8ex\hbox{$\,\buildrel < \over\sim\,$}}}
\def\gapprox{\hbox{\lower .8ex\hbox{$\,\buildrel > \over\sim\,$}}}
\def\gradient{\hbox{\raise .5ex\hbox{$\bigtriangledown$}}}


\def\reference{\parindent=0pt\tolerance=300}

\def\refindent{\par\hangindent=15pt\hangafter=1}
\def\refpaper#11#2,#3,#4,#5.{\refindent#11#2,#3,#4,#5}
\def\refedited#1in#2,#3)#4.{\refindent#1in#2,#3)#4}
\def\refbook#11#2,#3(#4)#5.{\refindent#11#2,#3(#4)#5}

\def\uncatcodespecials{\def\do##1{\catcode\lq##1=12 }\dospecials}
\newcount\lineno 

{\obeyspaces\global\let =\ } 
\def\setupverbatimno{\tt \lineno=0
     \def\par{\leavevmode\endgraf} \catcode`\`=\active
     \obeylines \uncatcodespecials \obeyspaces
     \everypar{\advance\lineno by1 \llap{\sevenrm\the\lineno\ \ }}}
{\obeyspaces\global\let =\ } 



{\obeyspaces\global\let =\ } 


\def\Romannumeral#1.{\uppercase\expandafter{\romannumeral#1}}
 
{
\catcode`\@=11
\global\def\Biggg#1{{\hbox{$\left#1\vbox to 25.pt{}\right.\n@space$}}}

}





\compact                     

\countnopagenumber=\number\pageno
{  
\coverpage

\vglue 1.0truein
\centerline{RADIOACTIVE DECAY ENERGY DEPOSITION IN SUPERNOVAE}
\centerline{AND THE EXPONENTIAL/QUASI-EXPONENTIAL BEHAVIOR}\nobreak
\centerline{OF LATE-TIME SUPERNOVA LIGHT CURVES}\nobreak
            
\vfill
\centerline{DAVID J.~JEFFERY
            \footnote{$^{\ApJstyle 1}$}{
            Department of Physics,
            University of Nevada, Las Vegas,
            4505 S. Maryland Parkway,
            Las Vegas, Nevada 89154-4002,
            email:  jeffery@physics.unlv.edu
            \hafskip
            }
            \footnote{$^{\ApJstyle 2}$}{
            Present address:
            Middle Tennessee State University,
            Department of Physics \&~Astronomy,
            Wiser-Patten Science Hall,
            1301 East Main Street,
            Murfreesboro, Tennessee  37132
            \hafskip
            }}

\oneskip
\centerline{Department of Physics, University of Nevada, Las Vegas}
\centerline{Las Vegas, Nevada 89154-4002}
\oneskip

\vfill

\vfill



\eject
} 

\specialpage
\pageno=2
\countspecial=\number\pageno
\counteqn=0
\centerline{ABSTRACT}\nobreak
\vskip\baselineskip\nobreak
     The radioactive decay energy (RDE) deposition in supernovae from the
decay chain
$^{\ApJstyle 56}$$\rm Ni\to^{\ApJstyle 56}$$\rm Co\to^{\ApJstyle 56}$$\rm Fe$
usually directly powers the ultraviolet/optical/infrared (UVOIR)
bolometric luminosity of supernovae in their quasi-steady state phase
until very late times.
The result for this phase is exponential/quasi-exponential UVOIR
bolometric light curves and often exponential/quasi-exponential broad band
light curves. 
A presentation is given of a simple, approximate, analytic
treatment of RDE deposition that provides
a straightforward understanding of the exponential/quasi-exponential
behavior of the UVOIR bolometric luminosity
and a partial understanding of the exponential/quasi-exponential behavior
of the broad band light curves.
The treatment reduces to using a normalized deposition function
$N_{\ApJstyle\rm Ni}^{\ApJstyle *}(t)$ as an analysis tool.
(The absolute deposition is determined by specifying the initial
$^{\ApJstyle 56}$Ni mass or fitting absolute supernova UVOIR bolometric
luminosity.)
The time evolution of $N_{\ApJstyle\rm Ni}^{\ApJstyle *}(t)$ is determined
by three time scales:   the half-lives of $^{\ApJstyle 56}$Ni and
$^{\ApJstyle 56}$Co, and a fiducial time parameter $t_{\ApJstyle 0}$ that
governs the time-varying $\gamma$-ray optical depth behavior of a supernova. 
The $t_{\ApJstyle 0}$ parameter can be extracted from a structural
supernova model, or by fitting either the RDE deposition curve from a
more detailed treatment of deposition or the observed UVOIR bolometric light
curve from the quasi-steady state phase of a supernova.
A $t_{\ApJstyle 0}$ parameter obtained from observations can provide
a constraint on the important physical parameters of the supernova.
The effective use of this constraint requires having an adequate
parameterized structural supernova model.

    The $N_{\ApJstyle\rm Ni}^{\ApJstyle *}(t)$ function is used
to analyze the preliminary UVOIR bolometric light curve of
SN~Ic~1998bw (the possible cause of $\gamma$-ray burst GRB980425).
The SN~1998bw fiducial time $t_{\ApJstyle 0}$ is found to
be $134.42$~days and a prediction
is made for the evolution of the SN~1998bw RDE deposition curve
and quasi-steady state UVOIR bolometric light curve out to day~1000 after
the explosion.
A crude estimate (perhaps a factor of a few too small) 
of the SN~1998bw mass obtained from a parameterized
core-collapse model and $t_{\ApJstyle 0}=134.42$~days is
$4.26\,M_{\ApJstyle\odot}$.
As further examples of the simple analytic treatment, the RDE
deposition and luminosity evolution of SN~Ia~1992A and SN~II~1987A have
also been examined.


     The simple analytic treatment of RDE deposition has actually
existed for 20 years at least without, apparently, being discussed at length.
The main value of this paper is the explicit, detailed, general presentation
of this analytic treatment.
\vskip 3\baselineskip\nobreak

\noindent {\it Subject Headings:}
supernovae:  general ---
supernovae:  individual (SN 1992A, SN 1987A, SN 1998bw) ---
radiative transfer ---
$\gamma$-rays:   theory

\vfill\eject

\specialpage

\centerline{1.\ \ INTRODUCTION}\nobreak
\vskip\baselineskip\nobreak
     Radioactive beta decay energy (RDE) from radioactive 
$^{\ApJstyle 56}$Ni synthesized in the explosion and its daughter
$^{\ApJstyle 56}$Co is a major source of supernova luminosity.
In Type~Ia supernovae (SNe~Ia), RDE powers the whole observable
luminosity from
a few days after explosion (e.g., Harkness 1991)
until as far as observations extend (i.e., about 944 days
after explosion for the case of SN~1992A) or so it seems
(Cappellaro~\etal 1997). 
In core-collapse supernovae (i.e., SNe~II, Ib, Ic), RDE is the main source
of luminosity starting from a few days to some tens of days after explosion
(e.g., Woosley 1988;  Young, Baron, \&~Branch 1995)
and perhaps lasting until of order 1000 days after explosion
(e.g., Suntzeff~\etal 1992).
Before the RDE becomes dominant,
shock heat left over from the explosion is the main source.
The transition time between shock heat and RDE powering depends on
the parameters of the supernova, most importantly mass, kinetic energy,
density profile, and composition.
Circumstellar interaction and perhaps pulsar action can also supply
energy to core-collapse supernovae.

     In this paper we provide a simple, analytic, normalized deposition
function to predict RDE deposition in supernovae as a function of time.
The treatment provides a straightforward understanding 
of the ultraviolet/optical/infrared (UVOIR) exponential/quasi-exponential
light curves and a partial understanding of the
exponential/quasi-exponential broad band light curves of supernovae
that occur tens to hundreds of days after explosion. 
The treatment is complementary to the detailed modeling of RDE deposition
usually performed by Monte Carlo calculations (e.g.,
H\"oflich, Khokhlov, \&~M\"uller 1992;
Swartz, Sutherland, \&~Harkness 1995)
or, at a less detailed
level, with grey (frequency-integrated) radiative transfer (by Monte
Carlos or by the radiative transfer equation)
(e.g., Colgate, Petschek, \&~Kriese 1980a, b;  Swartz~\etal 1995;
Cappellaro~\etal 1997;  Jeffery 1998a, b).

     The analytic treatment of RDE deposition has previously been
presented by Colgate~\etal (1980a, b), but only for a constant density
supernova model and only briefly.
We are not aware of any longer presentations.
Here we present the treatment in detail and in general:  i.e., we allow for any
sort of homologously expanding supernova model.
Special cases of supernova models are presented too.

     In \S~2 of this paper we briefly discuss the physics of RDE
deposition and exponential/quasi-exponential light curves. 
Section~3 develops our RDE deposition function and parameterized
structural models for SNe~Ia and core-collapse supernovae.
As examples of the simple analytic treatment of RDE deposition,
we examine the RDE deposition and luminosity evolution of
SN~Ia~1992A, SN~II~1987A, and SN~Ic~1998bw in \S\S~4, 5, and~6,
respectively. 
Conclusions appear in \S~7.
Appendix~A presents some useful analytic results for homologously
expanding supernova models with exponential density profiles.
\vskip 2\baselineskip

\centerline{2.\ \  THE RDE DEPOSITION}\nobreak
\centerline{AND EXPONENTIAL/QUASI-EXPONENTIAL LIGHT CURVES}\nobreak
\vskip\baselineskip\nobreak
     The RDE comes principally in the form of $\gamma$-rays and positrons
not counting neutrinos.
(Neutrinos almost entirely escape the supernova ejecta and do not
contribute significantly to the RDE deposition.)
The $\gamma$-ray energy that does not escape the supernova
is converted to fast electron kinetic energy almost
entirely by Compton scattering.
The positrons lose nearly all their kinetic energy slowing down by
collisions and then annihilate to form $\gamma$-rays. 
The lost kinetic energy and the annihilation $\gamma$-ray energy (if it 
does not escape) become 
fast electron kinetic energy also.
There is also a small amount of RDE
in X-rays from inner atomic orbital transitions
and in the kinetic energy of ejected fast atomic electrons: 
both forms being a direct consequence of the nuclear decay.
The X-ray energy (if it does not escape) is also converted to fast
electron kinetic energy by ionizations.
Although the fraction of decay energy in X-rays is very small,
it is likely to become significant at very late times when Compton optical
depth has become very small due to decreasing density, but X-ray optical
depth is still large because of the large X-ray opacity.
The ratio of the effective absorption opacities of the decay X-rays to the
decay $\gamma$-rays (using $0.01\,$MeV as the dividing line between X-rays
and $\gamma$-rays) is of order $10^{\ApJstyle 3}$--$10^{\ApJstyle 4}$.
The decay-ejected fast atomic electrons lose their kinetic energy to
other electrons, of course.
In this paper we will lump the positrons and ejected atomic
electrons together as positron-electron (PE) particles.

     The overwhelmingly dominant decay chain for the observable
epoch of most supernovae is
$^{\ApJstyle 56}$$\rm Ni\to^{\ApJstyle 56}$$\rm Co\to^{\ApJstyle 56}$$\rm Fe$
with \hbox{half-lives} of $6.077\pm0.012$~days and~$77.27\pm0.03$~days
for the first and second decays, respectively.
(Half-life data in this paper is from
Lawrence Berkeley National Laboratory Isotopes Project Web Data Base
1999, hereafter LBL.)
The first decay releases almost all its non-neutrino energy in the form of
$\gamma$-rays and the second in $\gamma$-rays and, in $19\,$\% of
the decays, in positrons (e.g., Browne \&~Firestone 1986;
Huo 1992).  
The $^{\ApJstyle 56}$Ni and $^{\ApJstyle 56}$Co $\gamma$-ray and
mean positron kinetic energy are in the
range $\sim 0.16$--$3.6\,$MeV (e.g., Browne \&~Firestone 1986;  Huo 1992).
The long-lived radioactive species $^{\ApJstyle 57}$Co (half-life
$271.79\pm0.09$ days), $^{\ApJstyle 55}$Fe (half-life $2.73\pm0.03$ years),
and $^{\ApJstyle 44}$Ti (half-life $63\pm3$ years) are likely to
become important at later times:  hundreds to thousands of days after
the explosion.
These species (or their short-lived parents) are synthesized in comparatively
small abundance in the explosion, but their long half-lives gives them
importance at very late times.

    The RDE converted into fast electron kinetic energy is said to be
deposited.
The fast electrons caused by the deposition ionize, excite, and
heat the low-ionization-state supernova plasma.
The deposited RDE is ultimately converted mainly into UVOIR radiation
that escapes the supernova and forms the
observed supernova light curves. 
(The complicated cascade process by which the fast electron kinetic
energy gets converted into other forms of energy is discussed by, e.g.,
Fransson [1994, p.~688ff] and Liu \&~Victor [1994].)
Because radioactive decay is exponential, there is a
{prima facie}        
reason to expect supernova light curves to decline with time at least partially
exponentially.
Historically, it was the exponential (really quasi-exponential) late-time
light curves of SNe~Ia that first suggested radioactive sources for these
supernovae (Baade~\etal 1956).

     Supernova light curves cannot be nearly exactly exponential in
most cases for a number of reasons. 
For core-collapse supernovae there is the aforesaid early shock heat source,
and circumstellar interaction and pulsar sources.
For all supernovae, the UVOIR diffusion time scale starts out much longer
than the dynamic time scale and the radioactive decay time scale.
Thus, deposited RDE and shock heat
energy is initially largely trapped and is released increasingly
rapidly as the supernova expands and decreases in density.
This early trapping rules out seeing the signature in the light curves
of the $^{\ApJstyle 56}$Ni decay entirely
and the early signature of the $^{\ApJstyle 56}$Co decay.
In the case of SNe~Ia, the transition to quasi-exponential light curve
decline at about 60 days after explosion (or about 40 days
after maximum light [i.e., $B$ maximum])
(e.g., Leibundgut 1988;  Leibundgut~\etal 1991b) shows that the
UVOIR trapping and the time period for processing
RDE into UVOIR emission have become small.
Thus a quasi-steady state has been established (or perhaps slightly
later at 70~days after explosion
[Pinto \&~Eastman 1996])  
in which at any instant the entire radiative transfer of the
supernova can be treated by a time-independent calculation to good accuracy.
In the case of core-collapse supernovae, the transition to
the quasi-steady state (exponential/quasi-exponential phase of the
light curves) can occur at various times depending on the nature
of progenitor, but for a massive progenitor typically at of order a
hundred days after explosion as evidenced by SN~1987A (e.g., Bouchet~\etal 1991;
Suntzeff~\etal 1991).

     The late-time light curves (i.e., those from after the
establishment of the quasi-steady state) are not usually truly
exponential despite the effectively instantaneous conversion of RDE to
UVOIR luminosity for two reasons.  
First, the $\gamma$-ray optical depth of the ejecta can cease to
be completely trapping 
before or at about the same time the UVOIR optical depth becomes small.
The escape of $\gamma$-rays from the ejecta adds a non-exponential factor to
the RDE deposition (as we will show in \S\S~3.1 and~3.2). 
Second, one observes light curves in particular frequency bands.
Only the late-time UVOIR bolometric light curve in the case of
complete $\gamma$-ray and PE particle trapping (or complete $\gamma$-ray escape
and complete PE particle trapping) is guaranteed to be exponential.
The individual bands receive the RDE after a more or less complicated
time-dependent distribution by atomic processes.

     Nevertheless, the late-time light curves of supernovae,
particularly SNe~Ia in the $B$ and $V$ bands, often appear very exponential
at least in some phases when they are not in fact exponential to 1st order
in those phases.
We define a 1st order exponential phase to one where
the logarithmic slope of a function or equivalently
its instantaneous half-life
(or instantaneous $e$-folding parameter) is constant to first order at
some point in the phase.
We use the expression quasi-exponential to describe phases of functions which
appear very exponential without being 1st order exponential.
We will adopt the convention that only light curve phases which are 1st order
exponential can be considered truly exponential phases. 

     With older light curve data, it was often possible to fit straight lines
(on semi-logarithmic plots) within the uncertainty to quasi-exponential
light curves.
For example, Doggett \&~Branch (1985) fit a straight line with
a half-life of 44 days to a compilation of SN~Ia blue magnitude data 
for days
$\sim 120$--$320$ after explosion:  the data had a dispersion of
order 1~magnitude (0.4~dex) about the line.
(Here ``blue'' refers to a mix of bands from various older and newer magnitude
systems that sample the blue side of the optical.)
Barbon, Cappellaro, \&~Turatto (1984) found a half-life of 50 days
for a line fit to another compilation of SN~Ia blue magnitude data for days 
$\sim 220$--$420$ after explosion:  the magnitudes had a dispersion of 
order 0.5~magnitude (0.2~dex) about the line. 
Kirshner \&~Oke (1975) fit a straight line to an $AB$ (roughly $B-0.2$)
light curve for SN~Ia~1972E with a half-life of 58~days for
$\sim 98$--$732$ days after explosion.
(We assume a rise time to maximum light of 18~days for SNe~Ia:  see
\S~4.)
Older late-time blue light curve data for SNe~Ia generally showed a half-life of
$\sim 56$ days (e.g., Rust, Leventhal, \&~McCall 1976).
This order of half-life originally suggested the spontaneous fission of
$^{254}$Cf (half-life $60.5\pm0.2$) as a dominant energy source
(Baade~\etal 1956).
The
$^{\ApJstyle 56}$$\rm Ni\to^{\ApJstyle 56}$$\rm Co\to^{\ApJstyle 56}$$\rm Fe$
beta decay chain for SNe~Ia
was put forward by
Pankey (1962) and Colgate \&~McKee (1969).
This RDE source was later accepted for core-collapse supernovae too.
 
     Very accurate modern data shows that SNe~Ia late-time light curves
are not truly exponential, but have slowly decreasing logarithmic decline rates
(i.e., increasing instantaneous half-lives)
(e.g., Cappellaro~\etal 1997).
(That the older SN~Ia decline half-lives cited above increase with
later or longer time coverage is probably not a coincidence.)
The increasing (instantaneous) half-life is easily understood qualitatively.
The half-life falls below the $^{\ApJstyle 56}$Co
half-life because of $\gamma$-ray escape.
But as the $\gamma$-ray escape increases with
decreasing optical depth, the fraction of RDE deposition from the
more strongly trapped PE particles increases and this eventually
drives the half-life back toward the
$^{\ApJstyle 56}$Co half-life asymptotically.
Colgate~\etal (1980a, b) quantitatively demonstrated the
increasing RDE deposition half-life in numerical calculations.
More recently Cappellaro~\etal (1997) have done the same demonstration
with light curve fits to modern SN~Ia data.
In \S~3.2, we show how both decreasing and increasing RDE deposition
half-lives can arise analytically.
In \S~3.3, we show why the signature of increasing half-life 
is what one sees in the late-time light curves of SNe~Ia.
Both decreasing and increasing half-life phases can (but not necessarily
will) have a signature in core-collapse
supernova light curves (see \S~3.4).

     Positrons and very fast electrons can also move in supernova ejecta and
can escape.
They are in fact thought to be kept from traveling far by magnetic
fields:  possibly tangled magnetic fields.
Positron transport may be an important process in supernovae
and it is being
actively investigated (e.g., Colgate~\etal 1980a, b;  Chan \&~Lingenfelter
1993;  Cappellaro~\etal 1997;  Ruiz-Lapuente 1997;  Milne, The,
\&~Leising 1997;
Ruiz-Lapuente \&~Spruit 1998).
However, it is certain that positrons and fast electrons are
much more strongly trapped than $\gamma$-rays.
Since our interest in this paper is in studying the most basic part of
RDE deposition analytically, we make the simplifying assumption
that PE particle kinetic
energy is deposited local to the creation of the PE particles.

     At very late times, hundreds of days after explosion, the quasi-steady
state is predicted to breakdown (e.g., Axelrod 1980, p.~48;
Fransson \&~Kozma 1993;  Fransson, Houck, \&~Kozma 1996).
This is because the time scale for the fast electrons to lose energy and
time scale for recombination cease to be short compared
to the decay and dynamic time scales.
Thus deviation from true 1st order exponential decay or quasi-steady state
quasi-exponential decay in supernova light curves is to be expected
for this reason if no other.
There is also an observational problem at very late times.
At some point an ``infrared catastrophe'' should occur
(e.g., Axelrod 1980, p.~70;  Fransson~\etal 1996) in which the bulk
of the emission shifts from
the optical to the infrared where it is usually much less observable.
Thus even if the UVOIR bolometric luminosity remains
exponential/quasi-exponential,
the observable light curves might show sudden, sharp declines.
\vskip 2\baselineskip

\centerline{3.\ \  THE RDE DEPOSITION FUNCTION}
\vskip\baselineskip\nobreak
      In the present case we are interested only in the global RDE 
deposition per unit time (i.e., the deposition function)
for a supernova principally as a means for understanding supernova
light curve behavior.
First, we will present a fairly exact, general expression
$D_{\ApJstyle\rm g}(t)$ for deposition and show how it depends on time
(\S~3.1).
Then in \S~3.2 we will derive $N_{\ApJstyle\rm Ni}^{\ApJstyle *}(t)$:
an approximate, analytic, normalized
deposition function for $^{\ApJstyle 56}$Ni and its daughter
$^{\ApJstyle 56}$Co.
It is $N_{\ApJstyle\rm Ni}^{\ApJstyle *}(t)$ and its subcomponent the absorption
function $f(x)$ that are useful for analytic insight into
the RDE deposition and light curve behavior.
In \S~3.3, we posit a parameterized SN~Ia model
and in \S~3.4 a parameterized core-collapse supernova model to which
$N_{\ApJstyle\rm Ni}^{\ApJstyle *}(t)$ is applied.
\vskip\baselineskip

\centerline{3.1. \it The Function $D_{\ApJstyle\rm g}(t)$}\nobreak
\vskip .5\baselineskip\nobreak
     A great thing about $\gamma$-ray transfer in supernovae
is that the non-scattering component of the source function (i.e., the
radioactive decay source function) is independent of $\gamma$-ray transfer.
This means that detailed $\gamma$-ray transfer with Compton
scattering can be treated by a non-iterative Monte Carlo calculation.
But a second great thing is that the actual complicated process of
Compton scattering and photon degradation in energy can be approximated
by a grey, pure absorption opacity treatment with an uncertainty
of only a few percent at most in local and global energy absorption
(Colgate~\etal 1980a, b;
Sutherland \&~Wheeler 1984;
Ambwani \&~Sutherland 1988;
Swartz~\etal 1995;
Jeffery 1998a, b).
This means that $\gamma$-ray transfer and absorption can be reduced to
simple integration and that is what we will do.
We also will assume that the $\gamma$-ray transfer time can
be treated as time-independent:
i.e., the time scale for $\gamma$-ray transfer is much shorter than
the supernova dynamical time scale and decay time scale.
This assumption can begin to fail hundreds of days after explosion
(e.g., Jeffery 1998b).

    The radiative transfer of X-rays must be treated separately
from $\gamma$-ray transfer, since X-rays face a much higher absorption
opacity than $\gamma$-rays.
But since the small amount of energy in X-rays is only important at very
late times (hundreds of days after explosion) we will not consider them
further in this section:  i.e., we will not include them in
the $D_{\ApJstyle\rm g}(t)$ or $N_{\ApJstyle\rm Ni}^{\ApJstyle *}(t)$
functions.

    With aforesaid simplifications an expression for $D_{\ApJstyle\rm g}(t)$
suitable for
our purposes can just be written down from simple radiative transfer:
$$ D_{\ApJstyle\rm g}(t)
 =\sum_{\ApJstyle i}\int d^{\ApJstyle 3}r\,\rho(\vec r,t)
    \epsilon_{\ApJstyle i}
   \left\{ f_{\ApJstyle \rm PE}
         +f_{\ApJstyle \rm ph}\oint{d\Omega\over4\pi}\,
              \left[1-\exp\left(-\tau\right)\right] \right\}_{\ApJstyle i}
                                     \,\, ,  \eqno(\eqn)$$
where 
the sum is over radioactive species $i$,
the first integral is over all volume,
$\rho$ is mass density,
$\epsilon_{\ApJstyle i}$ is the radioactive species RDE 
   production (not counting neutrinos which are all assumed to escape)
   per unit time per unit mass,
$\rho\epsilon_{\ApJstyle i}$ is the same except per unit volume, 
$f_{\ApJstyle\rm PE}$ is the fraction of RDE (not counting
   neutrino energy) that goes into
fast PE kinetic energy (assumed to be all locally deposited),
$f_{\ApJstyle\rm ph}$ is the fraction of RDE (counting positron
   annihilation energy, but not counting
   neutrino energy) that goes into photons,
the second integral averages over all solid angle for each point $\vec r$,
$\tau$ is the effective $\gamma$-ray absorption opacity optical depth in
  a given direction from $\vec r$ to the surface of the supernova
  (i.e., the optical depth of the beam path), 
and
$\left(1-e^{\ApJstyle -\tau}\right)$ is the absorption probability
for photons that travel from $\vec r$ to the surface.

     Supernovae after very early times are in homologous expansion where
the velocities of all mass elements are constants, and the internal gas
and bulk kinetic energy lost to $P\,dV$ work is negligible.
(Exceptions to homologous expansion probably exist and have important
consequences [e.g., Woosley 1988], but we will not consider these
in this paper.)
The radial position $\vec r$ of a mass element in homologous expansion is
given by
$$  \vec r=\vec v t    \,\, ,      \eqno(\eqn)$$
where $\vec v$ is the mass element velocity and $t$ is the
time since explosion:  the initial radii of the mass elements are negligible
after very early times.
Thus velocity can be used as a comoving coordinate.
The density at any velocity decreases as $t^{\ApJstyle -3}$.
Therefore we can write
$$ \rho(\vec r,t)=\rho(\vec v,t)=\rho_{\ApJstyle 0}(\vec v)
     \left({t_{\ApJstyle 0}\over t}\right)^{\ApJstyle 3} \,\, , \eqno(\eqn)$$
where $\rho_{\ApJstyle 0}(\vec v)$ is the density at $\vec v$
at a fiducial time $t_{\ApJstyle 0}$.
Note we are not assuming spherical symmetry for the supernova.

     Using the homologous condition equations~(\showeqn -1) and~(\showeqn -0),
equation~(\showeqn -2) can be rewritten as
$$ D_{\ApJstyle\rm g}(t)=\sum_{\ApJstyle i}
  \int d^{\ApJstyle 3}v\,t_{\ApJstyle 0}^{\ApJstyle 3}
   \rho_{\ApJstyle 0}(\vec v)\epsilon_{\ApJstyle i}
   \left\{ f_{\ApJstyle \rm PE}
         +f_{\ApJstyle \rm ph}\oint{d\Omega\over4\pi}\,
              \left[1-\exp\left(-\tau\right)\right] \right\}_{\ApJstyle i}
               \,\, ,  \eqno(\eqn)$$
where the first integral is now over all velocity space volume.
The deposition function now can be seen to depend on time in only two
ways.
First, through $\epsilon_{\ApJstyle i}$ which in general is a sum of
the exponential terms for radioactive species $i$.
For beta decay from a species synthesized in the explosion
$$ \epsilon_{\ApJstyle i}=
    X_{\ApJstyle i}^{\ApJstyle\rm ini}C_{\ApJstyle i}
         \exp\left(-t/t_{\ApJstyle e,i}\right)  \,\, , \eqno(\eqn)$$
where $X_{\ApJstyle i}^{\ApJstyle\rm ini}$ is initial mass fraction of
species $i$,
$C_{\ApJstyle i}$ is an energy generation rate coefficient,
and $t_{\ApJstyle e,i}$ is $e$-folding time (i.e.,
half-life divided by $\ln[2]$).
For beta decay from a species whose parent was synthesized in the explosion
$$ \epsilon_{\ApJstyle i}=X_{\ApJstyle {\rm pa}(i)}^{\ApJstyle\rm ini}
  B_{\ApJstyle i}
   \left[\exp\left(-t/t_{\ApJstyle e,i}\right)
         -\exp\left(-t/t_{\ApJstyle e,{\rm pa}(i)}\right)\right] 
                                                      \,\, , \eqno(\eqn)$$
where
${\rm pa}(i)$ identifies the radioactive parent of radioactive species $i$,
$X_{\ApJstyle {\rm p}(i)}^{\ApJstyle\rm ini}$ is initial mass fraction of
the parent,
$B_{\ApJstyle i}$ is an energy generation rate coefficient,
$t_{\ApJstyle e,i}$ is $e$-folding time for species $i$,
and
$t_{\ApJstyle e,{\rm pa}(i)}$ is $e$-folding time for the parent.
The $C$ and $B$ coefficients are given by
$$ C_{\ApJstyle i}={  Q_{\ApJstyle {\rm ph+PE,}i}
   \over m_{\ApJstyle\rm amu} A_{\ApJstyle i} t_{\ApJstyle e,i} }
\qquad{\rm and}\qquad
 B_{\ApJstyle i}={ Q_{\ApJstyle {\rm ph+PE,}i}
     \over m_{\ApJstyle\rm amu} A_{\ApJstyle {\rm pa}(i)} }
     { 1\over \left(t_{\ApJstyle e,i}-t_{\ApJstyle e,{\rm pa}(i)}\right) }
                                             \,\, ,  \eqno(\eqn)$$
where $Q_{\ApJstyle {\rm ph+PE,}i}$ is the mean photon plus fast PE 
kinetic energy per decay,
$A_{\ApJstyle i}$ is atomic mass, 
$A_{\ApJstyle {\rm pa}(i)}$ is the parent's atomic mass, 
and $m_{\ApJstyle\rm amu}$ is the atomic mass unit (amu).
For our simplified deposition treatment introduced in \S~3.2, we consider
only $^{\ApJstyle 56}$Ni and
$^{\ApJstyle 56}$Co decays since these are overwhelming the most important
until very late times, hundreds of days after explosion.
The parameters for $^{\ApJstyle 56}$Ni and $^{\ApJstyle 56}$Co decays are
given in Table~1.

     The second time dependence of $D_{\ApJstyle\rm g}(t)$ comes through
the optical depths of the beam paths: 
$$ \tau=\int_{\ApJstyle\rm em}^{\ApJstyle\rm sur}ds\,
          \kappa\rho\left[\vec v(s),t\right]
       =\left({t_{\ApJstyle 0}\over t}\right)^{\ApJstyle 2}
        \int_{\ApJstyle\rm em}^{\ApJstyle\rm sur}dv_{\ApJstyle s}\,
           t_{\ApJstyle 0}
\kappa\rho_{\ApJstyle 0}\left[\vec v\left(v_{\ApJstyle s}\right)\right] 
                                                   \,\, , \eqno(\eqn)$$
where we have used the homologous expansion condition equations~(\showeqn -6)
and~(\showeqn -5) to get the second
expression, ``em'' and ``sur'' specify the point of emission and
surface, respectively, in either space or velocity,
$s$ and $v_{\ApJstyle s}$ are beam path length in physical space and
velocity space, respectively,
and $\kappa$ is the effective absorption opacity.
Note the beam path is general:  i.e., not usually radial.
As can be seen from equation~(\showeqn -0),
the optical depth between any two points in velocity
space decreases as $t^{\ApJstyle -2}$. 

     Swartz~\etal (1995) showed for the $\gamma$-rays important in
supernovae that
$$ \kappa\approx\kappa_{\ApJstyle\rm c}\mu_{\ApJstyle e}^{\ApJstyle -1}
                                              \,\, .         \eqno(\eqn)$$
The overwhelmingly dominant $\gamma$-ray opacity in supernovae is
Compton scattering (e.g., Swartz~\etal 1995;  Jeffery 1998b).
Thus the opacity is largely determined by the number density of electrons
per unit mass and this accounts for the factor of the inverse 
of the mean atomic mass per electron
$$ \mu_{\ApJstyle e}^{\ApJstyle -1}=
  \sum_{\ApJstyle i}{X_{\ApJstyle i}Z_{\ApJstyle i}\over A_{\ApJstyle i}}
                                                  \,\, ,       \eqno(\eqn)$$
where the sum is over all elements $i$,
$X_{\ApJstyle i}$ is mass fraction of element $i$,
$Z_{\ApJstyle i}$ is nuclear change (since all electrons are approximately
free electrons to the $\gamma$-rays important in supernovae),
and
$A_{\ApJstyle i}$ is again atomic mass.
For $^{\ApJstyle 56}$Co,
the $\kappa_{\ApJstyle\rm c}$ parameter ranges between
$\sim 0.05$ to $0.065\,{\rm cm^{\ApJstyle 2}\,g^{\ApJstyle -1}}$
and
for $^{\ApJstyle 56}$Ni, between
$\sim 0.06$ to $0.1\,{\rm cm^{\ApJstyle 2}\,g^{\ApJstyle -1}}$
(Swartz~\etal 1995;  Jeffery 1998a, b).
The variation of the $\kappa_{\ApJstyle\rm c}$ parameter depends 
on the global optical depth structure of the supernova:
the low values are for the optically thin limit;  the large values for
the optically thick limit.
(Note $\kappa_{\ApJstyle\rm c}$ actually has a very weak composition
dependence too that we neglect here.) 
We will not try to incorporate the weak, but complex, global optical
depth structure dependence (and hence time dependence)
of $\kappa_{\ApJstyle\rm c}$ 
or variations in $\mu_{\ApJstyle e}$ in our formalism
directly.
We will simply try to choose appropriate values of
$\kappa$ for the particular cases we examine
and treat those values as constants for all locations and epochs.

     Equation~(\showeqn -6) would be vastly simplified if we could
replace the location-, direction- and time-dependent optical depth by a
mean optical depth that was only time-dependent.
Given all our assumptions about $\gamma$-ray transfer,
an exact mean $\bar\tau$ would be obtained from 
$$ 1-\exp\left(-\bar\tau\right)=
{\displaystyle \sum_{\ApJstyle i}
  \int d^{\ApJstyle 3}v\,
   \rho_{\ApJstyle 0}(\vec v)\epsilon_{\ApJstyle i}
         \oint{d\Omega\over4\pi}\,
              \left[1-\exp\left(-\tau\right)\right]_{\ApJstyle i}
\over \displaystyle
 \sum_{\ApJstyle i}
  \int d^{\ApJstyle 3}v\,
   \rho_{\ApJstyle 0}(\vec v)\epsilon_{\ApJstyle i}
}                                             \,\, .  \eqno(\eqn)$$
Evaluating this exact mean $\bar\tau$ is, of course, tantamount to
solving the exact deposition problem. 
We note that $\bar\tau$ has a complex time dependence through the
$\epsilon_{\ApJstyle i}$'s and the exponentials of the individual 
optical depths, not the simple $t^{\ApJstyle -2}$ time dependence of
an individual optical depth given by equation~(\showeqn -3).
In the optically thin limit equation~(\showeqn -0) reduces to
$$ \bar\tau=
{\displaystyle \sum_{\ApJstyle i}
  \int d^{\ApJstyle 3}v\,
   \rho_{\ApJstyle 0}(\vec v)\epsilon_{\ApJstyle i}
         \oint{d\Omega\over4\pi}\, \tau_{\ApJstyle i}
\over \displaystyle
 \sum_{\ApJstyle i}
  \int d^{\ApJstyle 3}v\,
   \rho_{\ApJstyle 0}(\vec v)\epsilon_{\ApJstyle i}
}                                             \,\, .  \eqno(\eqn)$$
In this limiting case when there is a only a single radioactive species,
$\bar\tau$ does have the simple $t^{\ApJstyle -2}$ time dependence
since the individual $\tau_{\ApJstyle i}$'s occur linearly in
equation~(\showeqn-0) and
the radioactive decay time dependence cancels. 
\vskip\baselineskip

\centerline{3.2. \it The Function
            $N_{\ApJstyle\rm Ni}^{\ApJstyle *}(t)$}\nobreak
\vskip .5\baselineskip\nobreak
     So far we have developed, given the assumptions we have made,
an accurate general expression for RDE deposition,
$D_{\ApJstyle\rm g}(t)$ (see \S~3.1).  
The normalized version of $D_{\ApJstyle\rm g}(t)$ restricted to
having only $^{\ApJstyle 56}$Ni at time zero is denoted
$N_{\ApJstyle\rm Ni}(t)$.
We will now develop an approximate version of $N_{\ApJstyle\rm Ni}(t)$,
denoted $N_{\ApJstyle\rm Ni}^{\ApJstyle *}(t)$, by
approximating some of
the components of equation~(\showeqn -8) for $D_{\ApJstyle\rm g}(t)$
(restricted to $^{\ApJstyle 56}$Ni at time zero)
and normalizing.
Since only a trace of $^{\ApJstyle 56}$Co is produced in a supernova
explosion
and other radioactive species are unimportant until hundreds of days
after the explosion, the assumption that there is
only $^{\ApJstyle 56}$Ni at time zero is excellent until 
very late times. 
The absolute deposition in a supernova model is, of course,
predicted from $N_{\ApJstyle\rm Ni}^{\ApJstyle *}(t)$ or
$N_{\ApJstyle\rm Ni}(t)$ with the scale set by specifying the initial
$^{\ApJstyle 56}$Ni mass or by fitting the absolute UVOIR luminosity
of an observed supernova.
 
     We assume that $^{\ApJstyle 56}$Ni decays to negligible abundance while 
the optical depth is still very large:  viz., the
$^{\ApJstyle 56}$Ni $\gamma$-rays are completely trapped. 
This is a fairly good assumption for all supernovae.
It begins to fail for SNe~Ia which become optically thin rather quickly.
But by the time the SN~Ia quasi-steady state phase (which is our main interest)
has begun, the $^{\ApJstyle 56}$Ni is negligible.
Given the complete trapping assumption for $^{\ApJstyle 56}$Ni $\gamma$-rays,
we replace $1-\exp(-\tau)$ for $^{\ApJstyle 56}$Ni by 1.

     Note that the effective absorption opacity
and thus optical depth scale for
$^{\ApJstyle 56}$Ni $\gamma$-rays are larger by a factor of about 1.5
in the optically thick limit than those for $^{\ApJstyle 56}$Co (see \S~3.1).
But in qualitative discussion below, we will treat the two optical
depth scales as being same.

      For $^{\ApJstyle 56}$Co, we
approximate the time-, location-, and direction-dependent optical depth $\tau$
given by equation~(\showeqn -4) by a location- and direction-independent
characteristic optical depth
$$ \tau_{\ApJstyle\rm ch}=\tau_{\ApJstyle\rm ch,0}
      \left({t_{\ApJstyle 0}\over t}\right)^{\ApJstyle 2}
      =x^{\ApJstyle -2} \,\, ,                          \eqno(\eqn)$$
where $\tau_{\ApJstyle\rm ch,0}$ is the characteristic optical depth
at the fiducial time $t_{\ApJstyle 0}$ (i.e., is the fiducial [characteristic]
optical depth) and
$$ x={1\over \sqrt{\tau_{\ApJstyle\rm ch,0}} }
     \left({t\over t_{\ApJstyle 0}}\right)                    \eqno(\eqn)$$
is a reduced time.

     Since the optically thin $\bar\tau$
has the same time dependence as $\tau_{\ApJstyle\rm ch}$
(assuming only $^{\ApJstyle 56}$Ni at time zero),
we make the optically thin
$\bar\tau$ our choice for $\tau_{\ApJstyle\rm ch}$ for
SNe~Ia and other low-mass/rapidly-expanding supernovae. 
These supernovae can be well observed in the $\gamma$-ray optically thin phase.
If we can establish the optically thin $\bar\tau$ exactly, then
$N_{\ApJstyle\rm Ni}^{\ApJstyle *}(t)$ will agree exactly with
$N_{\ApJstyle\rm Ni}(t)$ in the optically thin limit. 
For massive and/or slowly expanding
core-collapse supernovae, the fully optically thin phase may be
later than observations or later than the period in which
the simple $N_{\ApJstyle\rm Ni}(t)$ behavior is unperturbed by
RDE deposition from sources other than $^{\ApJstyle 56}$Ni
and $^{\ApJstyle 56}$Co.  
For such supernovae, $\tau_{\ApJstyle\rm ch}$ should probably
be chosen to be most characteristic of the supernova epoch one is
trying to model.
In order to determine the $\tau_{\ApJstyle\rm ch}$ one needs to
posit supernova models.
We do this for SNe~Ia and core-collapse supernovae in \S\S~3.3 and~3.4,
respectively.

     With the aforesaid approximations and dividing through by its value
at $t=0$ (which is a maximum), equation~(\showeqn -10) reduces to
our approximate normalized deposition function
$$ N_{\ApJstyle\rm Ni}^{\ApJstyle *}(t)=
   \exp\left(-t/t_{\ApJstyle e,{\rm Ni}}\right)  
  +G\left[\exp\left(-t/t_{\ApJstyle e,{\rm Co}}\right)
      -\exp\left(-t/t_{\ApJstyle e,{\rm Ni}}\right)\right]f[x(t)]
                       \,\, , \eqno(\eqn)$$
where
$G=B_{\ApJstyle\rm Co}/C_{\ApJstyle\rm Ni}=0.184641$ (to more digits
than significant for numerical consistency) 
and
$$ f(x)\equiv f_{\ApJstyle\rm PE} + f_{\ApJstyle\rm ph}\left[
  1-\exp\left(-x^{\ApJstyle -2}\right)\right]  \eqno(\eqn)$$
is what we call the absorption function (for $^{\ApJstyle 56}$Co).
The values of $f_{\ApJstyle\rm PE}$ and $f_{\ApJstyle\rm ph}$
for $^{\ApJstyle 56}$Co are given in Table~1.
Note that $N_{\ApJstyle\rm Ni}^{\ApJstyle *}(t)$ has only one
free parameter $t_{\ApJstyle 0}\sqrt{\tau_{\ApJstyle\rm ch,0}}$ which
relates reduced and real time through equation~(\showeqn -2).
This parameter reduces to just $t_{\ApJstyle 0}$ because we
will always set $\tau_{\ApJstyle\rm ch,0}=1$ since this conveniently
makes $t_{\ApJstyle 0}$ roughly the time of transition between
the optically thick and thin epochs.

     Figure~1a shows a logarithmic plot with the deposition function
$N_{\ApJstyle\rm Ni}^{\ApJstyle *}(t)$ for a range of
$t_{\ApJstyle 0}$ values (with $\tau_{\ApJstyle\rm ch,0}=1$ of course).
Figure~1b shows a logarithmic plot of the absorption function
$f(x)$ and three exponential fits (which are linear on a logarithmic plot)
that we discuss below.

    In Figure~1a, the complete $\gamma$-ray trapping deposition curve is the
curve $N_{\ApJstyle\rm Ni}^{\ApJstyle *}(t)$ approaches as
$t_{\ApJstyle 0}\to\infty$.
After the $^{\ApJstyle 56}$Ni contribution becomes negligible the
complete trapping curve decays with
the $^{\ApJstyle 56}$Co half-life.
The complete $\gamma$-ray escape deposition curve is the curve that 
$N_{\ApJstyle\rm Ni}^{\ApJstyle *}(t)$ approaches as $t_{\ApJstyle 0}\to0$.
The complete escape curve is only a complete escape curve for
$^{\ApJstyle 56}$Co $\gamma$-rays:  it is a complete trapping curve for 
$^{\ApJstyle 56}$Ni $\gamma$-rays.
The late-time asymptotic limit of the complete escape curve and all the
curves, except for the complete trapping curve, is just
$Gf_{\ApJstyle\rm PE}\exp\left(-t/t_{\ApJstyle e,{\rm Co}}\right)$ which also
decays with the $^{\ApJstyle 56}$Co half-life. 
Note that the smaller $t_{\ApJstyle 0}$, the more rapidly the ejecta
becomes optically thin and faster the $N_{\ApJstyle\rm Ni}^{\ApJstyle *}(t)$
approaches the complete escape curve.
We display the complete trapping and escape curves
on all subsequent deposition figures for convenient reference.
 
    The $N_{\ApJstyle\rm Ni}^{\ApJstyle *}(t)$ curves for
$t_{\ApJstyle 0}\lapprox t_{\ApJstyle e,{\rm Ni}}$~days are not
physically realistic.
Supernovae with such small $t_{\ApJstyle 0}$ values (and there is
no evidence that there are such supernovae) would become optically
thin before the $^{\ApJstyle 56}$Ni contribution had become negligible.
We have assumed that $^{\ApJstyle 56}$Ni becomes negligible before
the complete trapping phase is over.
We merely display the small $t_{\ApJstyle 0}$ cases to help
demonstrate the trend with varying $t_{\ApJstyle 0}$. 

     An exact exponential cannot be a sum of two terms, unless they have the
same $e$-folding parameter.
Thus $N_{\ApJstyle\rm Ni}^{\ApJstyle *}(t)$ cannot be an exact exponential.
However, it very closely approximates an exponential in some cases.
To dispense with it at once, consider the unphysical case (given
our assumptions) where $t_{\ApJstyle 0}\lapprox t_{\ApJstyle e,{\rm Ni}}$.
In this case, the second term of equation~(\showeqn -1) decreases
quickly (after a very short growth phase from zero) and for awhile
is small relative to the first term
(a $^{\ApJstyle 56}$Ni exponential).
Thus, there will period of nearly exponential decay with
almost the $^{\ApJstyle 56}$Ni half-life before
the $^{\ApJstyle 56}$Ni exponential 
rather abruptly becomes negligible and $N_{\ApJstyle\rm Ni}^{\ApJstyle *}(t)$
closely approaches its late-time asymptotic limit.

     The $^{\ApJstyle 56}$Ni nearly-exponential phase of
$N_{\ApJstyle\rm Ni}^{\ApJstyle *}(t)$ is actually not a 1st order
exponential phase (see definition in \S~2) although it can be very
exponential for small enough $t_{\ApJstyle 0}$ as Figure~1a shows.
A sum of exponentials with different $e$-folding parameters can only
be considered a 1st order exponential if all the exponential terms,
except the dominant exponential term, are
exponentially small compared to the dominant term. 
By being exponentially small, we mean that an exponential term
can be approximated by its value at infinity where it is a constant
zero to all orders in a series expansion about infinity.
The non-$^{\ApJstyle 56}$Ni exponential terms in the
$^{\ApJstyle 56}$Ni nearly-exponential phase are never really exponentially
small compared to the $^{\ApJstyle 56}$Ni exponential term even
for $t_{\ApJstyle 0}\to0$. 
In particular, it is easy to show that the instantaneous $e$-folding time of
the complete escape curve (the $t_{\ApJstyle 0}\to0$ case of
$N_{\ApJstyle\rm Ni}^{\ApJstyle *}(t)$) is never a first order constant
until the $^{\ApJstyle 56}$Ni exponentials are exponentially small. 

     Given that $^{\ApJstyle 56}$Ni is exponentially small, the
sufficient condition for $N_{\ApJstyle\rm Ni}^{\ApJstyle *}(t)$ being
a 1st order exponential is that $f(x)$ be either a 1st order constant or
a 1st order exponential itself.
The $f(x)$ function has two stationary points:  a maximum of 1 at $x=0$ and
a minimum of $f_{\ApJstyle\rm PE}$ at $x=\infty$.
These give what we will call the early and late 1st order exponential phases,
respectively.
Both, of course, have the half-life of $^{\ApJstyle 56}$Co.

     In order to realize the early 1st order exponential phase
the complete trapping phase for $^{\ApJstyle 56}$Co $\gamma$-rays
must continue after the $^{\ApJstyle 56}$Ni contribution is exponentially
small:
i.e., $f(x)$ must not differ significantly from 1 until after
the $^{\ApJstyle 56}$Ni contribution is exponentially small. 
(Note
$\left|\log[f(x)]\right|\leq0.001$ for $x\leq0.4$
and
$f(x)>0.99$ for $x\leq0.46$.)
In Figure~1a, only the 
$N_{\ApJstyle\rm Ni}^{\ApJstyle *}(t)$ deposition curve for
$t_{\ApJstyle 0}=200$~days has
a noticeable early 1st order exponential phase.
An early 1st order exponential phase may or may not have a direct signature
in light curves depending on whether or not
it lasts into the quasi-steady state period.

     After the $^{\ApJstyle 56}$Ni contribution has become 
exponentially small or after the early 1st order exponential phase
if it exists,
$N_{\ApJstyle\rm Ni}^{\ApJstyle *}(t)$ will decline more rapidly than
an exponential with $^{\ApJstyle 56}$Co half-life until very late times
(on the $t_{\ApJstyle 0}$ time scale)
(as Fig.~1a shows)
because of increasing $\gamma$-ray escape from the ejecta.
But at very late times the $\gamma$-ray escape probability approaches 1
and $1-\exp\left(-x^{\ApJstyle -2}\right)\to x^{\ApJstyle -2}\to0$.
Thus $f(x)$ will asymptotically approach
its minimum value at infinity, $f_{\ApJstyle\rm PE}=0.0320$.
In the asymptotic limit,
deposition will be at $^{\ApJstyle 56}$Co decay rate and  
entirely due to the decay PE particle kinetic energy
which we assume to be entirely locally deposited:  see \S~2 for a caveat
about this assumption.
(Recall also that we are neglecting X-rays.) 
As the asymptotic limit is approached one has a
late 1st order exponential phase as the curves with smaller $t_{\ApJstyle 0}$
values in Figure~1a suggest.
The approach to the late 1st order exponential phase is
slow on the $t_{\ApJstyle 0}$ time scale
because the fractional contribution of $\gamma$-rays at late times
declines only as $(t_{\ApJstyle 0}/t)^{\ApJstyle 2}$.
On the time scale of the $^{\ApJstyle 56}$Co half-life the approach
can be slow or fast, of course, as Figure~1a shows.

     Although there is no sharp transition to the late 1st order
exponential phase,
the time when the $\gamma$-ray and PE deposition are equal
can be taken as a conventional transition $x_{\ApJstyle\rm tr}$.
When the ratio of PE to $\gamma$-ray deposition is $R$, the reduced time 
(assuming the $^{\ApJstyle 56}$Ni contribution to deposition is negligible) is
$$ x(R)=
\sqrt{-1\over\ln\left[1-f_{\ApJstyle\rm PE}/
                 \left(f_{\ApJstyle\rm ph}R\right)\right]}
\approx\sqrt{f_{\ApJstyle\rm ph}R\over f_{\ApJstyle\rm PE} }
                                             \,\, .\eqno(\eqn)$$
For $R=1$, we have $x_{\ApJstyle\rm tr}\approx5.454099$.
At $x_{\ApJstyle\rm tr}$, $f(x)$ is $0.301030$~dex
(or a factor of $2$)   
above its asymptotic value and
at $2x_{\ApJstyle\rm tr}$, $0.098007$~dex (or a factor of $\sim 1.25$).

      The slow approach to the late 1st order exponential phase gives rise
to a continuum of quasi-exponential phases as Figure~1a suggests. 
As an example
(chosen for reason:  see \S~4),  
we have fitted an exponential 
$K_{\ApJstyle\rm coef}\exp\left(-x/x_{\ApJstyle e}\right)$
to $f(x)$ in the $x$-range $[2,3]$.
The fit is shown in Figure~1b and its parameters are given in Table~2.
As can be seen in Figure~1b the fit closely matches $f(x)$ over
the $x$-range $[2,3]$:  the discrepancy between the two is
always less than $0.01$~dex or about $2.5\,$\%  
in the $x$-range $[2,3]$.

     Would discrepancies from exponential behavior arising from
the curvature of the $\ln\left[f(x)\right]$ function be
seen in late-time supernova light curves?
%
Late-time light curve data for the various bands
from before circa 1980 often had
uncertainties of order $0.5$ magnitudes or $0.2$~dex.
Clearly, discrepancies of $\sim 0.01$~dex or even
$\sim 0.1$~dex between late-time light curves and
exponential fits  
would be hard to notice or assess for significance.  
As one observed for longer, larger discrepancies could arise.
But the later the light curve, the fainter the supernova, the more
the uncertainty in the observation. 
Modern, accurate late-time light curves for SNe~Ia do show discrepancies
from exponential behavior that are a least partly due to the
non-exponential nature of $f(x)$ in its slow approach to
its late asymptotic limit (e.g., Cappellaro~\etal 1997;
see also \S~4). 
The logarithmic decline rates of the light curves
become smaller with time.
It is now recognized that the behavior of
late-time SN~Ia light curves from say day~60 after the explosion
until very late times is only quasi-exponential.

     Besides the early and late 1st order exponential phases,
$N_{\ApJstyle\rm Ni}^{\ApJstyle *}(t)$ predicts that there can be a
third 1st order exponential phase.
This is when $f(x)$ itself becomes 1st order exponential.
In Figure~1b, it can be seen that $\ln[f(x)]$ has an inflection point
near $x=1$ where the logarithmic slope becomes 1st order constant
and $f(x)$ becomes a 1st order exponential.
Assuming the $^{\ApJstyle 56}$Ni contribution has become exponentially small,
then before the inflection point
the RDE deposition logarithmic decline rate increases (i.e.,
the instantaneous half-life decreases) and after, the RDE deposition
logarithmic decline rate decreases
(i.e., the instantaneous half-life increases).

     The inflection point $x_{\ApJstyle\rm infl}$
can be solved for numerically from an iteration relation
obtained from the second derivative of $\ln[f(x)]$:
$$  x_{\ApJstyle i}=\sqrt{\left({2\over3}\right)
     \left[1+{f_{\ApJstyle\rm ph}
              \exp\left(-x_{\ApJstyle i-1}^{\ApJstyle -2}\right)
              \over f\left(x_{\ApJstyle i-1}\right) }
     \right] }                         \,\, ,         \eqno(\eqn)$$
where $i$ is iteration number.
One obtains $x_{\ApJstyle\rm infl}\approx1.040765$
and $f\left(x_{\ApJstyle\rm infl}\right)\approx0.615465$.
The parameters for an exponential fitted to $f(x)$ at the inflection point
are given in Table~2.
In the $x$-range $[0.8,1.35]$, the maximum deviation of $f(x)$
from the inflection point fit is
$\sim 0.004$~dex ($0.01$ magnitudes).
In the $x$-range $[0.56,1.82]$, however, the maximum deviation
has grown to $\sim 0.04$~dex ($0.1$ magnitudes).

     An inflection point 1st order exponential phase will
be realized in RDE deposition if 
$^{\ApJstyle 56}$Ni abundance has become exponentially small
by (for the sake of definiteness) $x\approx1.35$ (and, of course, provided
that $N_{\ApJstyle\rm Ni}^{\ApJstyle *}(t)$ adequately describes
deposition).
In Figure~1a, the $N_{\ApJstyle\rm Ni}^{\ApJstyle *}(t)$ deposition
curves for $t_{\ApJstyle 0}$ values 200, 100, and~60~days all show
inflection point 1st order exponential phases.
The $t_{\ApJstyle 0}=30$~day curve does not obviously show one.
The $^{\ApJstyle 56}$Ni fraction of RDE production (not deposition)
falls below $50\,$\% by day~18, $20\,$\% by day~30,
$7.6\,$\% by day~40 ($x=4/3$ for $t_{\ApJstyle 0}=30$~day),
and $1\,$\% of the RDE at day~60. 
From these numbers, it is clear we would not expect much of an inflection
point 1st order exponential phase for $t_{\ApJstyle 0}\lapprox30$~days.

     A direct
signature of a 1st order exponential RDE deposition in the light curves
appears only if the supernova is in quasi-steady state phase.
If there is an inflection point 1st order exponential phase entirely
in the quasi-steady state phase, 
observers measuring a UVOIR bolometric light curve or a tracer
of the same in a largish region about
the inflection point would probably infer a slower exponential
decline than that given by the exponential fit at
the inflection point.
This is because the finite uncertainty in their data would hide
some of the logarithmic absorption function curvature. 
In Table~2 we give narrow (n), middling (m), and broad (b) inflection region
fits to $f(x)$.
The maximum discrepancy of these fits from $f(x)$ are
$\sim 0.014$~dex ($0.035$ magnitudes),
$\sim 0.028$~dex ($0.07$ magnitudes),
and
$\sim 0.04$~dex ($0.1$ magnitudes).
All but the very best modern supernova photometry might be unable to
show systematic deviations of $0.035$ or $0.07$ magnitudes from a straight
fit to light curves.
Thus a UVOIR bolometric light curve or tracer
would probably appear very exponential for a period significantly
longer than that of the true 1st order exponential behavior.

    The parameters for the exponential fits to $f(x)$ that we have found
of particular interest in this paper are collected in Table~2.
In addition we have included in Table~2 the half-lives of RDE deposition
from strictly $^{\ApJstyle 56}$Co decay that result from the product
of the $^{\ApJstyle 56}$Co decay exponential with
the $f(x)$ function exponential fits for the two fiducial
supernova models we present in \S\S~3.3 and~3.4.
The models are needed to convert the half-life of a fitted exponential
into a real time half-life $t_{\ApJstyle 1/2}^{\ApJstyle\rm fit}$.
The half-life of a model's RDE deposition,
$t_{\ApJstyle 1/2}^{\ApJstyle\rm mod}$, is then given by
$$   t_{\ApJstyle 1/2}^{\ApJstyle\rm mod}=
     { t_{\ApJstyle 1/2}^{\ApJstyle\rm fit}t_{\ApJstyle\rm 1/2,Co}
       \over
       t_{\ApJstyle 1/2}^{\ApJstyle\rm fit}+t_{\ApJstyle\rm 1/2,Co} } \,\, , 
                                                          \eqno(\eqn)$$
where $t_{\ApJstyle\rm 1/2,Co}$ is the $^{\ApJstyle 56}$Co half-life.

     Finally, we should take stock of the accuracy of 
$N_{\ApJstyle\rm Ni}^{\ApJstyle *}(t)$ in comparison to
$N_{\ApJstyle\rm Ni}(t)$.
First, we note that our characteristic optical depth with its simple
time dependence cannot in general be the real mean optical of the ejecta
discussed in \S~3.1.
Thus in general $N_{\ApJstyle\rm Ni}^{\ApJstyle *}(t)$ cannot be exactly
$N_{\ApJstyle\rm Ni}(t)$ at all times.
For SNe~Ia and other low-mass/rapidly-expanding supernovae, we have already
chosen $\tau_{\ApJstyle\rm ch}$ to be the optically
thin $\bar\tau$.
Insofar as the optically thin $\bar\tau$ can be determined exactly,
$N_{\ApJstyle\rm Ni}^{\ApJstyle *}(t)$ with this choice of
$\tau_{\ApJstyle\rm ch}$ will be exact in the optically thin limit.
On the other hand, in the optically thick limit at very early times,
$N_{\ApJstyle\rm Ni}^{\ApJstyle *}(t)$ and $N_{\ApJstyle\rm Ni}$ should
be nearly exactly the same for any $\tau_{\ApJstyle\rm ch}$ of order of
the true mean optical depth since virtually all $\gamma$-rays are
locally trapped.

      One can plausibly expect that if $N_{\ApJstyle\rm Ni}^{\ApJstyle *}(t)$
is exactly correct in two opposite time limits, then it interpolates to good
accuracy in the intervening transition epoch when $\tau_{\ApJstyle\rm ch}$
is of order 1. 
Nevertheless, $N_{\ApJstyle\rm Ni}^{\ApJstyle *}(t)$ can be expected
to be at its worst in this period.
The reason is that the real individual optical depths of a supernova
will probably range over values larger and smaller than 1 giving
rise to complex location and direction varying $\gamma$-ray escape and
trapping behavior that a simple
optical depth prescription cannot easily mimic exactly.
Without having the real $\bar\tau$ for the transition epoch (with its in
general complex time dependence), deviations of
$N_{\ApJstyle\rm Ni}^{\ApJstyle *}(t)$ from
$N_{\ApJstyle\rm Ni}(t)$ will occur.
Of course, a 1st order exponential phase
corresponding to the single inflection point one predicted by 
$N_{\ApJstyle\rm Ni}^{\ApJstyle *}(t)$ can still exist.
If the supernova ejecta is smoothly varying in its properties,
a smooth transition in RDE deposition from early behavior
to a late 1st order exponential phase (with $^{\ApJstyle 56}$Co half-life
of course) can yield a single inflection point 1st order
exponential phase at some point provided the $^{\ApJstyle 56}$Ni
contribution becomes exponentially small soon enough. 
But if the ejecta has several high density clumps or shells, then
the RDE deposition curve might show complex behavior with 
multiple inflection point exponential phases
that occur as these clumps or shells make the
transition from optically thick to thin at different times.

     There is a case in which $N_{\ApJstyle\rm Ni}^{\ApJstyle *}(t)$ would
always be exact if the effective $\gamma$-ray absorption
opacity $\kappa$ were a true constant.
In the limit that the $^{\ApJstyle 56}$Ni is confined to
the exact center of a spherically symmetric supernova,
$\bar\tau$ is the exactly the radial optical depth to the center
at all epochs and hence $\bar\tau$ always has the simple
$t^{\ApJstyle -2}$ time dependence.  
The choice of the radial optical depth from the center
for $\tau_{\ApJstyle\rm ch}$
would then cause $N_{\ApJstyle\rm Ni}^{\ApJstyle *}(t)$
to be exactly $N_{\ApJstyle\rm Ni}(t)$.
Core-collapse supernovae are expected to have their 
$^{\ApJstyle 56}$Ni confined to the central region unless there
is very extensive mixing.
Thus, the exact case of $N_{\ApJstyle\rm Ni}^{\ApJstyle *}(t)$
could actually be approached if $\kappa$ were constant.
Unfortunately, dependence of $\kappa$ on the optical depth
structure of supernovae and hence on time, albeit weak, is not negligible.
\vskip\baselineskip

\centerline{3.3. \it A Parameterized SN~Ia Model}\nobreak
\vskip .5\baselineskip\nobreak
      In order to determine the optically thin $\bar\tau$ (our
choice for $\tau_{\ApJstyle\rm ch}$ for
SNe~Ia and other low-mass/rapidly-expanding supernovae)
and relate real time $t$ and reduced time $x$,
we need a structural supernova model.
In this section we will specify a simple parameterized
structural model for SNe~Ia. 

      Spherically symmetric hydrodynamic calculations of SN~Ia explosions
often (but not always)
produce models with density profiles that are very exponential (i.e.,
inverse exponential) with velocity after homologous expansion has set in.
For example, the well regarded Chandrasekhar mass SN~Ia models W7 (Nomoto,
Thielemann, \&~Yokoi 1984;  Thielemann, Nomoto, \&~Yokoi 1986),
DD4 (Woosley \&~Weaver 1994), and M36 (H\"oflich 1995, Fig.~10, but note that
the density is mislabeled as energy deposition) are quite exponential
with equivalent-exponential model $e$-folding velocities 
(see Appendix~A, eq.~[A10])
of about $2700\,{\rm km\,s^{\ApJstyle -1}}$,
$2750\,{\rm km\,s^{\ApJstyle -1}}$, and
$3000\,{\rm km\,s^{\ApJstyle -1}}$, respectively.  
Such nearly-exponential density profile models
have been quite successful in reproducing SN~Ia spectra 
(e.g., Jeffery~\etal 1992;  Kirshner~\etal
1993;  H\"oflich 1995;  Nugent~\etal 1995).
Therefore we will assume a spherically-symmetric, exponential
density profile model (i.e., an exponential model for brevity)
for our homologous epoch, parameterized SN~Ia model. 
In Appendix~A, we present a number of useful analytic results for 
exponential models
and give a prescription for exactly exponential models
(equivalent-exponential models) that can approximately replace
nearly exponential hydrodynamic explosion models.

     The density profile of an exponential model (for the homologous
epoch) is given by
$$ \rho(v,t)=\rho_{\ApJstyle \rm ce,0}
        \left({t_{\ApJstyle 0}\over t}\right)^{\ApJstyle 3}
        \exp\left(-v/v_{\ApJstyle e}\right)  
=\rho_{\ApJstyle \rm ce,0}
        \left({t_{\ApJstyle 0}\over t}\right)^{\ApJstyle 3}
        \exp\left(-z\right)                     \,\,  , \eqno(\eqn)$$
where $\rho_{\ApJstyle \rm ce,0}$ is the central density at
fiducial time $t_{\ApJstyle 0}$, $v$ is the radial velocity,
$v_{\ApJstyle e}$ is the $e$-folding velocity,
and $z$ is radial velocity or radial position in velocity space
in units of $v_{\ApJstyle e}$.
Substituting for density from equation~(\showeqn -0) into
equation~(8) and assuming the opacity $\kappa$ is constant,
we find for an exponential model that the $\gamma$-ray optical depth
from an emission point $z$ to the surface (which is at infinity) is 
$$ \tau=\tau_{\ApJstyle\rm ce,0}
\left({t_{\ApJstyle 0}\over t}\right)^{\ApJstyle 2}
\int_{\ApJstyle 0}^{\ApJstyle\infty}dz_{\ApJstyle s}\,\exp\left(-z'\right) 
                                                      \,\, , \eqno(\eqn)$$
where $\tau_{\ApJstyle\rm ce,0}$ is the radial optical depth to the
center at the fiducial time $t_{\ApJstyle\rm 0}$ (see eq.~[A16] in 
Appendix~A for the expression),
$z_{\ApJstyle s}$ is beam path velocity length in units of
the $e$-folding velocity,
and 
$$  z'=\sqrt{z^{\ApJstyle 2}+z_{\ApJstyle s}^{\ApJstyle 2}
           +2z z_{\ApJstyle s}\mu} \,\, .                       \eqno(\eqn)$$
The $\mu$ is the cosine of the angle at the emission point
between the outward radial direction and the beam propagation direction.
For a beam in the outward radial direction,
$\mu=1$ and the optical depth expression reduces to
$$ \tau_{\ApJstyle r}=\tau_{\ApJstyle\rm ce,0}
\left({t_{\ApJstyle 0}\over t}\right)^{\ApJstyle 2}
\exp\left(-z\right)                \,\, . \eqno(\eqn)$$
 
    Given that there is only $^{\ApJstyle 56}$Co (which was initially
$^{\ApJstyle 56}$Ni), 
equation~(12) for the optically thin $\bar\tau$ changes 
for our exponential SN~Ia model to 
$$ \bar\tau=\tau_{\ApJstyle\rm ce,0}
\left({t_{\ApJstyle 0}\over t}\right)^{\ApJstyle 2}q \,\, , \eqno(\eqn)$$
where the $q$ parameter is defined by
$$ q={\displaystyle
  \int_{\ApJstyle 0}^{\ApJstyle\infty}
    dz\,z^{\ApJstyle 2}
    \exp\left(-z\right)
    X_{\ApJstyle\rm Ni}^{\ApJstyle\rm ini}
    \int_{\ApJstyle -1}^{\ApJstyle 1}{d\mu\over2}\,
  \int_{\ApJstyle 0}^{\ApJstyle\infty}
    dz_{\ApJstyle s}\,\exp\left(-z'\right)   
\over
  \int_{\ApJstyle 0}^{\ApJstyle\infty}
    dz\,z^{\ApJstyle 2}
    \exp\left(-z\right)
    X_{\ApJstyle\rm Ni}^{\ApJstyle\rm ini}
    }                                             \,\, .  \eqno(\eqn)$$
Note that the initial $^{\ApJstyle 56}$Ni fraction is location-dependent
although time-independent, of course.

     A bit of analysis of equation~(\showeqn -0) shows that when all
$^{\ApJstyle 56}$Ni is concentrated in the center,
$q$ has its maximum value of 1 and
the optically thin $\bar\tau$, its maximum value of 
$\tau_{\ApJstyle\rm ce,0}\left(t_{\ApJstyle 0}/t\right)^{\ApJstyle 2}$.
The $q$ parameter, in fact, acts as measure of
$^{\ApJstyle 56}$Ni concentration.
The more the $^{\ApJstyle 56}$Ni is spread to low optical depth regions, 
the more important the small optical depth contributions are
in equation~(\showeqn -0), and the smaller $q$ becomes.
We generalize the $q$ parameter beyond the definition of
equation~(\showeqn -0) (which is specific to exponential models)
and consider $q$ as a general initial $^{\ApJstyle 56}$Ni concentration
parameter regardless of the exact supernova model.
The general $q$ is large for high concentration, small for low concentration.

     To study the behavior of the (exponential model) $q$ parameter
consider the special case that
the (initial) $^{\ApJstyle 56}$Ni exists only between velocity radii
$a$ and $b$ (in units of the $e$-folding velocity of course) and that it is a
constant there.
Then equation~(\showeqn -0) reduces to
$$ q={\displaystyle
  \int_{\ApJstyle a}^{\ApJstyle b}
    dz\,z^{\ApJstyle 2}
    \exp\left(-z\right)
    \int_{\ApJstyle -1}^{\ApJstyle 1}{d\mu\over2}\,
  \int_{\ApJstyle 0}^{\ApJstyle\infty}
    dz_{\ApJstyle s}\,\exp\left(-z'\right)   
\over
  \int_{\ApJstyle a}^{\ApJstyle b}
    dz\,z^{\ApJstyle 2}
    \exp\left(-z\right)
    }                                             \,\, .  \eqno(\eqn)$$
We have been unable to analytically solve the triple integral in the numerator
of equation~(\showeqn -0) or even find an accurate analytic approximation:  
crude approximate analytic solutions can be found for certain special cases.
The integrations in the numerator can, however, be easily done numerically.
The denominator can be solved analytically using equation~(A1) in Appendix~A. 

     Equation~(\showeqn -0) is actually heuristically useful for SNe~Ia
since hydrodynamic models often show the $^{\ApJstyle 56}$Ni is concentrated
in a distinct layer where its abundance is overwhelmingly dominant.
For example, model~W7 has an initial $^{\ApJstyle 56}$Ni mass fraction of 
over $0.5$ in the range $\sim 3375$--$9750\,{\rm km\,s^{\ApJstyle -1}}$.
In most of this range the mass fraction is $\sim 0.9$.
Outside of the range the mass fraction falls fairly quickly to very low
values.
Using the $e$-folding velocity $2700\,{\rm km\,s^{\ApJstyle -1}}$
which is in round numbers the equivalent-exponential model
$e$-folding velocity for model~W7 (see Appendix~A, eq.~[A10]), we obtain from
equation~(\showeqn -0) by numerical integration
                       $q=0.3591$.  
The $q$ value extracted from a full $\gamma$-ray transfer calculation
for the model~W7 (assuming model~W7 is an exponential model
with certain fiducial parameter values:  see the description below)
yields $q=0.3308$.
The difference is caused by the fact that model~W7 is not exactly
exponential in density (e.g., it has local density maxima at
$\sim  9200\,{\rm km\,s^{\ApJstyle -1}}$,
$\sim 13300\,{\rm km\,s^{\ApJstyle -1}}$,
$\sim 15000\,{\rm km\,s^{\ApJstyle -1}}$, and
cuts off sharply at $\sim 22000\,{\rm km\,s^{\ApJstyle -1}}$)
and the $^{\ApJstyle 56}$Ni mass fraction is not exactly constant inside
of a sharply defined layer. 

     If we set $a$ and $b$ in equation~(\showeqn -0) to, respectively,
zero and infinity
(i.e., an evenly-spread-$^{\ApJstyle 56}$Ni case),
we obtain a $q=0.33333326$ at the highest accuracy we
have computed.         
This value is so close to $1/3$ that it is likely that the actual
$q$ value for the evenly-spread-$^{\ApJstyle 56}$Ni case is
exactly $1/3$.
We have not, however, been able to find an analytic proof.
The fact that the $q$ value for the evenly-spread-$^{\ApJstyle 56}$Ni case
is not so different from the model~W7 and approximated model~W7 $q$ values
cited just above shows
that the interior $^{\ApJstyle 56}$Ni at large optical depth
dominates the deposition in this case.
Because the $1/3$ value is a simple rational number that
seems to be exactly correct for a particularly well defined system 
and is not so different from values likely to be obtained for more
realistic SN~Ia models, we adopt $1/3$ below as our fiducial $q$ value 
for our parameterized SN~Ia model. 

     From equation~(\showeqn -2) evaluated at the fiducial time
$t_{\ApJstyle 0}$ with $\tau_{\ApJstyle\rm ce,0}$
substituted for from equation~(A16) in Appendix~A, we
obtain an expression for a fiducial time in terms of a fiducial
characteristic optical depth $\tau_{\ApJstyle\rm ch,0}$ (chosen to be
the optically thin $\bar\tau$ at the fiducial time):
$$ t_{\ApJstyle 0}=\sqrt{ {M\over8\pi}
          {\kappa q\over\tau_{\ApJstyle\rm ch,0}} }
          \,{1\over v_{\ApJstyle e}}   \,\, .  \eqno(\eqn)$$
We will now rewrite equation~(\showeqn -0) in terms of fiducial
values.
At present Chandrasekhar mass C-O white dwarfs are favored SN~Ia
progenitors, and  so the fiducial mass is chosen to be
$1.38\,M_{\ApJstyle\odot}$.   
Since we want our expression to be as exactly valid as possible
in the optically thin phase, we set the fiducial
$\kappa$ to $0.025\,{\rm cm^{\ApJstyle 2}\,g^{\ApJstyle -1}}$:
a general good opacity value for all-metal ejecta
($\mu_{\ApJstyle e}\approx 2$) in the optically thin limit
(see \S~3.1).
We take $v_{\ApJstyle e}=2700\,{\rm km\,s^{\ApJstyle -1}}$
which in round numbers is the equivalent-exponential model
$e$-folding velocity for model~W7 (see Appendix~A, eq.~[A10]):  model~W7 is
representative of currently favored SN~Ia models.
The fiducial $q$ is set to $1/3$ for reasons given above.
As explained in \S~3.2, the fiducial $\tau_{\ApJstyle\rm ch,0}$ is 
set to 1 so that $t_{\ApJstyle 0}$ is conveniently
roughly the time of transition between
the optically thick and thin epochs. 
With these choices the expression for fiducial time becomes
$$\eqalignno{
t_{\ApJstyle 0}
  &= 40.895902\,\, {\rm days} \times   
   \sqrt{
\left({M\over 1.38\,M_{\ApJstyle\odot}}\right)
\left({\kappa\over 0.025\,{\rm cm^{\ApJstyle 2}\,g^{\ApJstyle -1}} }\right)
\left({1\over \tau_{\ApJstyle\rm ch,0} } \right)
\left({q\over1/3}\right)
         }                                                           &\cr
  &\qquad\times \left({2700\,{\rm km\,s^{\ApJstyle -1}} \over v_{\ApJstyle e} }
               \right)    \,\, .            &(\eqn)\cr}$$
For numerical consistency here and throughout this paper
we treat fiducial values and the  
solar mass unit $M_{\ApJstyle\odot}=1.9891\times10^{\ApJstyle 33}\,{\rm g}$
(Lide \&~Frederikse 1994, p.~14-2) as nearly exact numbers.
Our fiducial SN~Ia model is the one that has the fiducial
values and thus has fiducial time $t_{\ApJstyle 0}=40.895902$~days.

     To test our $N_{\ApJstyle\rm Ni}^{\ApJstyle *}(t)$ expression
and our parameterized SN~Ia model, we have fitted
$N_{\ApJstyle\rm Ni}^{\ApJstyle *}(t)$ to a normalized W7 RDE deposition
curve calculated using a grey radiative transfer procedure and the
actual model~W7.
The LS grey $\gamma$-ray transfer procedure (Jeffery 1998a, b) was used
for the calculation.
This is a free-parameter-free grey procedure
that accounts crudely for the variation of effective absorption opacity
with optical depth structure.
The LS procedure has an uncertainty of order $10\,$\% at most for
total RDE deposition:  the uncertainty vanishes in the optically
thick and thin limits.
For a fair comparison to
$N_{\ApJstyle\rm Ni}^{\ApJstyle *}(t)$, we included only
the $^{\ApJstyle 56}$Ni and $^{\ApJstyle 56}$Co radioactive sources
and turned off X-rays in the LS~procedure calculation.
Because of these limitations on the RDE deposition, we call the curve
computed by the LS procedure, the limited W7 deposition curve.

     The fitting was done by choosing a $t_{\ApJstyle 0}$ value
(with $\tau_{\ApJstyle\rm ch,0}=1$ as always)
that made the $N_{\ApJstyle\rm Ni}^{\ApJstyle *}(t)$ equal the
limited W7 deposition curve exactly on day~2000 after explosion. 
The day~2000 epoch is well into the optically thin phase,
and so the fitted $t_{\ApJstyle 0}$ reproduces
the time-varying optically thin $\bar\tau$ to within numerical accuracy. 
The fitted fiducial time $t_{\ApJstyle 0}=40.74$~days.
This model~W7 fiducial time is very close to the fiducial SN~Ia model fiducial
time 40.895902~days.
Assuming model~W7 is exactly an exponential model with the prescribed
fiducial parameters, except for the $t_{\ApJstyle 0}$ and $q$ values,
we invert equation~(\showeqn -0) using $t_{\ApJstyle 0}=40.74$~days
to obtain $q=0.3308$.
This $q$ cannot have exactly the interpretation dictated by the
$q$ equations~(\showeqn -3) and~(\showeqn -2) since model~W7 is not
exactly an exponential model.
However, in the general, if vague, interpretation of $q$ as a
concentration parameter, the $q$ from the fit has a useful meaning.
Model~W7 has roughly the same order of concentration of
$^{\ApJstyle 56}$Ni as the exact exponential models we considered above.

     Figures~2a and~2b display the limited W7 deposition curve and
the fitted $N_{\ApJstyle\rm Ni}^{\ApJstyle *}(t)$ curve to
day~100 and day~500 after the explosion, respectively.
The $N_{\ApJstyle\rm Ni}^{\ApJstyle *}(t)$ curve is in very close
agreement with the limited W7 deposition curve and consequently is
difficult to see on the Figure~2b scale.
The agreement is closest at early and late times as expected:
less than $\sim 0.001$~dex discrepancy before about day~10 and
after about day~240.
The only period when the discrepancy is $\gapprox 0.01$~dex the
day 15--73 period where the $N_{\ApJstyle\rm Ni}^{\ApJstyle *}(t)$ curve
rises above limited W7 deposition curve with the maximum
disrepancy being $\sim 0.07$~dex (about $18\,$\%) at day~33.
It was expected that the largest discrepancies would occur
during the transition from the optically thin to thick epochs:
i.e., during the phase centered on reduced time $x=1$ (see \S~3.2).
For model~W7 this is the period centered on about day~40 after the explosion. 
(Recall the uncertainty in the limited W7 deposition curve is
also largest in transition period, but it is no more than of order $10\,$\% 
percent.)

     We have also plotted on Figures~2a and~2b
the complete trapping and escape curves discussed in \S~3.2.
As in Figure~1a, the complete trapping curve shows the transition
between the early phase where declining $^{\ApJstyle 56}$Ni
and increasing $^{\ApJstyle 56}$Co provide the RDE and the later
phase where decreasing $^{\ApJstyle 56}$Co provides the RDE.
As we discussed in \S~3.2, $^{\ApJstyle 56}$Ni contribution is
truly exponentially small only by of order day~60
which is $\sim 1.5t_{\ApJstyle 0}$ for both the fiducial SN~Ia model and
model~W7.
Thus, these models cannot really have either an early-time nor
an inflection point 1st order exponential phase in RDE deposition.
For an inflection point 1st order exponential phase, the
$^{\ApJstyle 56}$Ni contribution must be exponentially
small by of order $\sim 1.35t_{\ApJstyle 0}$ at about the latest.
Figure~2a, examined closely, confirms the lack of
a first order exponential phase centered on
about $t_{\ApJstyle 0}=40$~days in the model deposition curves.
Since the fiducial SN~Ia model and model~W7 are reasonable models for
SNe~Ia, we conclude that SNe~Ia are very unlikely to have either
an early time or an inflection point 1st order exponential phase.

     Recall that the quasi-steady state of SNe~Ia does not start until
about day~60 after the explosion.
Thus even if the early or inflection point 1st order exponential
phases in RDE deposition existed, they would not have a direct signature
in the light curves.
Because SNe~Ia are likely already in the
post-inflection point epoch when the quasi-steady state phase begins,
one expects to see the instantaneous half-lives of their light curves
only increase with time (i.e., the logarithmic decline rates decrease
with time) in
the quasi-steady state phase at least until very late times.
And this is in fact what observations suggest happens
(e.g., Cappellaro~\etal 1997; see also \S~4).
Comparing the deposition curves to the complete escape curve in Figure~2b
shows how gradual the approach is to late 1st order exponential phase
on the time scale of observations which have very rarely gone beyond
500~days.

     The real time half-lives of the exponential fits to RDE deposition 
that result from the product of pure $^{\ApJstyle 56}$Co exponential decay and
the $f(x)$ function exponential fits discussed in \S~3.2 and put on the 
real time scale using the fiducial SN~Ia model $t_{\ApJstyle 0}$ are
shown in Table~2.
As we have seen above, the early-time and inflection point 1st order
exponential phases in RDE deposition cannot be realized for the fiducial
SN~Ia model nor model~W7 nor in all likelihood actual SNe~Ia.
Thus the exponential fits that correspond to those cases and
similarly the fit to the $x$-range $[0.23,0.53]$ cannot be realized.
The half-lives corresponding to those cases are presented
{pro forma} only.      
However, the exponential fits to RDE deposition
for the $x$-ranges $[2,3]$ and $[4,8]$, and
for the late-time limit (i.e., to the time when $x\to\infty$)
are good fits (after choosing the appropriate constant coefficient for
the exponential fit) to the 
$N_{\ApJstyle\rm Ni}^{\ApJstyle *}(t)$ and limited W7 deposition curves
since they occur after the $^{\ApJstyle 56}$Ni is exponentially small.
The $x$-range $[2,3]$ and $[4,8]$ fits correspond to the periods
$\sim 80$--$120$ and $\sim 160$--$320$~days, respectively, for both the
fiducial SN~Ia model and model~W7.
But these are only exponential fits to quasi-exponential phases.

     Finally, we note we just needed model~W7 itself not the
parameterized SN~Ia model in order to determine
$t_{\ApJstyle 0}$ by fitting.
The parameterized model did, however, predict model~W7's
$t_{\ApJstyle 0}$ value fairly
well from reasonably chosen fiducial values.
Furthermore, the parameterized model gives insight into how the
deposition is determined.
\vskip\baselineskip

\centerline{3.4. \it A Parameterized Core-Collapse Supernova Model}\nobreak
\vskip .5\baselineskip\nobreak
     Most supernovae Ia are fairly homogeneous in their behavior.
Core-collapse supernovae are much more heterogeneous.
Thus it is possible that the parameterized exponential SN~Ia model
presented in \S~3.3 may be fairly adequate for most or all
SNe~Ia.
But a parameterized core-collapse supernova model is likely to apply
only sometimes.
Here we will develop a parameterized model based on the spherically
symmetric
$16.2\,M_{\ApJstyle\odot}$ explosion model~10H of Woosley (1988):
about $14.5\,M_{\ApJstyle\odot}$ of the model are ejected and 
about $1.7\,M_{\ApJstyle\odot}$ are left in a compact remnant, presumably
a neutron star.
Model~10H is able to account for many of the features of SN~II~1987A. 

      The velocity profile of model~10H in the homologous epoch
is very roughly describable as linear with respect to interior mass
over most of the ejecta mass:  the most linear region extends
over the range $\sim 4\,M_{\ApJstyle\odot}$--$\sim 15\,M_{\ApJstyle\odot}$ 
corresponding to $\sim 1000$--$4250\,{\rm km\,s^{\ApJstyle -1}}$
(Woosley 1988, Fig.~12a).  
The rough linear relation between velocity and mass implies that
density goes roughly as the inverse square of velocity over most of
the ejecta mass. 
In the innermost region below $\sim 3\,M_{\ApJstyle\odot}$ (where most of
the mass was left in the compact remnant), the model~10H ejecta
density decreases inward roughly speaking (Woosley 1988, Fig.~28):
there is no density singularity at the center of the ejecta.

      For our parameterized model in the homologous epoch,
we assume all the mass is located between velocities $v_{\ApJstyle\rm a}$ and 
$v_{\ApJstyle\rm b}$, and that density goes exactly an inverse-square of
velocity in this interval.
Thus total mass is given by
$$  M=
4\pi\int_{\ApJstyle v_{\ApJstyle\rm a}}^{\ApJstyle v_{\ApJstyle\rm b}}dv\,
       v^{\ApJstyle 2}
       t^{\ApJstyle 3}
       \rho_{\ApJstyle 0}
       \left({t_{\ApJstyle 0}\over t}\right)^{\ApJstyle 3}
       \left({v_{\ApJstyle 0}\over v}\right)^{\ApJstyle 2} 
= 4\pi \rho_{\ApJstyle 0}t_{\ApJstyle 0}^{\ApJstyle 3}
             v_{\ApJstyle 0}^{\ApJstyle 2}
     \left(v_{\ApJstyle\rm b}-v_{\ApJstyle\rm a}\right) \,\, , \eqno(\eqn)$$
where $\rho_{\ApJstyle 0}$ is a fiducial density at a fiducial radial
velocity $v_{\ApJstyle 0}$ and fiducial time $t_{\ApJstyle 0}$.
The radial $\gamma$-ray optical depth from a velocity $v$ to the surface
is given by
$$ \tau(v)=\int_{\ApJstyle v}^{\ApJstyle v_{\ApJstyle\rm b}}dv\,t\kappa
       \rho_{\ApJstyle 0}
       \left({t_{\ApJstyle 0}\over t}\right)^{\ApJstyle 3}
       \left({v_{\ApJstyle 0}\over v}\right)^{\ApJstyle 2}    
    =t^{\ApJstyle -2}\kappa\rho_{\ApJstyle 0}
                      t_{\ApJstyle 0}^{\ApJstyle 3}
                      v_{\ApJstyle 0}^{\ApJstyle 2}
     \left({1\over v}-{1\over v_{\ApJstyle\rm b}}\right)
                                                \,\, , \eqno(\eqn)$$
where we have assumed $\kappa$ is a constant.

     We now parameterize the characteristic optical depth of the
ejecta $\tau_{\ApJstyle\rm ch}$ by the expression
$$\tau_{\ApJstyle\rm ch}
        =\tau\left(v_{\ApJstyle\rm a}/q\right) \,\, , \eqno(\eqn)$$
where $q$ is again a concentration parameter.
The larger $q$, the smaller $v_{\ApJstyle\rm a}/q$, the larger
$\tau_{\ApJstyle\rm ch}$, and the more concentrated the
initial $^{\ApJstyle 56}$Ni must be in the central region of the ejecta.
The chosen parameterization of $\tau_{\ApJstyle\rm ch}$ is, of course, based
on the idea that actual mean optical depth will depend on the overall
supernova parameters in roughly the same way as a radial optical depth
from some velocity to the surface.

     We note that if all the initial $^{\ApJstyle 56}$Ni is concentrated
just at the innermost physical layer (i.e., at $v_{\ApJstyle\rm a}$),
the mean optical depth for $\gamma$-ray deposition is larger than
$\tau\left(v_{\ApJstyle\rm a}\right)$.
A beam of $\gamma$-rays from a point at $v_{\ApJstyle\rm a}$ must cross
a larger optical depth than $\tau\left(v_{\ApJstyle\rm a}\right)$, except
in the outward and inward radial directions.
Thus $q$ can be larger than 1.
This contrasts with our parameterized SN~Ia model where 1 was the maximum value
of $q$ (see \S~3.3). 

     Taking $\tau\left(v_{\ApJstyle\rm a}/q\right)$ for 
$\tau_{\ApJstyle\rm ch}$ and using equations~(\showeqn -2) and~(\showeqn -1),
we can now solve for the time when
$\tau_{\ApJstyle\rm ch}$ is the characteristic optical depth:
$$ t=\sqrt{ {M\over4\pi} {\kappa\over \tau_{\ApJstyle\rm ch}}
            {\left[1-\left(v_{\ApJstyle\rm a}/v_{\ApJstyle\rm b}\right)
                     \left(1/q\right) \right]
             \over
             \left[1-\left(v_{\ApJstyle\rm a}/v_{\ApJstyle\rm b}\right)\right]
             }
            {q\over v_{\ApJstyle\rm a} v_{\ApJstyle\rm b} }
          }        \,\, . \eqno(\eqn)$$

     We will rewrite equation~(\showeqn -0) in terms
of fiducial values guided mainly by model~10H and SN~II~1987A. 
For mass we use $15\,M_{\ApJstyle\odot}$ which is roughly the model~10H
ejected mass.
The most linear region of model~10H cited above contains most, but
not almost all, the model~10H mass.
Thus it is difficult to select $v_{\ApJstyle\rm a}$ and
$v_{\ApJstyle\rm b}$ based model~10H alone without considerable
arbitrariness.
But to fit the quasi-steady state UVOIR bolometric light curve of SN~1987A
(see \S~5),
we select the values $v_{\ApJstyle\rm a}=700\,{\rm km\,s^{\ApJstyle -1}}$
and $v_{\ApJstyle\rm b}=5000\,{\rm km\,s^{\ApJstyle -1}}$
which give a good fit (along with our other fiducial values) while being
consistent with what one might estimate from model~10H (Woosley 1988,
Fig.~12a).
The fit is not uniquely good, of course.
In model~10H, the $^{\ApJstyle 56}$Ni is confined to the lowest
ejecta velocities (Woosley 1988, Fig.~4).
This suggests $q$ to be of order 1.
It is clear, however, that some $^{\ApJstyle 56}$Ni was mixed outward
perhaps as far as halfway or more in mass fraction through the
ejecta (e.g., Nomoto~\etal 1994a, p.~528ff, and references therein).
This suggests a $q$ parameter less than 1, maybe much less.
But for the sake of definiteness we take the fiducial $q=1$.
We again choose $\tau_{\ApJstyle\rm ch,0}=1$ so that our
fiducial time $t_{\ApJstyle 0}$ is conveniently roughly
the time of transition between the optically thick and thin epochs
(see \S~3.2).

    Model~10H has of order $10\,M_{\ApJstyle\odot}$ of metals
and helium and of order $5\,M_{\ApJstyle\odot}$ of hydrogen-rich material
which is presumably solar or sub-solar in metal (Woosley 1988, Figs.~4 and~17).
For a mixed composition of this matter using 
Table~1 of Jeffery (1998a) or Table~2 of Jeffery (1998b),
we find $\kappa\approx0.04\,{\rm cm^{\ApJstyle 2}\,g^{\ApJstyle -1}}$
for the optically thick limit
and $\kappa\approx0.03\,{\rm cm^{\ApJstyle 2}\,g^{\ApJstyle -1}}$
for the optically thin limit.
The transition, broadly speaking, from optically thick to optically
thin phase of SN~1987A spans whole of the SN~1987A quasi-steady state UVOIR
bolometric light curve (see \S~5). 
Thus we take average value of the limiting opacities
$0.035\,{\rm cm^{\ApJstyle 2}\,g^{\ApJstyle -1}}$ as our
fiducial $\kappa$.

      With the specified fiducial values, the expression for the
fiducial time is
$$\eqalignno{
t_{\ApJstyle 0}&= 563.969041\,\,{\rm days}  \cr
&\qquad\times\sqrt{ \left({M\over 15\,M_{\ApJstyle\odot} }\right)
  \left({\kappa\over 0.035\,{\rm cm^{\ApJstyle 2}\,g^{\ApJstyle -1}} }\right)
  \left({1 \over \tau_{\ApJstyle\rm ch,0}}\right)
  \left[{1-\left(v_{\ApJstyle\rm a}/v_{\ApJstyle\rm b}\right)\left(1/q\right)
         \over
         1-\left(v_{\ApJstyle\rm a}/v_{\ApJstyle\rm b}\right)
         }\right] 
           }        \cr
&\qquad\qquad\times
   \sqrt{
  \left({700\,{\rm km\,s^{\ApJstyle -1}} \over v_{\ApJstyle\rm a}}\right)
  \left({5000\,{\rm km\,s^{\ApJstyle -1}} \over v_{\ApJstyle\rm b}}\right)
  \left({q\over 1}\right)
          }   \,\, . &(\eqn)\cr}$$
The parameterized model with all the fiducial values chosen is our
fiducial core-collapse model and $563.969041$~days is its fiducial time.

     In Figure~3 we show the fiducial model
$N_{\ApJstyle\rm Ni}^{\ApJstyle *}(t)$ deposition curve.
The $t_{\ApJstyle 0}$ value evaluated for the fiducial model is
so large that $N_{\ApJstyle\rm Ni}^{\ApJstyle *}(t)$ will realize
all the 1st order exponential and quasi-exponential RDE deposition
phases discussed in \S~3.2, except, of course, for
the $^{\ApJstyle 56}$Ni nearly-exponential phase which requires small
$t_{\ApJstyle 0}$.
This is because $^{\ApJstyle 56}$Ni contribution to the RDE deposition
will become exponentially small by of order day~60 after explosion (as
discussed in \S~3.2 and as best seen in Figure~2a), and at this time the
ejecta is practically speaking still in the optically thick limit. 
The real time half-lives of the exponential fits 
that result from the product of pure $^{\ApJstyle 56}$Co exponential decay and
the $f(x)$ function exponential fits discussed in \S~3.2 and put on the 
real time scale with the fiducial core-collapse model $t_{\ApJstyle 0}$ are
shown in Table~2.
All these fits (when provided with appropriate constant coefficients)
are good fits to the
$N_{\ApJstyle\rm Ni}^{\ApJstyle *}(t)$ deposition curve in this case.

     Actual large-mass and/or slowly-expanding core-collapse supernovae
may never get much beyond the early or inflection point 1st order
exponential phases of $N_{\ApJstyle\rm Ni}^{\ApJstyle *}(t)$-like
RDE deposition.
The reason is that these phases could be so long that long-lived
radioactive species may become important or the quasi-steady
state period may end before the phases end. 
Supernova SN~1987A, for example, which we discuss in \S~5 did not
get past (or much past anyway) the inflection point 1st order phase before
both long-lived radioactive species and time-dependent effects in the
energy processing from decay to UVOIR emission became important.
On the other hand, small mass and/or quickly expanding core-collapse
supernovae (probably most SNe~Ib and~Ic) reach lower optical depth sooner
and may get well past the first two 1st order exponential phases
before $N_{\ApJstyle\rm Ni}^{\ApJstyle *}(t)$-like RDE deposition ends.
In \S~6 we consider SN~Ic~1998bw as a possible example of this case.
\vskip 2\baselineskip

\centerline{4.\ \  SN Ia 1992A}\nobreak
\vskip\baselineskip\nobreak
     As an example of how far simple RDE deposition calculations
can take one in understanding SN~Ia light curves we present a
comparison in Figures~4a (days~0---250) and~4b (days~0---1000)
between two normalized RDE deposition curves
calculated using the LS procedure (see \S~3.3)
for model~W7 and the $V$ light curve of normal SN~Ia~1992A.
The first of the deposition curves (solid line) is a full W7 deposition
curve calculated with deposition from all the significant radioactive
species in model~W7 and with X-ray deposition included.
In Table~3 we summarize the significant radioactive decays for
model~W7.
The other deposition curve (dotted line) is just the limited W7 deposition
curve presented in \S~3.3.
This curve is, for our present interest, virtually the same
as the $N_{\ApJstyle\rm Ni}^{\ApJstyle *}(t)$
curve fitted to model~W7 (see \S~3.3).
The full and limited curves are in close agreement until very late
times (see Fig.~4b).


     At very early times, $^{\ApJstyle 55}$Co and $^{\ApJstyle 57}$Ni
contribute significantly to the full curve although this contribution
is almost invisible on the scale of Figure~4a.  
(The full curve is normalized by the RDE deposition of the limited curve
at time zero, and so rises above 1 until about day~1.3.)
At time zero, $^{\ApJstyle 55}$Co and $^{\ApJstyle 57}$Ni
contribute about $20\,$\% of the deposition.
Their rapid decay rate is the reason for their significant contribution
despite the fact that they have only trace abundance.
But, of course, their rapid decay rate causes their contribution to
become negligible quickly:  down to
to $\sim 6\,$\% by day~3 and $\sim 1\,$\% by day~8.
After the $^{\ApJstyle 55}$Co and $^{\ApJstyle 57}$Ni abundances
have become exponentially small,
$^{\ApJstyle 56}$Ni and $^{\ApJstyle 56}$Co, and then 
$^{\ApJstyle 56}$Co alone
overwhelmingly dominate the full curve
because $^{\ApJstyle 56}$Ni is produced in bulk in SNe~Ia unlike
the other radioactive species.
In model~W7 the initial
$^{\ApJstyle 56}$Ni abundance is $0.58\,M_{\ApJstyle\odot}$
and this is a typical value calculated for SNe~Ia.
The long-lived radioactive species, however, become more important as time
passes.
By day~400 the long-lived radioactive species
are contributing about $3\,$\% and the $^{\ApJstyle 56}$Co X-rays
about $1\,$\% to the full curve RDE deposition.
At about day~930 after explosion the long-lived radioactive species 
become dominant:  the $^{\ApJstyle 56}$Co X-rays
still supplying only about $1\,$\% of the deposition.


     At day~1000 the breakdown of the full curve's deposition is as follows.
Photons contribute $29.2\,$\% and PE particle kinetic energy $70.8\,$\%.
The X-rays contribute $25.9\,$\% and the $\gamma$-rays only
$3.3\,$\%.
%
%
The $^{\ApJstyle 56}$Co, $^{\ApJstyle 57}$Co,
and $^{\ApJstyle 55}$Fe contribute $38.9\,$\%,
$56.3\,$\%, and $4.6\,$\%, respectively.
The $^{\ApJstyle 44}$Ti and its short-lived daughter
$^{\ApJstyle 44}$Sc contribute only $0.16\,$\%.
Because of $^{\ApJstyle 44}$Ti's long half-life, however,
$^{\ApJstyle 44}$Ti/$^{\ApJstyle 44}$Sc will dominate
the RDE deposition by about day~5835.  

      It should be emphasized that the predictions of radioactive
species abundance by model~W7 are uncertain on the grounds of both nuclear
reaction rates and the explosive nucleosynthesis.
In the case of $^{\ApJstyle 56}$Ni, the uncertainty is probably small
since explosive production of order $0.6\,M_{\odot}$ of 
$^{\ApJstyle 56}$Ni in normal SNe~Ia is well supported by a variety
of empirical evidence.
The abundances of trace radioactive species do not have nearly such strong
support from observations and are much less certain.

      Ideally one would like to compare the RDE deposition curves
to SN~Ia UVOIR bolometric light curves for the quasi-steady state phase
which begins at $\sim 60$~days after explosion.
So far, however, no adequate late-time UVOIR bolometric light curves for
SNe~Ia have been extracted.
The $V$ light curve, however, appears to be a reasonable ersatz.
It has been suggested that the bolometric correction to the $V$ light curve
from about 100~days and up to maybe 600~days after explosion
is $\lapprox 0.1$---$0.2$ magnitudes and is relatively constant
(e.g., Cappellaro~\etal 1997).
Therefore, as an example, we have fit the $V$ light curve of normal
SN~Ia~1992A (Cappellaro~\etal 1997) to our deposition curves.
The $V$ light curve for SN~1992A is one of the best observed for SNe~Ia
and has the latest observation of a SN~Ia ever:  an HST datum
from 926 days after the $B$~maximum
taken by the SINS team (e.g., Kirshner~\etal 1993) and reduced by
Cappellaro~\etal (1997).

     To fit the SN~1992A $V$ light curve we have assumed a rise time to
maximum light (taken to be the $B$~maximum by convention) of $18$~days.
This rise time is suggested by observational evidence.
The fairly normal SN~Ia~1990N was discovered at a very early
phase $17.5\pm1$~days before maximum light (Leibundgut~\etal 1991a).
And there is empirically estimated rise time to maximum light
of $17.6\pm0.5$~days for SN~1994D (another fairly normal
SN~Ia) (Vacca \&~Leibundgut 1996).
Additionally, observations of cosmologically remote
SNe~Ia (which are usually discovered well before maximum
light) suggest a typical SN~Ia rise time of $\sim 17$--$18$ days
(Nugent~\etal 1999).
Variation from 18~days by a few days are expected for some
SNe~Ia, but this will not much affect a fit of late-time light curves to
theoretical predictions.
The vertical level of the $V$~light curve was determined by fitting
the $V$ light curve by eye to the deposition curves in the range
from about day~300 to day~400. 

     The first thing to notice in Figures~4a and~4b is that the fit
from day~51 on is rather good.
(The period before about day~60 [i.e., before the
quasi-steady phase] is that of significant UVOIR diffusion time, and
so the $V$ light curve in that period is not expected to trace the
RDE deposition directly.)
The second is, of course, the discrepancies in the post-day-51 period.
The $V$~data from the days~51---109 after the explosion
are systematically too high by of order $0.1$~dex.
The $V$~data from days~207---451 have a dispersion about the RDE deposition
curves of order $0.05$~dex.
The $V$ data point from day~944 (i.e., from 926 days after maximum light)
is in a phase in which the two RDE deposition curves have diverged.
With its estimated uncertainty of $0.12$~dex, this point is 
somewhat inconsistent with both of the two deposition curves.

     Reasons for the discrepancies are easily found.
First, the $V$ light curve may well not be a sufficiently adequate tracer
of the UVOIR bolometric light curve.
This is probably at least partially the cause of the systematic difference
between the days~51---109 and days~207---451 behavior.
On the other hand we can improve the overall fit by replacing 
model~W7 with fiducial time $t_{\ApJstyle 0}=40.74$~days 
with a model with fiducial time $t_{\ApJstyle 0}=47$~days: 
the days~51--109 region is better fit;
the days~207-451 region is fit only slightly worse.
Perhaps, a model with a $t_{\ApJstyle 0}$ value significantly
larger than the $t_{\ApJstyle 0}$ value of model~W7 is needed.

     The dispersion of the day~207---451 data is consistent
with the probable uncertainty in these data of $\lapprox 0.1$~dex
(Cappellaro~\etal 1997).

     That the day~944 $V$ datum is close to the RDE deposition curves
at all is remarkable.
The prediction for SNe~Ia is that the infrared catastrophe 
should set in $\sim 400$--$600$ days after explosion
(e.g., Axelrod 1980, p.~70, 133;
Fransson~\etal 1996;  see also \S~2).
This means that a $V$ light curve fitted to the RDE deposition curve
in the early quasi-steady state phase is expected to fall well below
the RDE deposition curve later on when most of the UVOIR bolometric
emission shifts to the infrared.
Fransson~\etal (1996) suggested that perhaps clumping in the ejecta
or more optical emission from recombination cascades might prevent the
infrared catastrophe by keeping strong emission in the optical.
In addition to the infrared catastrophe, the quasi-steady state
should breakdown after about day~600 (e.g., Axelrod 1980, p.~48).
A third factor is that PE particle escape could reduce the
UVOIR bolometric light curve below the complete PE particle trapping
prediction we have made (see \S~2).

    Of course, not too much weight should be placed on a single isolated
$V$ datum even if it is accurate.
However, there is also one very late $AB$ band magnitude ($AB\approx B-0.2$)
for SN~Ia~1972E from day~732 (again taking the rise time to maximum
as 18~days) that is consistent within uncertainty with quasi-exponential
decline from the day~98--434 period (Kirshner \&~Oke 1975).
(Note the $B$ light curve is not considered as good a tracer of the
UVOIR bolometric
light curve as the $V$ light curve, but $V$ data beyond day~434 is not
available for SN~1972E.)
Somehow a combination of effects seem to keep the very late-time SN~Ia light
curves (out to about day~950) quasi-exponential or nearly so with respect
to early late-time behavior.
 
     We have only considered one SN~Ia here as an example.
There certainly are variations between late-time $V$ light curves from
different SNe~Ia (e.g., Cappellaro~\etal 1997).
However, there are certain average SN~Ia observational results
that can be compared to our RDE deposition curves.
Leibundgut (1988) and Leibundgut~\etal (1991b) presented template
light curves for SNe~Ia that extend into in the early quasi-steady
state phase.
The half-life of an exponential fitted to the day~80--120 period
of their $V$ template is 29.4 days.
The mean SN~Ia light $V$ light curve of Doggett \&~Branch (1985) drawn
from older data has a half-life of 52 days for the period of about
160--320 days after explosion.
These half-lives are in fair agreement with the half-lives
of 29.23~days and 54.95~days for exponential fits to the RDE
deposition for the $x$-ranges $[2,3]$ (approximately days~80--120)
and $[4,8]$ (approximately days~160--320), respectively,
for our fiducial SN~Ia model (see Table~2).
The corresponding half-lives for model~W7 
$N_{\ApJstyle\rm Ni}^{\ApJstyle *}(t)$ curve (which is very close
agreement with our fiducial model curve) (see \S~3.3)
are 29.16~days and 54.88~days, respectively.
The good agreement between the mean observational half-lives and
the half-lives of the exponential fits to our fiducial SN~Ia model and
model~W7 $N_{\ApJstyle\rm Ni}^{\ApJstyle *}(t)$ RDE deposition curves
and the fairly good fit of the model~W7 RDE deposition curves
to the SN~1992A $V$ light curve in Figures~4a and~4b
suggest that model~W7 and similar
models are adequate average models for SNe~Ia in respect
to RDE deposition.

     Cappellaro~\etal (1997) have done a more extensive, detailed
analysis of late-time SN~Ia $V$ light curves and have included positron
transport in their calculations.
Probably their most notable conclusion was that even normal SNe~Ia
could come from a range of masses.
The greater masses are needed for the slowest declining SNe~Ia:
i.e., those whose $V$ light curves approach the late 1st order
exponential phase most slowly.
That their slowest declining SNe~Ia, by one measure at least (viz.,
the width of nebular phase iron-peak element emission lines), seem
to have the lowest central concentration of initial $^{\ApJstyle 56}$Ni
makes the need for mass variation seem all the stronger.
Recall that lower concentration of $^{\ApJstyle 56}$Ni 
(implying smaller $q$ and $t_{\ApJstyle 0}$) aids $\gamma$-ray
escape, and so hastens the approach to the late 1st order
exponential phase (see \S~3.3).

     Cappellaro~\etal (1997), however, did not perform a detailed
NLTE treatment of the conversion of RDE into UVOIR emission.
Such treatments have been done (e.g., Axelrod 1980;  Ruiz-Lapuente~\etal 1995;
Ruiz-Lapuente 1997;
Liu~\etal 1997;  Liu, Jeffery, \&~Schultz 1998), but limitations,
particularly in atomic data input, have limited the conclusions.
More detailed, less limited NLTE treatments and better late-time
observations of SNe~Ia would greatly aid in determining the SN~Ia
nature.
\vskip 2\baselineskip

\centerline{5.\ \  SN II 1987A}\nobreak
\vskip\baselineskip\nobreak
     Many detailed studies of the late-time emission of
SN~II~1987A have been done and these have made considerable progress
in understanding this object (e.g.,
Woosley 1988;
Bouchet~\etal 1991;
Suntzeff~\etal 1991, 1992;
Li, McCray, \&~Sunyaev 1993;
Fransson \&~Kozma 1993;
Nomoto~\etal 1994a;
Fransson~\etal 1996;
Suntzeff 1998,
and references therein).
Here we simply wish to show that the $N_{\ApJstyle\rm Ni}^{\ApJstyle *}(t)$
RDE deposition curve for our fiducial core-collapse model (see \S~3.4)
with fiducial time $t_{\ApJstyle 0}=563.969041$~days
adequately accounts for the early late-time (i.e., the
quasi-steady state) UVOIR bolometric light curve of SN~1987A.

      We plot the fiducial curve in Figures~5a and~5b along with
the UVOIR bolometric light curve of SN~1987A (Bouchet~\etal 1991;
Suntzeff~\etal 1991).
We have vertically adjusted the SN~1987A curve to fit
the fiducial curve in day 130--200 period.
The explosion epoch of SN~1987A is, of course, solidly fixed by 
the SN~1987A 
neutrino burst (Bionta~\etal 1987;  Hirata~\etal 1987).
For the day~134--432 period we have only plotted the
very high accuracy bolometric data derived from spectrophotometry
(Bouchet~\etal 1991). 

      As can be seen from Figure~5a, SN~1987A very suddenly
entered the quasi-steady state phase where RDE deposition equals
the UVOIR bolometric light curve.
The transition of the SN~1987A curve
from steep decline to slow decline took place over
the day 120--128 period (Suntzeff \&~Bouchet 1991).
The early slow decline phase of the SN~1987A curve is in fact 
the manifestation of the early 1st order
exponential phase of RDE deposition that we discussed in \S~3.2.
This can be seen by noting that the deposition curve in Figure~5a
only slowly diverges from the complete trapping curve and only
well after the $^{\ApJstyle 56}$Ni deposition phase has ended. 
Bouchet~\etal (1991) give the SN~1987A curve half-life for the day~134--300
period to be $76.0\pm0.2$~days.
For the fiducial model, the day~134--300 period corresponds in reduced
time to the $x$-range $[0.238,0.532]$.
The half-life for an exponential fit to the fiducial curve for the $x$-range
$[0.23,0.53]$ is 75.91~days (see Table~2). 
The observed and fiducial model half-lives are in good agreement.

     The fiducial curve fits the SN~1987A curve very well from the
beginning of the quasi-steady state phase until about day~700.
Of course, we chose the parameter values for the fiducial model
in order to get this good fit (see \S~3.4). 
In reality, however, we have only shown that the
$N_{\ApJstyle\rm Ni}^{\ApJstyle *}(t)$ function with a well chosen
$t_{\ApJstyle 0}$ value gives a good fit.
Thus, the underlying picture of $^{\ApJstyle 56}$Co RDE deposition
in a medium of increasing transparency is certainly adequate.
However, our parameterized core-collapse model may not itself be very
adequate to account for SN~1987A.
But even if it were exactly right for SN~1987A, we would still not be able to
determine the right parameter values without more constraints than the
SN~1987A curve.
Model~10H suggests some other constraints, but this is not sufficient to find a
unique fit since among other things model~10H does not exactly
correspond to our parameterized core-collapse supernova model (see \S~3.4).

     After $\sim 800$ days after explosion, the UVOIR bolometric light curve of
SN~1987A starts declining distinctly less steeply.
(This change is not obvious without showing post-day-1000 data.)
The change is adequately accounted for by $^{\ApJstyle 57}$Co RDE deposition
becoming
important (e.g., Suntzeff~\etal 1992) and the ending of quasi-steady state
of the ejecta by the ionization freeze-out (e.g., Fransson \&~Kozma 1993;
Fransson~\etal 1996). 
These effects are, of course, not included in the 
$N_{\ApJstyle\rm Ni}^{\ApJstyle *}(t)$ function.
\vskip 2\baselineskip

\centerline{6.\ \ SN Ic 1998bw}\nobreak
\vskip\baselineskip\nobreak
     Supernova SN~1998bw is a very interesting object because 
it is a possible cause of the $\gamma$-ray burst GRB980425 which
occurred on 1998 April~25.91 (Galama~\etal 1998;  Iwamoto~\etal 1998).
It was also remarkable just as a SN~Ic in being unusually bright and
having very fast outer ejecta (Iwamoto~\etal 1998).
And the supernova showed net intrinsic polarization (Kay~\etal 1988)
suggesting large asymmetry (H\"oflich, Wheeler, \&~Wang 1998).
Intrinsic polarization, however, is not it seems an unusual feature of
core-collapse supernovae (Wang, Wheeler, \&~H\"oflich 1998).

     Yet another remarkable aspect of SN~1998bw is that its
$BVI$ light curves for $\sim 60$--$185$ days after explosion (almost
certainly entirely in the quasi-steady state phase) are
very exponential, but do not have the $^{\ApJstyle 56}$Co
half-life (McKenzie \&~Schaefer 1999).
(Here we assume that GRB980425 sets the date of explosion.
From early-time light curve analysis of SN~1998bw,
Iwamoto~\etal [1998] find that
the explosion epoch agrees with GRB980425 to within $+0.7$/$-2$ days.)
As discussed in \S~3.2, there is a continuum of quasi-exponential phases
between the inflection point 1st order exponential phase
and the late 1st order exponential phase as $t\to\infty$.
The half-lives determined for exponential fits to the quasi-exponential
phases will not be the $^{\ApJstyle 56}$Co half-life.
At first one might suppose that the SN~1998bw light curves for 
the day~60--185 period are from a quasi-exponential phase. 
The light curves, however, appear truly exponential, not
just quasi-exponential.
Over the day~60--185 period, the data with uncertainty of
$\lapprox 0.1$~magnitudes ($\lapprox 0.04$~dex) shows perhaps
only a slight trace of curvature on a semi-logarithmic plot
and that trace has been deemed insignificant
(McKenzie \&~Schaefer 1999, Fig.~1).

     It may be in the SN~1998bw case that the light curves
are showing the signature of the inflection point 1st order exponential phase
of RDE deposition that occurs near $t=t_{\ApJstyle 0}$ (see \S~3.2).
Militating against this idea is the fact that the light curve half-lives
are not all the same.
The $B$, $V$, and $I$ half-lives are $53.4\pm0.8$, $40.9\pm0.7$,
and $41.6\pm0.7$~days, respectively.
If $V$ and $I$ dominated the emission, then one could tentatively
conclude that they trace a UVOIR bolometric light curve with a half-life of
about $41$~days.
But spectra through day~136 show that the flux in the $B$ band is comparable
to that in the $VI$ bands (Patat~\etal 1999).
The sum of different exponentials is not exactly exponential.
Nevertheless it is difficult to believe that one can get close to exponentials
in three bands, two of them with almost the same half-life, without
the driving RDE deposition being nearly an exponential.
We will assume that the UVOIR bolometric light curve for the day~60--185 period
is nearly an exponential with the UVOIR bolometric light curve half-life
$44$~days estimated by McKenzie \&~Schaefer (1999).

      Figure~6a shows a schematic SN~1998bw UVOIR bolometric light curve
for the day~60--185 period with half-life 44~days, a fitted
$N_{\ApJstyle\rm Ni}^{\ApJstyle *}(t)$ RDE deposition curve,
and the complete trapping curve.
The $N_{\ApJstyle\rm Ni}^{\ApJstyle *}(t)$ curve has been
fitted to the schematic curve by the following procedure.
The vertical level of the schematic SN~1998bw curve was determined by a
least-squares fit to the $N_{\ApJstyle\rm Ni}^{\ApJstyle *}(t)$ curve.
The $N_{\ApJstyle\rm Ni}^{\ApJstyle *}(t)$ curve was then varied by
varying $t_{\ApJstyle 0}$ (which is $N_{\ApJstyle\rm Ni}^{\ApJstyle *}(t)$'s
only free parameter when $\tau_{\ApJstyle\rm ch,0}$ is set to 1 as it
always is) in order to minimize the maximum deviation between the
$N_{\ApJstyle\rm Ni}^{\ApJstyle *}(t)$ curve and the schematic SN~1998bw
curve.
The final fitted $N_{\ApJstyle\rm Ni}^{\ApJstyle *}(t)$ curve has a
fiducial time $t_{\ApJstyle 0}=134.42$~days (for $\tau_{\ApJstyle\rm ch,0}=1$).
The maximum deviation of fit is 0.024~dex ($0.06$ magnitudes).
McKenzie \&~Schaefer (1999) find no evidence for systematic deviations
from exponential behavior in the broad band light curves over
the day~60--185 period greater than $0.02$~dex ($0.05$~magnitudes).
If the actual UVOIR bolometric light curve of SN~1998bw has the same closeness
to exact exponential behavior as the observed broad band light curves,
then our fitted $N_{\ApJstyle\rm Ni}^{\ApJstyle *}(t)$ curve has
about as much deviation from exponential behavior has can be tolerated. 
The maximum deviations of $N_{\ApJstyle\rm Ni}^{\ApJstyle *}(t)$ curves with
fiducial times 90, 120, 150, and~180~days
are 0.12~dex, 0.048~dex, 0.061~dex, and 0.11~dex, respectively.
None of these curves could be tolerated given the supposed exponential
behavior of the UVOIR bolometric light curve. 

     Our fitted $t_{\ApJstyle 0}=134.42$~days puts the inflection point
of the RDE deposition at day~139.90 (see \S~3.2).
Thus, as we could have anticipated, a best fit to a highly exponential
region of
UVOIR bolometric light curve that does not have the $^{\ApJstyle 56}$Co
half-life can force the fitted region to be in the inflection point 1st order
exponential region of the $N_{\ApJstyle\rm Ni}^{\ApJstyle *}(t)$ function.
That the $N_{\ApJstyle\rm Ni}^{\ApJstyle *}(t)$ approximation
tends to be weakest when time is of order $t_{\ApJstyle 0}$ (see \S~3.2) is
a weakness of our analysis.
But a single inflection point 1st order exponential region is expected even
in a more sophisticated treatment of RDE deposition provided the
supernova ejecta is fairly smoothly varying and the $^{\ApJstyle 56}$Ni
contribution becomes exponentially small soon enough (see \S~3.2). 

     In Figure~6b we predict RDE deposition curve and quasi-steady state
UVOIR bolometric
light curve out to day~1000 using $N_{\ApJstyle\rm Ni}^{\ApJstyle *}(t)$
with $t_{\ApJstyle 0}=134.42$~days.
The schematic SN~1998bw curve is again displayed.
The actual RDE deposition curve will probably differ from the prediction
due to long-lived radioactive species;  the UVOIR bolometric light curve
due to those species and time-dependent effects (see \S~2).
The infrared catastrophe (see \S~2) does not in itself cause
deviations from $N_{\ApJstyle\rm Ni}^{\ApJstyle *}(t)$-like
behavior in the UVOIR bolometric light curve, but it
does make it harder to measure the UVOIR bolometric light curve.
The predicted RDE deposition curve's approach to the
complete escape curve is sufficiently slow that non-exponential behavior
in the UVOIR bolometric light curve would be difficult to detect without
high quality data out to several hundreds of days after explosion. 

     Taking our fitted $N_{\ApJstyle\rm Ni}^{\ApJstyle *}(t)$ curve at
face value, we can make an estimate of the ejecta mass by inverting
an expression for $t_{\ApJstyle 0}$ for a given parameterized model.
We will assume the parameterized core-collapse model of \S~3.4 and invert
equation~(\showeqn -0) for mass.
Recall that this parameterized model assumes that the bulk of the
ejecta has density going as the inverse square of velocity.
Some support for this density profile for outer ejecta is provided by
Branch (1999) who found that it worked well in a parameterized LTE analysis
of the photospheric epoch spectra of SN~1998bw.

     We now need to estimate the model parameters.
From Iwamoto~\etal (1998) it is known that SN~1998bw ejecta extend
out to velocities $\gapprox 28,000\,{\rm km\,s^{\ApJstyle -1}}$.
The analysis of Branch (1999) suggests ejecta velocities
$\gapprox 60,000\,{\rm km\,s^{\ApJstyle -1}}$ and that the inverse-square
density profile extends even that far.
We will be more conservative and assume that the vast bulk of the ejecta
is confined to an inner high-mass core as in model~10H upon which we based
our parameterized core-collapse model.
The estimated half-width of one emission line of the
[O~I]~$\lambda\lambda6300,6364$   
emission above the estimated
pseudo-continuum of the day~136 nebular spectrum of SN~1998bw (Patat~\etal 1999)
corresponds to a velocity of order $20000\,{\rm km\,s^{\ApJstyle -1}}$.
We assume $20000\,{\rm km\,s^{\ApJstyle -1}}$
is the outer velocity $v_{\ApJstyle\rm b}$ of the high-mass core.
Given that the analysis of Branch (1999) and that the pseudo-continuum is
ill-defined, $20000\,{\rm km\,s^{\ApJstyle -1}}$ could be a factor of 2 or 3
too small.
For the inner velocity $v_{\ApJstyle\rm a}$ of the high-mass core,
we have no strong evidence.
But if there was a large hollow at the center of the ejecta the
unblended emission lines would tend to be flat-topped (e.g., Mihalas 1978,
p.~477;  Jeffery \&~Branch 1990, p.~190).
The [O~I]~$\lambda\lambda6300,6364$ 
emission in the day~136 spectrum has a very sharp peak (which is probably
mostly due to the strongest line), and we estimate that
a low-mass, low-density center effectively setting 
$v_{\ApJstyle\rm a}\gapprox2000\,{\rm km\,s^{\ApJstyle -1}}$ is unlikely.
We will just assume $v_{\ApJstyle\rm a}=700\,{\rm km\,s^{\ApJstyle -1}}$
as we did for our fiducial core-collapse model.
For the concentration factor $q$ we choose $1$ implying that the
$^{\ApJstyle 56}$Ni was concentrated in the innermost regions of
the high-mass core.
Note that Iwamoto~\etal (1998) concluded that large-scale mixing of
the $^{\ApJstyle 56}$Ni was needed to fit the fast rise of
the estimated pre-maximum SN~1998bw UVOIR bolometric light curve.
The real $q$ could be significantly less than 1. 
For $\kappa$ we choose $0.028\,{\rm cm^{\ApJstyle 2}\,g^{\ApJstyle -1}}$.
This an appropriate value for all-metal ejecta in the transition phase between
optically thick and thin limits as
determined from Table~1 of Jeffery (1998a) or Table~2 of Jeffery (1998b).
(SNe~Ic are believed to have no significant hydrogen or helium.)
In extracting our $t_{\ApJstyle 0}$ value from the fit to the
schematic SN~1998bw
curve we have already set $\tau_{\ApJstyle\rm ch,0}=1$ as we always do.

     The expression for mass for our parameterized core-collapse model is
$$\eqalignno{
M&=15\,M_{\ApJstyle\odot}\times
\left({t_{\ApJstyle 0}\over 563.969041\,\,{\rm days}}\right)^{\ApJstyle 2}
\left({0.035\,{\rm cm^{\ApJstyle 2}\,g^{\ApJstyle -1}}\over\kappa }\right) \cr
&\qquad \times \left({\tau_{\ApJstyle\rm ch,0}\over 1}\right)
  \left[{1-\left(v_{\ApJstyle\rm a}/v_{\ApJstyle\rm b}\right)
         \over
         1-\left(v_{\ApJstyle\rm a}/v_{\ApJstyle\rm b}\right)\left(1/q\right)
        }\right]                                    \cr
&\qquad\qquad\times
  \left({v_{\ApJstyle\rm a} \over 700\,{\rm km\,s^{\ApJstyle -1}} } \right)
  \left({v_{\ApJstyle\rm b} \over 5000\,{\rm km\,s^{\ApJstyle -1}} } \right)
  \left({1\over q}\right)
                                          \,\, . &(\eqn)\cr}$$
With the chosen parameter values, we obtain a
mass of $4.26\,M_{\ApJstyle\odot}$.
Even given that our parameterized core-collapse model is appropriate
for SN~1998bw, this value could be factor of few too small:
$v_{\ApJstyle\rm b}$ could be 2 or 3 times what we assume and $q$ could
significantly less than 1.
The optimum $v_{\ApJstyle\rm a}$ value could also be different from
what we have assumed by a factor of 2 or more either up or down. 
But since our parameterized core-collapse model may not be appropriate,
our mass estimate may be only order of magnitude in any case.

      Other mass determinations or lower limits on mass for SN~1998bw have
been given.
For example, Iwamoto~\etal (1998) find that
a $13.8\,M_{\ApJstyle\odot}$ carbon-oxygen
star explosion model can reproduce fairly well 
estimated early-time SN~1998bw UVOIR bolometric light curve.
And Branch (1999) from his photospheric epoch parameterized LTE analysis
estimated that SN~1998bw had
$\sim 6\,M_{\ApJstyle\odot}$ above $7000\,{\rm km\,s^{\ApJstyle -1}}$.
There is still, however, considerable debate about the SN~1998bw mass and
structure.
Given an adequate parameterized structural model for the supernova,
the $t_{\ApJstyle 0}$ parameter obtained
by fitting the UVOIR bolometric light curve from the quasi-steady state phase
would help to constrain the supernova's mass and/or other parameters.
Since SN~1998bw may be highly asymmetric, the determination
of an adequate parameterized structural model may be difficult.
\vskip 2\baselineskip

\centerline{7.\ \  CONCLUSIONS}\nobreak
\vskip\baselineskip\nobreak
     We have given a presentation of a simple, approximate, analytic
treatment of RDE deposition in supernovae from the decay chain
$^{\ApJstyle 56}$$\rm Ni\to^{\ApJstyle 56}$$\rm Co\to^{\ApJstyle 56}$$\rm Fe$.
The treatment provides
a straightforward understanding of the exponential/quasi-exponential
behavior of the UVOIR bolometric luminosity
and a partial understanding of the exponential/quasi-exponential behavior
of the broad band light curves.
The treatment reduces to using the normalized
$N_{\ApJstyle\rm Ni}^{\ApJstyle *}(t)$ deposition function (see \S~3.2)
as an analysis tool.
(The absolute deposition is determined by specifying the initial
$^{\ApJstyle 56}$Ni mass or fitting absolute UVOIR supernova luminosity.)
The time evolution of $N_{\ApJstyle\rm Ni}^{\ApJstyle *}(t)$ is determined
by three time scales:   the half-lives of $^{\ApJstyle 56}$Ni and
$^{\ApJstyle 56}$Co, and a fiducial time parameter $t_{\ApJstyle 0}$ that
governs the $\gamma$-ray optical depth behavior of a supernova.
The $t_{\ApJstyle 0}$ parameter can be extracted from a structural
supernova model, and we have shown examples of how this is done in \S\S~3.3
and~3.4.
It can also be obtained by fitting
the RDE deposition curve from a more detailed treatment of deposition
as we did in \S~3.3 or by fitting to an observed
UVOIR bolometric light curve from the quasi-steady state phase of a supernova
as we did in \S\S~5 and~6. 
A $t_{\ApJstyle 0}$ parameter extracted from observations can provide
a constraint on the important physical parameters of a supernova.
Effective use of this constraint, however, requires having an adequate
parameterized structural supernova model.

     The $N_{\ApJstyle\rm Ni}^{\ApJstyle *}(t)$ function is used
to analyze the preliminary UVOIR bolometric light curve of
SN~1998bw (the possible cause of GRB980425) (\S~6).
The SN~1998bw fiducial time $t_{\ApJstyle 0}$ is found to be $134.42$~days
and a prediction
is made for the evolution of the SN~1998bw RDE deposition curve
and quasi-steady state UVOIR bolometric light curve out to day~1000 after
the explosion.
A crude estimate (perhaps a factor of a few too small)
of the SN~1998bw mass obtained from a parameterized
core-collapse model and $t_{\ApJstyle 0}=134.42$~days is
$4.26\,M_{\ApJstyle\odot}$.
As further examples of the simple analytic treatment, the RDE deposition
and luminosity evolution
of SN~Ia~1992A and SN~II~1987A have also been examined (see \S\S~4 and~5).

     The simple analytic treatment of RDE deposition has
actually existed at least since 
Colgate~\etal (1980a, b), but has not hitherto been given a detailed or
general presentation as far as we know. 
The main value of this paper is the explicit, detailed, general presentation
of this analytic treatment.
\vskip 2\baselineskip

   This work was supported by 
   the Department of Physics of the University of Nevada, Las Vegas.
   I thank Enrico Cappellaro for providing me with the $V$~light curve
   of SN~1992A
   and David Branch for his comments. 
\vskip 2\baselineskip

\counteqn=0
\centerline{APPENDIX A}\nobreak
\vskip\baselineskip\nobreak
\centerline{THE EXPONENTIAL DENSITY PROFILE MODEL}\nobreak
\centerline{FOR HOMOLOGOUSLY EXPANDING SUPERNOVAE}\nobreak
\vskip\baselineskip\nobreak
      As discussed in \S~3.3, hydrodynamic explosion models for 
SNe~Ia often have density profiles in the homologous expansion epoch 
that are close to exponentials (i.e., inverse exponentials)
as functions of radial velocity.
In this appendix we present some useful analytic results for exponential
density profile models for homologously expanding supernovae.

     First, we note the following general integral solution that is
useful in developing the analytic results:
$$
I_{\ApJstyle n}(z)=
\int_{\ApJstyle z}^{\ApJstyle\infty}dz'\,
 \left(z'\right)^{\ApJstyle n}\exp\left(-z'\right)
=\exp\left(-z\right)
   \sum_{\ApJstyle k=0}^{\ApJstyle n} {n!\over k!}z^{\ApJstyle k} \,\, ,
                                                 \eqno({\rm A}\eqn) $$
where $n\geq0$ is an integer.

     Next recall that the radius $r$ of a mass element in homologous
expansion after the initial radii have become insignificant is
given by
$$ r=vt                   \,\,  ,                \eqno({\rm A}\eqn) $$
where $v$ is the mass element's radial velocity and
$t$ is the time since explosion.
Recall also that the element's density at any velocity
declines as $t^{\ApJstyle -3}$.
We will use radial velocity as comoving radial coordinate
and define a dimensionless radial coordinate $z$ by
$$               z=v/v_{\ApJstyle e}    \,\, ,       \eqno({\rm A}\eqn)$$
where $v_{\ApJstyle e}$ is the $e$-folding velocity of an exponential
model's density profile.
The expression for the density profile can then
be written 
$$ \rho(v,t)=\rho_{\ApJstyle \rm ce,0}
        \left({t_{\ApJstyle 0}\over t}\right)^{\ApJstyle 3}
        \exp\left(-z\right)   \,\,  , \eqno({\rm A}\eqn)$$
where $\rho_{\ApJstyle \rm ce,0}$ is the central density at
fiducial time $t_{\ApJstyle 0}$.

     Using the equations~(A\showeqn -3)--(A\showeqn -0),
the expression for mass exterior to radius $z$ for
an exponential model is 
$$\eqalignno{
M(z)&=4\pi t^{\ApJstyle 3}
        \int_{\ApJstyle v}^{\ApJstyle\infty}dv\,v^{\ApJstyle 2}
             \rho(v,t)
=M\exp(-z)\left(1+z+{1\over2}z^{\ApJstyle 2}\right)\,\, , &({\rm A}\eqn)\cr}$$
where $M=8\pi\rho_{\ApJstyle\rm ce,0}
  \left(v_{\ApJstyle e}t_{\ApJstyle 0}\right)^{\ApJstyle 3}$ is total mass.
Using the same equations, the expression for
kinetic energy exterior to radius $z$ is
$$\eqalignno{
E(z)&=
4\pi t^{\ApJstyle 3}
   \int_{\ApJstyle v}^{\ApJstyle\infty}dv\,{v^{\ApJstyle 4}\over2}\rho(v,t)\cr
&=6Mv_{\ApJstyle e}^{\ApJstyle 2}
\exp(-z)\left(1+z+{1\over2}z^{\ApJstyle 2}
                 +{1\over6}z^{\ApJstyle 3}+{1\over24}z^{\ApJstyle 4}\right)
                                          \,\, ,       &({\rm A}\eqn)\cr}$$
where
$E=48\pi\rho_{\ApJstyle\rm ce,0}
   \left(v_{\ApJstyle e}t_{\ApJstyle 0}\right)^{\ApJstyle 3}
   v_{\ApJstyle e}^{\ApJstyle 2}
  =6Mv_{\ApJstyle e}^{\ApJstyle 2}$
is total kinetic energy. 

      It is often useful to have expressions for central density
$\rho_{\ApJstyle\rm ce,0}$,
total kinetic energy $E$,
and the ratio of total kinetic energy to total mass $E/M$ in terms
of fiducial parameter values.
Since model~W7 (Nomoto, Thielemann, \&~Yokoi 1984;
Thielemann, Nomoto, \&~Yokoi 1986) is a widely-used standard SN~Ia model
whose density closely approximates an exponential, we will use it as
a basis for the fiducial values.
Model~W7 is a Chandrasekhar mass model, and so has total mass
$M=1.38\,M_{\ApJstyle\odot}$.
The equivalent-exponential model $v_{\ApJstyle e}$ for model~W7 in round
numbers is $2700\,{\rm km\,s^{\ApJstyle -1}}$.
(We explain equivalent-exponential model below.)
For a fiducial time $t_{\ApJstyle 0}$, we choose 1~day for convenience
in simple calculations. 
(Note this fiducial time is not for the same purpose as the fiducial time
we use in the main text for optical depth evolution.)
Using
$M=1.38\,M_{\ApJstyle\odot}$,
$v_{\ApJstyle e}=2700\,{\rm km\,s^{\ApJstyle -1}}$,
and $t_{\ApJstyle 0}=1\,{\rm day}$ as fiducial values,
the desired expressions are
$$\eqalignno{
\rho_{\ApJstyle\rm ce,0}&
={M\over 8\pi
  \left(v_{\ApJstyle e}t_{\ApJstyle 0}\right)^{\ApJstyle 3} } \cr
&=0.860327\times10^{\ApJstyle -8}\,{\rm g\,cm^{\ApJstyle -3}}
  \times \left({M\over 1.38\,M_{\ApJstyle\odot}}\right)  \cr
&\qquad\times  \left({2700\,{\rm km\,s^{\ApJstyle -1}}
        \over v_{\ApJstyle e} }\right)^{\ApJstyle 3}
  \left({1\,{\rm day}\over t_{\ApJstyle 0}}\right)^{\ApJstyle 3} 
                                                \,\, ,  &({\rm A}\eqn)\cr 
E&=6Mv_{\ApJstyle e}^{\ApJstyle 2}
  = 1.20064\,{\rm foe}
  \times \left({M\over 1.38\,M_{\ApJstyle\odot}}\right)
  \left({v_{\ApJstyle e}\over 2700\,{\rm km\,s^{\ApJstyle -1}}
       }\right)^{\ApJstyle 2} 
              \,\, ,            &({\rm A}\eqn)\cr
\noalign{\noindent\rm and}
{E\over M}&=6 v_{\ApJstyle e}^{\ApJstyle 2}
             =4.37400\times 10^{\ApJstyle 17}\,{\rm ergs\,g^{\ApJstyle -1}}
             \times
             \left({v_{\ApJstyle e}\over 2700
                   \,{\rm km\,s^{\ApJstyle -1}} }\right)^{\ApJstyle 2}
                                                                 &\cr
             &= 0.870032 
            \,{{\rm foe}\,M_{\ApJstyle\odot}^{\ApJstyle -1}} \times 
  \left({v_{\ApJstyle e}\over 2700
     \,{\rm km\,s^{\ApJstyle -1}} }\right)^{\ApJstyle 2}
                      \,\, ,                   &({\rm A}\eqn) \cr}$$
where a foe (for ten to the fifty-one ergs) is a standard supernova
energy unit of $10^{\ApJstyle 51}\,{\rm ergs}$.
(Recall from the main text that for numerical consistency we treat
fiducial values and the 
solar mass unit $M_{\ApJstyle\odot}=1.9891\times10^{\ApJstyle 33}\,{\rm g}$
[Lide \&~Frederikse 1994, p.~14-2]
as nearly exact numbers.)

      Since SN~Ia hydrodynamic explosion models often have 
density profiles that are close to exponential, there is some interest
in specifying exponential models that in some ways 
are equivalent to those explosion models:  viz., equivalent-exponential models.
Two parameters are needed to specify an equivalent-exponential model.
The explosion model's central density and an $e$-folding velocity
drawn from a fit to the explosion model's density profile may not be good
choices for these parameters:
the central density may not be representative of the overall explosion model
and good criteria for obtaining the fitted $e$-folding velocity need
to be specified.
We will instead choose the explosion model's total mass $M$ and total kinetic
energy $E$ as the parameters.
Together these parameters should yield an equivalent-exponential model
that globally is much like the original explosion model:  the closer the
explosion model is to an exponential model, the better the likeness, of course. The expression for the equivalent-exponential model $e$-folding velocity
follows from equation~(A\showeqn -1) or~equation~(\showeqn -0):
$$  v_{\ApJstyle e}=\sqrt{{1\over6}{E\over M}}
                   =2700\,{\rm km\,s^{\ApJstyle -1}}\times
                    \sqrt{ \left({E\over 1.20064\,{\rm foe} }\right)
                           \left({1.38\,M_{\ApJstyle\odot} \over M}\right)
                         }                  \,\,   .  \eqno({\rm A}\eqn)$$
Given $v_{\ApJstyle e}$, the central density of the equivalent-exponential
model is obtainable from equation~(A\showeqn -3).

     We have calculated the equivalent-exponential model $v_{\ApJstyle e}$ for
model~W7 to be $2724\,{\rm km\,s^{\ApJstyle -1}}$.
Since the calculation depends a bit on how one integrates the
kinetic energy over the finite number of zones that make up model~W7,
there is no real reason for using exactly $2724\,{\rm km\,s^{\ApJstyle -1}}$
as a fiducial $v_{\ApJstyle e}$ value.
Thus, above we chose the round number $2700\,{\rm km\,s^{\ApJstyle -1}}$.

     It is useful to specify a few other exponential model results.
The central atom density $n_{\ApJstyle\rm at}$ is given by
$$\eqalignno{
n_{\ApJstyle\rm at}&={ \rho_{\ApJstyle\rm ce,0}
                       \over m_{\ApJstyle\rm amu} \mu_{\ApJstyle\rm at} }\cr
                   &=0.925179\times10^{\ApJstyle 14}
                       \,{\rm cm^{\ApJstyle -3}}  \cr
&\qquad \times \left({56\over \mu_{\ApJstyle\rm at}}\right)
\left({M\over 1.38\,M_{\ApJstyle\odot}}\right)
  \left({2700\,{\rm km\,s^{\ApJstyle -1}}
      \over v_{\ApJstyle e} }\right)^{\ApJstyle 3}
  \left({1\,{\rm day}\over t_{\ApJstyle 0}}\right)^{\ApJstyle 3} 
                                                \,\, , &({\rm A}\eqn)\cr}$$
where $m_{\ApJstyle\rm amu}=1.6605402(10)\times10^{\ApJstyle -24}\,{\rm g}$
(Lide \&~Frederikse 1994, p.~1-1:  uncertainty in the last digits is given in
the brackets) is the atomic mass unit (amu) 
and $\mu_{\ApJstyle\rm at}$
is the mean atomic mass, and where the second expression is in terms
of fiducial values.
The mean atomic mass is defined by
$$ \mu_{\ApJstyle\rm at}^{\ApJstyle -1}=
  \sum_{\ApJstyle i}{X_{\ApJstyle i}\over A_{\ApJstyle i}}
                                                  \,\, , \eqno({\rm A}\eqn)$$
where the sum is over all elements $i$,
$X_{\ApJstyle i}$ is the mass fraction of element $i$,
and $A_{\ApJstyle i}$ is element $i$'s atomic mass.
Since the center of model~W7 is dominated by stable $^{\ApJstyle 56}$Fe
(even at time zero) and this may be typical of SNe~Ia, we chose the fiducial
value of $\mu_{\ApJstyle\rm at}$ to be 56, the whole number atomic mass of
$^{\ApJstyle 56}$Fe.
Note that even if $^{\ApJstyle 56}$Fe (or $^{\ApJstyle 56}$Ni or
$^{\ApJstyle 56}$Co) do not dominate the center, iron peak
elements (with $A$ values fairly close to 56) almost certainly do.

     The central free electron density $n_{\ApJstyle\rm e}$ is given by
$$\eqalignno{
n_{\ApJstyle e}&={ \rho_{\ApJstyle\rm ce,0}
                   \over m_{\ApJstyle\rm amu}\tilde\mu_{\ApJstyle\rm e} }\cr
               &=0.925179\times10^{\ApJstyle 14}
                       \,{\rm cm^{\ApJstyle -3}}  \cr
&\qquad\times\left({56\over \tilde\mu_{\ApJstyle\rm e}}\right)
\left({M\over 1.38\,M_{\ApJstyle\odot}}\right)
  \left({2700\,{\rm km\,s^{\ApJstyle -1}}
     \over v_{\ApJstyle e} }\right)^{\ApJstyle 3} 
  \left({1\,{\rm day}\over t_{\ApJstyle 0}}\right)^{\ApJstyle 3}
                                               \,\, , &({\rm A}\eqn)\cr}$$
where $\tilde\mu_{\ApJstyle e}$ is the mean atomic mass per free electron
and where the second expression is again in terms of fiducial values.
The mean atomic mass per free electron is defined by
$$ \tilde\mu_{\ApJstyle e}^{\ApJstyle -1}=
  \sum_{\ApJstyle i}{\tilde X_{\ApJstyle i}\tilde Z_{\ApJstyle i}
                    \over A_{\ApJstyle i}}     \,\, ,  \eqno({\rm A}\eqn)$$
where the sum is over all ions $i$,
$\tilde X_{\ApJstyle i}$ is the mass fraction of ion $i$,
$\tilde Z_{\ApJstyle i}$ is the charge on ion $i$,
and $A_{\ApJstyle i}$ is ion $i$'s atomic mass.
Note that $\tilde\mu_{\ApJstyle e}$ is not the same as 
$\mu_{\ApJstyle e}$, the mean atomic mass per electron defined by equation~(10)
in \S~3.1:  $\mu_{\ApJstyle e}$ accounts for all electrons,
not just free electrons. 
The ionization stage of the center of ejecta probably varies strongly
as a function of time. 
But since the central iron-peak elements are likely to be at least singly
ionized until very late times, it seems most convenient just to choose the
singly ionized state of an $A=56$ species as the fiducial ionization.
Calculations suggest that the central iron will be
mostly singly ionized at about day~300 after explosion
(e.g., Ruiz-Lapuente~\etal 1995;  Liu~\etal 1998).

      For a constant opacity $\kappa$, the radial optical depth from 
radius $z$ to infinity is
$$\eqalignno{
 \tau(z)
      &=\kappa\rho_{\ApJstyle \rm ce,0} v_{\ApJstyle e}t
       \left({t_{\ApJstyle 0}\over t}\right)^{\ApJstyle 3}
       \exp(-z)                                                   
      ={\kappa M \over 8\pi v_{\ApJstyle e}^{\ApJstyle 2}
          t^{\ApJstyle 2}  }
        \exp(-z)                                                \cr
      &=\tau_{\ApJstyle\rm ce,0}
       \left({t_{\ApJstyle 0}\over t}\right)^{\ApJstyle 2}\exp(-z)
                    \,\,  ,  &({\rm A}\eqn)\cr}$$
where $\tau_{\ApJstyle\rm ce,0}$, defined by
$$ \tau_{\ApJstyle\rm ce,0}=
    {\kappa M \over 8\pi v_{\ApJstyle e}^{\ApJstyle 2}
           t_{\ApJstyle 0}^{\ApJstyle 2}  } \,\, ,         \eqno({\rm A}\eqn)$$
is the radial optical depth to the center at the fiducial time
$t_{\ApJstyle 0}$.
The constant opacity $\gamma$-ray 
and free electron (i.e., Thomson) radial optical depths from the center to
infinity in terms of the fiducial values are 
$$\eqalignno{
\tau_{\ApJstyle\rm ce,0}^{\ApJstyle \gamma}&=5017.42     \cr
&\qquad\times
 \left(\kappa \over 0.025\,{\rm cm^{\ApJstyle 2}\,g^{\ApJstyle -1}}\right)
\left({M\over 1.38\,M_{\ApJstyle\odot}}\right)            \cr
&\qquad\qquad\times  \left({2700\,{\rm km\,s^{\ApJstyle -1}}
      \over v_{\ApJstyle e} }\right)^{\ApJstyle 2}
  \left({1\,{\rm day}\over t_{\ApJstyle 0}}\right)^{\ApJstyle 2}
                                            &({\rm A}\eqn)\cr
\noalign{\noindent\rm and}
\tau_{\ApJstyle\rm ce,0}^{\ApJstyle e}&=1435.77\times
\left(56 \over \tilde\mu_{\ApJstyle e}\right)
\left({M\over 1.38\,M_{\ApJstyle\odot}}\right)    \cr
&\qquad \times\left({2700\,{\rm km\,s^{\ApJstyle -1}}
      \over v_{\ApJstyle e} }\right)^{\ApJstyle 2}
  \left({1\,{\rm day}\over t_{\ApJstyle 0}}\right)^{\ApJstyle 2}
                                   \,\, ,         &({\rm A}\eqn)\cr}$$
respectively.
For the fiducial value of $\kappa$ for $\gamma$-rays we chose
$0.025\,{\rm cm^{\ApJstyle 2}\,g^{\ApJstyle -1}}$ which is a good
general value for the
effective absorption opacity for $^{\ApJstyle 56}$Co $\gamma$-rays
for an all-metal medium in the optically thin limit
(Swartz~\etal 1995;  Jeffery 1998a, b;  see also \S~3.1).
The electron opacity is given by
$$  \kappa_{\ApJstyle e}={\sigma_{\ApJstyle e}\over
                          m_{\ApJstyle\rm amu}\tilde\mu_{\ApJstyle e} }
                        ={0.40062033\over \tilde\mu_{\ApJstyle e} }
                                   \,\, ,         \eqno({\rm A}\eqn)$$
where $\sigma_{\ApJstyle e}
=0.66524616(18)\times10^{\ApJstyle -24}\,{\rm cm^{\ApJstyle 2}}$
(Lide \&~Frederikse 1994, p.~1-2:  uncertainty in the last digits is given in
the brackets) is the Thomson cross section.
For the fiducial value of $\tilde\mu_{\ApJstyle e}$ we again chose 56
for niceness even though the ionization state and composition of
ejecta vary widely with velocity location and the ionization state
with time also.
\vskip 2\baselineskip

\vfill\eject

\specialpage
{                                 
\centerline{REFERENCES}
\baselineskip=20pt
\vskip\baselineskip
\reference

\refpaper Ambwani, K., \&~Sutherland, P. G. 1988, ApJ, 325, 820.



\refindent Axelrod, T. S. 1980, Ph.D.~thesis, Univ.~of California,
             Santa Cruz  




\refpaper Baade, W., Burbidge, G. R., Hoyle, F., Burbidge, E. M.,
            Christy, R. F., \&~Fowler, W. A. 1956, PASP, 68, 296.

\refpaper Barbon, R., Cappellaro, E., \&~Turatto, M. 1984, A\&A, 135, 27.
\refpaper Bhat, M. R. 1992, Nuclear Data Sheets, 67, 195.

\refpaper Bionta, R. M.,~\etal
                1987, in Phys. Rev. Letters, 58, 1494.


\refpaper Bouchet, P., Phillips, M. M., Suntzeff, N. B., Gouiffes, C.,
            Hanuschik, R. W., \&~Wooden, D. H. 1991, A\&A, 245, 490.
\refindent Branch 1999, in Supernovae and Gamma-Ray Bursts, ed.~M.~Livio,
             in press, astro-ph/9906168
\refbook Browne, E., \&~Firestone, R. B. 1986, Table of Radioactive
          Isotopes (New York:  John Wiley \& Sons, Inc.).





\refpaper Cappellaro, E., Mazzali, P. A., Benetti, S., Danziger, I. J.,
            Turatto, M., Della Valle, M., \&~Patat, F. 1997, A\&A, 328, 203.


\refpaper Chan, K. W., \&~Lingenfelter, R. E. 1993, ApJ, 405, 614.



\refpaper Colgate, S. A., \&~McKee, C. 1969, ApJ, 157, 623.

\refedited Colgate, S.~A., Petschek, A.~G., \&~Kriese, J.~T. 1980a, in AIP 
             Conference Proceedings, No.~63:  Supernova Spectra,
             ed.~R.~Meyerott \&~G.~H.~Gillespie (New York:  American
             Institute of Physics), 7.

\refpaper Colgate, S.~A., Petschek, A.~G., \&~Kriese, J.~T. 1980b, ApJ, 237,
            L81.  




\refpaper Doggett, J. B., \&~Branch, D. 1985, AJ, 90, 2303.
\refedited Fransson, C. 1994, in Supernovae:
             Session~LIV of the
             Les~Houches \'Ecole d'\'Et\'e de Physique Th\'eorique,
             ed.~S.~A.~Bludman, R.~Mochkovitch, \&~J.~Zinn-Justin
             (Amsterdam:  North-Holland), 677.

\refedited Fransson, C., Houck, J., \&~Kozma, C. 1996, in Supernovae
             and Supernova Remnants,
             ed.~McCray, R.
             (Cambridge:  Cambridge University Press), 211.

\refpaper Fransson, C., \&~Kozma, C. 1993, ApJ, 408, L25.




\refpaper Galama, T. J.,~\etal 1998, Nature, 395, 670.





\refindent Harkness, R. P. 1991, in {ESO/EIPC Workshop:  SN~1987A
             and Other Supernovae,} ed.~I.~J.~Danziger \&~K.~Kj\"ar (Garching:
             ESO), 447





\refpaper Hirata, K.,~\etal
              1987, in Phys. Rev. Letters, 58, 1490.




\refpaper H\"oflich, P.  1995, ApJ, 443, 89.



\refpaper H\"oflich, P., Khokhlov, A., \&~M\"uller, E. 1992, A\&A, 259, 549.



\refindent H\"oflich, P., Wheeler, J. C., \&~Wang, L. 1999, ApJ, submitted,
             astro-ph/9808086



\refpaper Huo, J. 1992, Nuclear Data Sheets, 67, 523.







\refpaper Iwamoto, K.,~\etal 1998, Nature, 395, 672.









\refindent Jeffery, D. J. 1998a,
             in Stellar Evolution, Stellar Explosions, and Galactic
             Chemical Evolution:  Proc. 2nd Oak Ridge Symposium on
             Atomic \&~Nuclear
             Astrophysics, ed.~A.~Mezzacappa
             (Bristol:  Institute of Physics Publishing), 687,
             astro-ph/9802229

\refindent Jeffery, D. J. 1998b, astro-ph/9811356

\refedited Jeffery, D. J., \&~Branch, D. 1990, in Jerusalem Winter
             School for Theoretical Physics, Vol.~6, Supernovae,
             ed.~J.~C.~Wheeler, T.~Piran, \&~S.~Weinberg (Singapore:  World
             Scientific), 149.   

\refpaper Jeffery, D. J., Leibundgut, B., Kirshner, R. P., Benetti, S.,
             Branch, D., \&~Sonneborn, G. 1992, ApJ, 397, 304.




\refindent Kay, L. E., Halpern, J. P., Leighly, K. M., Heathcote, S.,
             \&~Magalhaes, A. M. 1998, IAU Circ., No.~6969










\refpaper Kirshner, R. P.,~\etal 1993, ApJ, 415, 589.

\refpaper Kirshner, R.~P., \&~Oke, J.~B. 1975, ApJ, 200, 574.








\refindent Lawrence Berkeley National Laboratory Isotopes Project Web Data Base
           1999, http://isotopes.lbl.gov/ (LBL)

\refindent Leibundgut, B. 1988, Ph.D.~thesis, Univ.~of Basel


\refpaper Leibundgut, B., Kirshner, R. P., Filippenko, A. V., Shields, J. C.,
             Foltz, C. B., Phillips,~M.~M., \&~Sonneborn,~G. 1991a, ApJ, 371,
             L23. 





\refpaper Leibundgut, B., Tammann, G. A., Cadonau, R., \&~Cerrito, D. 1991b,
            A\&AS, 89, 537.  

\refpaper Li, H., McCray, R., \&~Sunyaev, R. A. 1993, ApJ, 419, 824.

\refbook Lide, D. R., \&~Frederikse, H. P. R. (ed.) 1994,
          CRC Handbook of Chemistry
          and Physics (Boca Raton:  CRC Press).




\refpaper Liu, W., Jeffery, D. J., \&~Schultz, D. R. 1998,
             ApJ, 494, 812.

\refpaper Liu, W., Jeffery, D. J., Schultz, D. R., Quinet, P.,
             Shaw, J., \&~Pindzola, M. S. 1997,
             ApJ, 489, L141.

\refpaper Liu, W., \&~Victor, G. A. 1994, ApJ, 435, 909.
\refindent McKenzie, E. H., \&~Schaefer, B. E. 1999, PASP, in press










\refbook Mihalas, D. 1978, Stellar Atmospheres (San Francisco:  Freeman).



\refedited Milne, P. A., The, L.-S., Leising, M. D. 1997, in
             Proc.~The Fourth Compton Symposium, ed.~C.~D.~Dermer,
             M.~S.~Strickman, \&~J.~D.~Kurfess (New York:
             American Institute of Physics Press), 1022, astro-ph/9707111.













\refedited Nomoto, K., Shigeyama, T., Kumagai, S., Yamaoka, H.,
             \&~Suzuki, T. 1994a, in Supernovae:
             Session~LIV of the
             Les~Houches \'Ecole d'\'Et\'e de Physique Th\'eorique,
             ed.~S.~A.~Bludman, R.~Mochkovitch, \&~J.~Zinn-Justin
             (Amsterdam:  North-Holland), 489.

\refpaper Nomoto, K., Thielemann, F.-K., \&~Yokoi, K. 1984, ApJ, 286, 644.

\refedited Nomoto, K., Yamaoka, H., Shigeyama, T., Kumagai, S.,
             \&~Tsujimoto, T. 1994b, in Supernovae:
             Session~LIV of the
             Les~Houches \'Ecole d'\'Et\'e de Physique Th\'eorique,
             ed.~S.~A.~Bludman, R.~Mochkovitch, \&~J.~Zinn-Justin
             (Amsterdam:  North-Holland), 199.


\refpaper Nugent, P., Baron, E., Hauschildt, P. H., \&~Branch, D.
            1995, ApJ, 441, L33.


\refindent Nugent, P.,~\etal 1999, in preparation









\refindent Pankey, T., Jr. 1962, Ph.D.~thesis, Howard University


\refindent Patat, F.,~\etal 1999, in preparation
\refindent Pinto, P. A., \&~Eastman, R. G.  1996, astro-ph/9611195
\refedited Ruiz-Lapuente, P. 1997,
             in Proc.~NATO ASI on
             Thermonuclear Supernovae, ed.~P.~Ruiz-Lapuente,
             R.~Canal, \&~J.~Isern
             (Dordrecht:  Kluwer), 681, astro-ph/9604094.





\refpaper Ruiz-Lapuente, P., Kirshner, R. P., Phillips, M. M., Challis, P. M.,
             Schmidt, B. P., Filippenko, A. V., \&~Wheeler, J. C. 1995, ApJ,
             439, 60.




\refindent Ruiz-Lapuente, P., \&~Spruit, H. C. 1998, ApJ, 500, 360,
             astro-ph/9711248

\refpaper Rust, B. W., Leventhal, M., \&~McCall, S. L. 1976, Nature,
            262, 118. 
\refindent Suntzeff, N. B. 1998, in SN 1987A:  Ten Years After:
              The Fifth CTIO/ESO/LCO Workshop, ed.~M.~M.~Phillips
              \&~N.~B.~Suntzeff (Provo:  Astr.~Soc.~of the Pacific),  
              in press 

\refindent Suntzeff, N. B., \&~Bouchet, P. 1991, in Supernovae:
             The Tenth Santa Cruz
             Workshop in Astronomy and Astrophysics, ed.~S.~E. Woosley
             (New~York:  \hbox{Springer-Verlag}), 3.


\refpaper Suntzeff, N. B., Phillips, M. M., Depoy, D. L., Elias, J. H.,
            \&~Walker, A. R. 1991, AJ, 102, 1118.

\refpaper Suntzeff, N. B., Phillips, M. M., Elias, J. H., Depoy, D. L., 
            \&~Walker, A. R. 1992, ApJ, 384, L33.



\refpaper Sutherland, P. G., \&~Wheeler, J. C. 1984, ApJ, 280, 282.

\refpaper Swartz, D. A., Sutherland, P. G., \&~Harkness, R. P. 1995,
            ApJ, 446, 766.



\refpaper Thielemann, F.-K., Nomoto, K., \&~Yokoi, K. 1986, A\&A, 158, 17.





\refpaper Vacca, W. D., \&~Leibundgut, B.  1996, ApJ, 471, L37.




\refindent Wang, L., Wheeler, J. C., \&~H\"oflich, P. 1998, in SN 1987A:
              Ten Years After:
              The Fifth CTIO/ESO/LCO Workshop, ed.~M.~M.~Phillips
              \&~N.~B.~Suntzeff (Provo:  Astr.~Soc.~of the Pacific),  
              in press 
\refpaper Woosley, S. E. 1988, ApJ, 330, 218.







\refedited Woosley, S. E., \&~Weaver, T. A. 1994, in Supernovae:
             Session~LIV of the
             Les~Houches \'Ecole d'\'Et\'e de Physique Th\'eorique,
             ed.~S.~A.~Bludman, R.~Mochkovitch, \&~J.~Zinn-Justin
             (Amsterdam:  North-Holland), 63.


\refpaper Young, T. R., Baron, E., \&~Branch, D. 1995, ApJ, 449, L51.


}   

\specialpage
\centerline{FIGURE LEGENDS}
\vskip\baselineskip

{                 
\parindent=0pt

FIG. 1a.---The $N_{\ApJstyle\rm Ni}^{\ApJstyle *}(t)$ deposition curves
          for a range of $t_{\ApJstyle 0}$ values.
          The complete trapping curve is a
          $N_{\ApJstyle\rm Ni}^{\ApJstyle *}(t)$ curve with 
          $t_{\ApJstyle 0}=\infty$.
          The complete escape curve is a
          $N_{\ApJstyle\rm Ni}^{\ApJstyle *}(t)$ curve with 
          $t_{\ApJstyle 0}=0$.
          The complete trapping and escape curves are shown on
          all subsequent deposition curve figures for convenient reference.

FIG. 1b.---The absorption function $f(x)$ (for $^{\ApJstyle 56}$Co) and
          exponential
          fits to its inflection point, the point at infinity, and
          the $x$-range [2,3].
          The fit to the point at infinity is just the constant asymptote
          which $f(x)$ approaches as $x\to\infty$. 

FIG. 2a.---The normalized deposition curve for model~W7 calculated
          with only $^{\ApJstyle 56}$Ni and $^{\ApJstyle 56}$Co
          and no X-rays (i.e., the limited W7 deposition curve) and
          the fitted analytic $N_{\ApJstyle\rm Ni}^{\ApJstyle *}(t)$
          W7 deposition curve.

FIG. 2b.---The same as Fig.~2a, except extending to day~500.

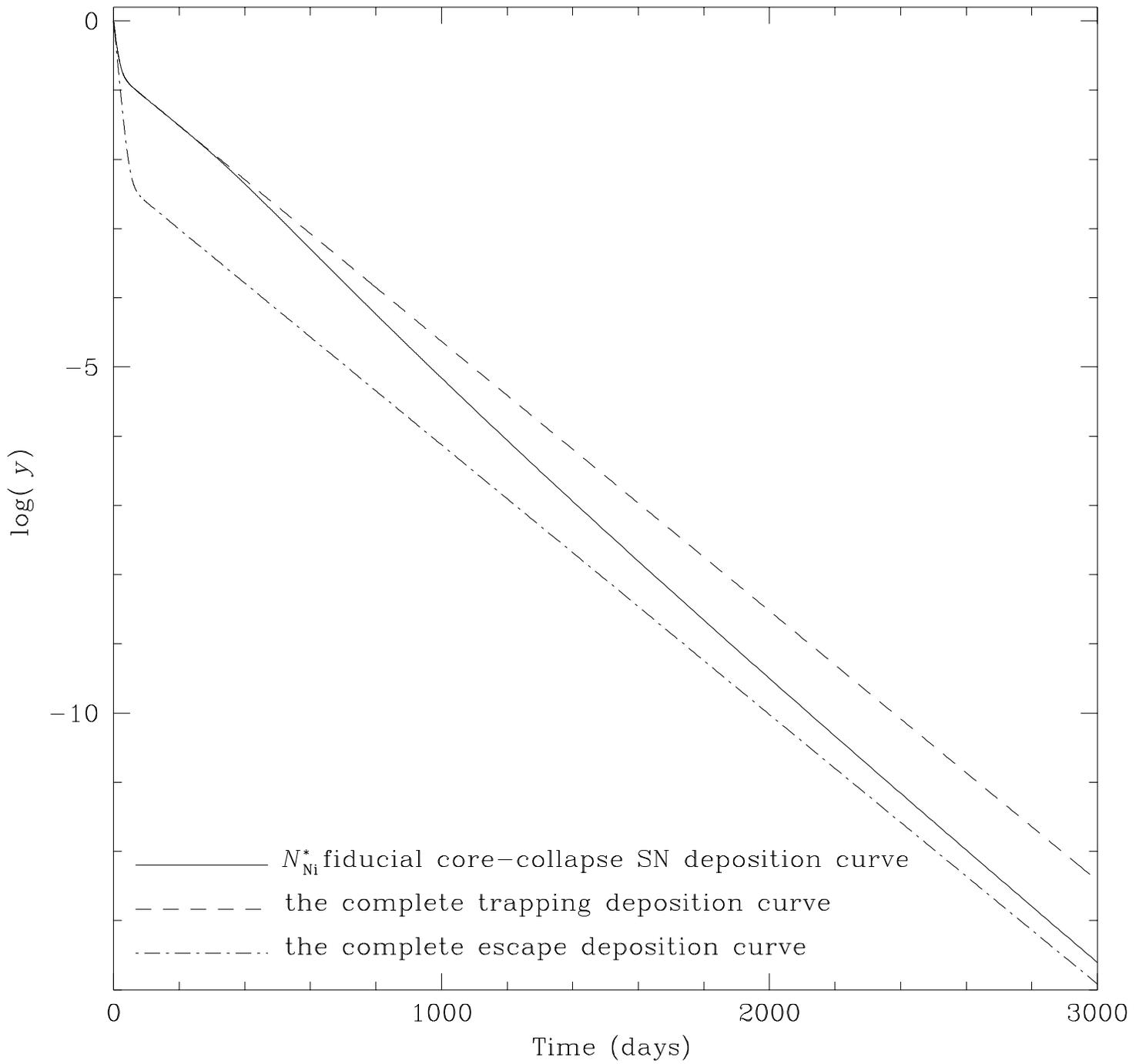
FIG. 3.---The $N_{\ApJstyle\rm Ni}^{\ApJstyle *}(t)$ fiducial core-collapse
          supernova (SN) deposition curve. 

FIG. 4a.---The normalized deposition curve for model~W7 calculated
          with all important radioactive species and X-rays included
          (i.e., the full W7 deposition curve),
          the limited W7 deposition curve,
          and the $V$ light curve of SN~Ia~1992A.
          The $V$ light curve has been vertically shifted to
          fit the deposition curves in roughly the day~300--400 period
          (which is shown in Fig.~4b).
          The fit is simply determined by eye.

FIG. 4b.---The same as Fig.~4a, but extending to day~1000.

FIG. 5a.---The UVOIR bolometric light curve of SN~II~1987A compared
           to the $N_{\ApJstyle\rm Ni}^{\ApJstyle *}(t)$ fiducial
           core-collapse supernova (SN) deposition curve.
          The SN~1987A curve has been vertically shifted to
          fit the deposition curve in the day~130--200 period.
          The fit is simply determined by eye.

FIG. 5b.---The same as Fig.~5a, but extending to day~1000.

FIG. 6a.---The schematic SN~Ic~1998bw UVOIR bolometric light curve
           (half-life 44~days)
           and a fitted $N_{\ApJstyle\rm Ni}^{\ApJstyle *}(t)$ 
           deposition curve with fiducial time $t_{\ApJstyle 0}=134.42$~days.

FIG. 6b.---The same as Fig.~6a, except extending to day~1000.

}                 


{  
\specialpage
\vskip\baselineskip
\vglue -.8truein

\centerline{TABLES}

\vskip\baselineskip

{  

\baselineskip=20pt
\tabskip=50pt minus 50pt  
\newdimen\digitwidth
\setbox0=\hbox{\rm0}
\digitwidth=\wd0
\catcode\lq?=\active
\def?{\kern\digitwidth}

\centerline{TABLE 1}
\centerline{PARAMETERS FOR THE RADIOACTIVE DECAYS}
\centerline{OF $^{\ApJstyle 56}$Co AND $^{\ApJstyle 56}$Ni}
\vskip\baselineskip\hrule\smallskip\hrule\medskip
\halign to \hsize{#\hfil    &#\hfil &#\hfil \cr
Parameter   &$^{\ApJstyle 56}$Co &$^{\ApJstyle 56}$Ni    \cr
\noalign{\medskip\hrule\medskip}

$t_{\ApJstyle 1/2}$ (days)    &77.27(3)    &6.077(12)    \cr

$t_{\ApJstyle e}={t_{\ApJstyle 1/2}/\displaystyle \ln(2)}$ (days)
                              &111.48(4)   &8.767(17)    \cr

$Q_{\ApJstyle\rm total}$ (MeV)   &4.566(2)  &2.135(11)      \cr

$Q_{\ApJstyle\rm ph+PE}$ (MeV)   &3.74(4)  &1.729(17)      \cr

$Q_{\ApJstyle\rm ph}$ (MeV)      &3.62(4)  &1.723(17)      \cr

$f_{\ApJstyle\rm ph}$            &0.968(14) &0.996(14)     \cr

$f_{\ApJstyle\rm PE}$            &3.20(16)$-$2  &3.99(18)$-$3  \cr

$C$ ($\rm ergs\,s^{\ApJstyle -1}\,g^{\ApJstyle -1}$)
                              &6.70(7)+9  &3.94(4)+10 \cr

$B$ ($\rm ergs\,s^{\ApJstyle -1}\,g^{\ApJstyle -1}$)
                              &7.27(7)+9  & ---           \cr
} 
\medskip\hrule
\vskip\baselineskip
     NOTE.---The values have been taken or derived
from LBL, Huo 1992, and
Browne \&~Firestone 1986.  
We have put the uncertainties in the last digits of the parameters 
in brackets and have written $\times 10^{\ApJstyle\pm k}$ as
$\pm k$.

     The parameters are defined as follows:
$t_{\ApJstyle 1/2}$ is half-life,
$t_{\ApJstyle e}$ is $e$-folding time,
$Q_{\ApJstyle\rm total}$ is the total energy (including
  neutrino energy) per decay,
$Q_{\ApJstyle\rm ph+PE}$ is the mean photon plus PE (positron
  and decay-ejected atomic electron) kinetic
  energy per decay,
$Q_{\ApJstyle\rm ph}$ is the mean photon energy per decay,
$f_{\ApJstyle\rm ph}$ is the fraction of $Q_{\ApJstyle\rm ph+PE}$
  in photon energy,
$f_{\ApJstyle\rm PE}$ is the fraction of $Q_{\ApJstyle\rm ph+PE}$
  in PE kinetic energy,
and $C$ and $B$ are energy generation coefficients specified in
  the text (see \S~3.1).
The $\gamma$-ray energy from positron annihilation is included
in $Q_{\ApJstyle\rm ph+PE}$ and $Q_{\ApJstyle\rm ph}$. 
We assume that the neutrinos simply escape the supernova ejecta
and make no contribution to the RDE deposition.
\vfill
\supereject
}   

{  

\baselineskip=12pt
\tabskip=50pt minus 50pt  
\newdimen\digitwidth
\setbox0=\hbox{\rm0}
\digitwidth=\wd0
\catcode\lq?=\active
\def?{\kern\digitwidth}

\centerline{TABLE 2}
\centerline{EXPONENTIAL FITS}
\centerline{TO THE ABSORPTION FUNCTION $f(x)$}
\vskip\baselineskip\hrule\smallskip\hrule\medskip
\halign to \hsize{#\hfil   &#\hfil &\hfil#\hfil
                           &\hfil#\hfil &\hfil#\hfil  \cr
Fit at x/over $\left[x_{\ApJstyle\rm a},x_{\ApJstyle\rm b}\right]$
                          &$K_{\ApJstyle\rm coef}$ &$x_{\ApJstyle 1/2}$ 
                            &$t_{\ApJstyle 1/2}^{\ApJstyle\rm fid,Ia}$
                            &$t_{\ApJstyle 1/2}^{\ApJstyle\rm fid,CC}$ \cr
\quad Max.~Error (dex)
      in $\left[x_{\ApJstyle\rm a},x_{\ApJstyle\rm b}\right]$ 
      &   &                 &(days) &(days) \cr
\noalign{\medskip\hrule\medskip}

$x=0$                   &1         &$\infty$  &77.27   &77.27     \cr
\quad 0.001 in $[0,0.40]$                                         \cr
\quad 0.04 in $[0,0.64]$                                          \cr
\noalign{\vskip\baselineskip}

$[0.23,0.53]$           &1.027959  &7.659517  &61.98   &75.91     \cr
\quad 0.0033 in $[0.23,0.53]$                                     \cr
\noalign{\vskip\baselineskip}

$x=x_{\ApJstyle\rm infl}\approx 1.040765$
                        &1.950777  &0.625347  &19.21   &63.38     \cr
\quad 0.004 in $[0.8,1.35]$                                       \cr
\quad 0.04  in $[0.56,1.82]$                                      \cr
\noalign{\vskip\baselineskip}

$[0.54,1.98]$ about $x_{\ApJstyle\rm infl}$ (n)
                        &1.694608  &0.705462  &21.01   &64.70     \cr
\quad 0.014 in $[0.54,1.98]$                                      \cr
\noalign{\vskip\baselineskip}

$[0.42,2.46]$ about $x_{\ApJstyle\rm infl}$ (m)
                        &1.540518  &0.770151  &22.38   &65.60      \cr
\quad 0.028 in $[0.42,2.46]$                                       \cr
\noalign{\vskip\baselineskip}

$[0.33,2.82]$ about $x_{\ApJstyle\rm infl}$ (b)
                        &1.441097  &0.822176  &23.43   &66.23      \cr
\quad 0.04 in $[0.33,2.82]$                                        \cr
\noalign{\vskip\baselineskip}

$[2,3]$                 &0.804766  &1.149675  &29.23   &69.04      \cr
\quad 0.01 in $[2,3]$                                              \cr
\noalign{\vskip\baselineskip}

$[4,8]$                 &0.151079  &4.653005  &54.95   &75.06      \cr
\quad 0.037 in $[4,8]$                                             \cr
\noalign{\vskip\baselineskip}

$x=\infty$              &0.0320    &$\infty$  &77.27   &77.27      \cr
\quad 0.098007 in $[2x_{\ApJstyle\rm tr},\infty]$                  \cr
} 
\medskip\hrule
{          
\baselineskip=20pt
\vskip\baselineskip
     NOTE.---The exponential function
$K_{\ApJstyle\rm coef}\exp\left(-x/x_{\ApJstyle e}\right)$
(where $x_{\ApJstyle e}$ is the $e$-folding reduced time)
has been fitted to the absorption function $f(x)$.
The fits are either to $f(x)$'s value and its logarithmic slope at
a given point $x$ or are fits to $f(x)$ over a given $x$-range
$\left[x_{\ApJstyle\rm a},x_{\ApJstyle\rm b}\right]$.
The point fits are to the asymptotic behavior of $f(x)$ at
zero and infinity, and to $f(x)$ at its inflection point
$x_{\ApJstyle\rm infl}$.
A fit to an $x$-range was chosen so as to
reproduce $f(x)$ close to optimally over that range. 

     The first column gives the point or $x$-range of the fit in
the first line and in subsequent lines the maximum error of the fit (or
the maximum deviation of $f(x)$ from the fitted exponential)
in specified ranges.
The second column gives the coefficient $K_{\ApJstyle\rm coef}$
of the fitted exponential,
The third column gives the reduced time half-life of the exponential:
$x_{\ApJstyle 1/2}=x_{\ApJstyle e}\ln(2)$.
The fourth and fifth columns give the half-lives (in real time)
of the product of the $^{\ApJstyle 56}$Co decay exponential and the fitted
exponential for the fiducial SN~Ia model and fiducial core-collapse
(CC) supernova model posited in \S\S~3.3 and~3.4, respectively.

}   

\vfill
\supereject
}   

{  

\baselineskip=20pt
\tabskip=50pt minus 50pt  
\newdimen\digitwidth
\setbox0=\hbox{\rm0}
\digitwidth=\wd0
\catcode\lq?=\active
\def?{\kern\digitwidth}

\centerline{TABLE 3}
\centerline{SUMMARY OF RADIOACTIVE DECAYS IMPORTANT IN MODEL W7}
\vskip\baselineskip\hrule\smallskip\hrule\medskip
\halign to \hsize{#\hfil   &#\hfil   &\hfil#\hfil 
                                     &\hfil#\hfil
                                     &\hfil#\hfil\cr
\noalign{\noindent Radioactive species and initial mass}
Decay   &$t_{\ApJstyle 1/2}$   &$Q_{\ApJstyle\rm total}$ 
                               &$Q_{\ApJstyle\rm ph+PE}$
                               &$Q_{\ApJstyle\rm ph}$ \cr
        &                      &(MeV) &(MeV) &(MeV) \cr 
\noalign{\medskip\hrule\medskip}
\noalign{\noindent $
^{\ApJstyle 44}{\rm Ti}\quad 1.8\times10^{\ApJstyle -5}\,M_{\ApJstyle\odot}
{\rm ;\qquad}
^{\ApJstyle 44}{\rm Sc}\quad 1.8\times10^{\ApJstyle -9}\,M_{\ApJstyle\odot}
$}
$^{\ApJstyle 44}$$\rm Ti\to^{\ApJstyle 44}$$\rm Sc$
        &63(3) years     &0.2675(19)?  &0.1493(15)    &0.1384(14) \cr
$^{\ApJstyle 44}$$\rm Sc\to^{\ApJstyle 44}$$\rm Ca$
        &0.1636(3) days  &3.6533(19)?  &2.73(23)??    &2.13(23)??   \cr
\noalign{\vskip\baselineskip}
\noalign{\noindent $
^{\ApJstyle 55}{\rm Co}\quad 4.5\times10^{\ApJstyle -3}\,M_{\ApJstyle\odot}
{\rm ;\qquad}
^{\ApJstyle 55}{\rm Fe}\quad 1.4\times10^{\ApJstyle -3}\,M_{\ApJstyle\odot}
$}
$^{\ApJstyle 55}$$\rm Co\to^{\ApJstyle 55}$$\rm Fe$
        &0.7304(13) days &3.4513(4)??  &2.43(4)???    &2.00(4)???  \cr
$^{\ApJstyle 55}$$\rm Fe\to^{\ApJstyle 55}$$\rm Mn$
        &2.73(3) years   &0.23138(10)  &0.0054(3)?    &0.00163(5) \cr
\noalign{\vskip\baselineskip}
\noalign{\noindent $
^{\ApJstyle 56}{\rm Ni}\quad 0.58\,M_{\ApJstyle\odot}
{\rm ;\qquad}
^{\ApJstyle 56}{\rm Co}\quad 6.1\times10^{\ApJstyle -5}\,M_{\ApJstyle\odot}
$}
$^{\ApJstyle 56}$$\rm Ni\to^{\ApJstyle 56}$$\rm Co$
        &6.077(12) days  &2.135(11)??  &1.729(17)?  &1.723(17)?    \cr
$^{\ApJstyle 56}$$\rm Co\to^{\ApJstyle 56}$$\rm Fe$
        &77.27(3) days   &4.566(2)???  &3.74(4)???   &3.62(4)???   \cr
\noalign{\vskip\baselineskip}
\noalign{\noindent $
^{\ApJstyle 57}{\rm Ni}\quad 2.15\times10^{\ApJstyle -2}\,M_{\ApJstyle\odot}
{\rm ;\qquad}
^{\ApJstyle 57}{\rm Co}\quad 8.1\times10^{\ApJstyle -4}\,M_{\ApJstyle\odot}
$}
$^{\ApJstyle 57}$$\rm Ni\to^{\ApJstyle 57}$$\rm Co$
        &1.4833(25) days &3.264(3)???  &2.07(3)???   &1.92(3)???   \cr
$^{\ApJstyle 57}$$\rm Co\to^{\ApJstyle 57}$$\rm Fe$
        &271.79(9) days  &0.8360(4)??  &0.1429(8)?   &0.1253(6)?   \cr
} 
\medskip\hrule
\vskip\baselineskip
     NOTE.---The masses of the radioactive species are from epoch just
after the model~W7 explosion (i.e., effectively time zero)
(Thielemann~\etal 1986;  Nomoto~\etal 1994b).
The nuclear data have been taken or derived from 
LBL, Huo 1992, Bhat 1992, and
Browne \&~Firestone 1986.  
We have put the uncertainties in the last digits of the parameters
in brackets.

     See the note to Table~1 for the definitions of the parameters.
\vfill
\supereject
}   

}  

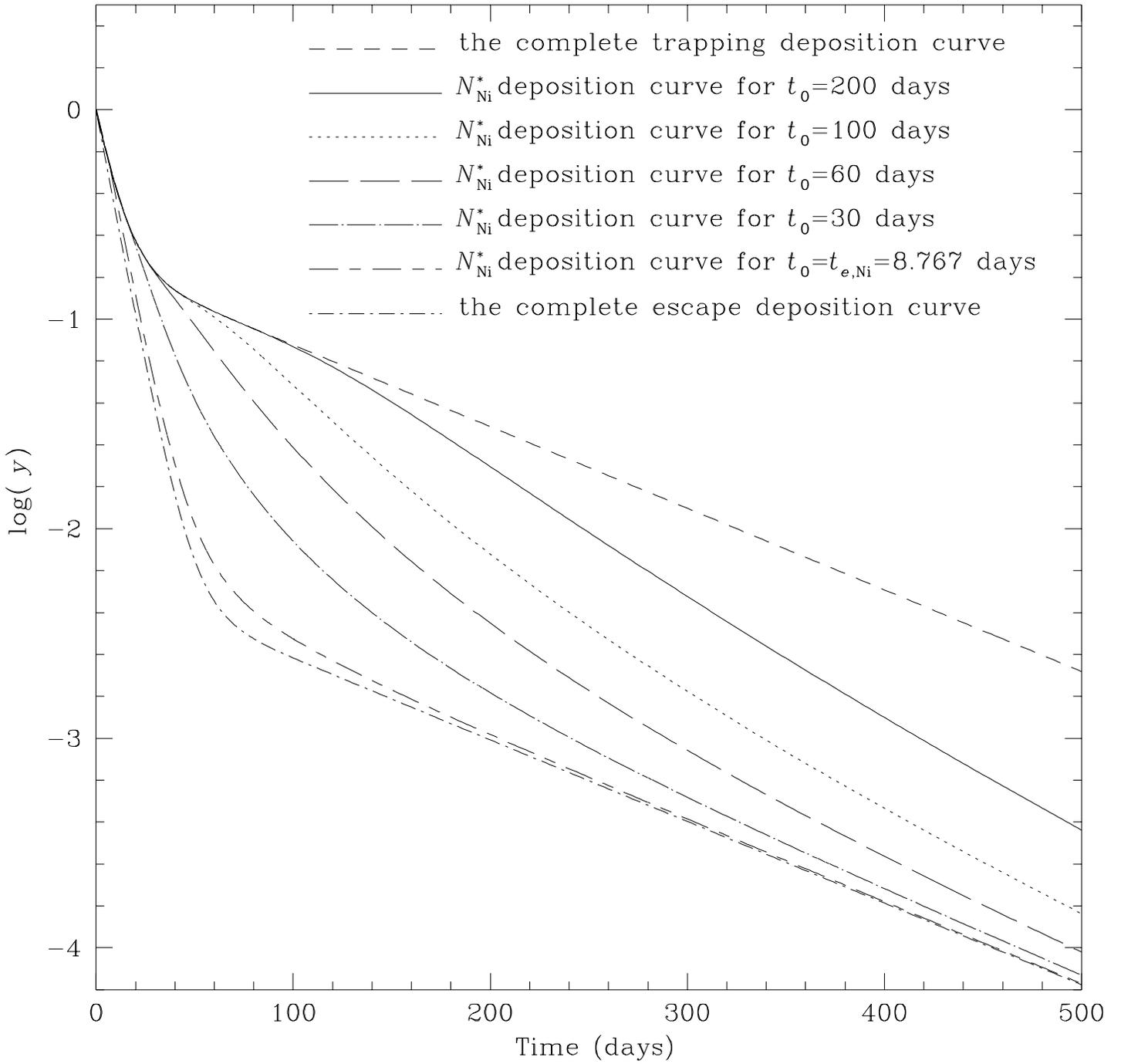
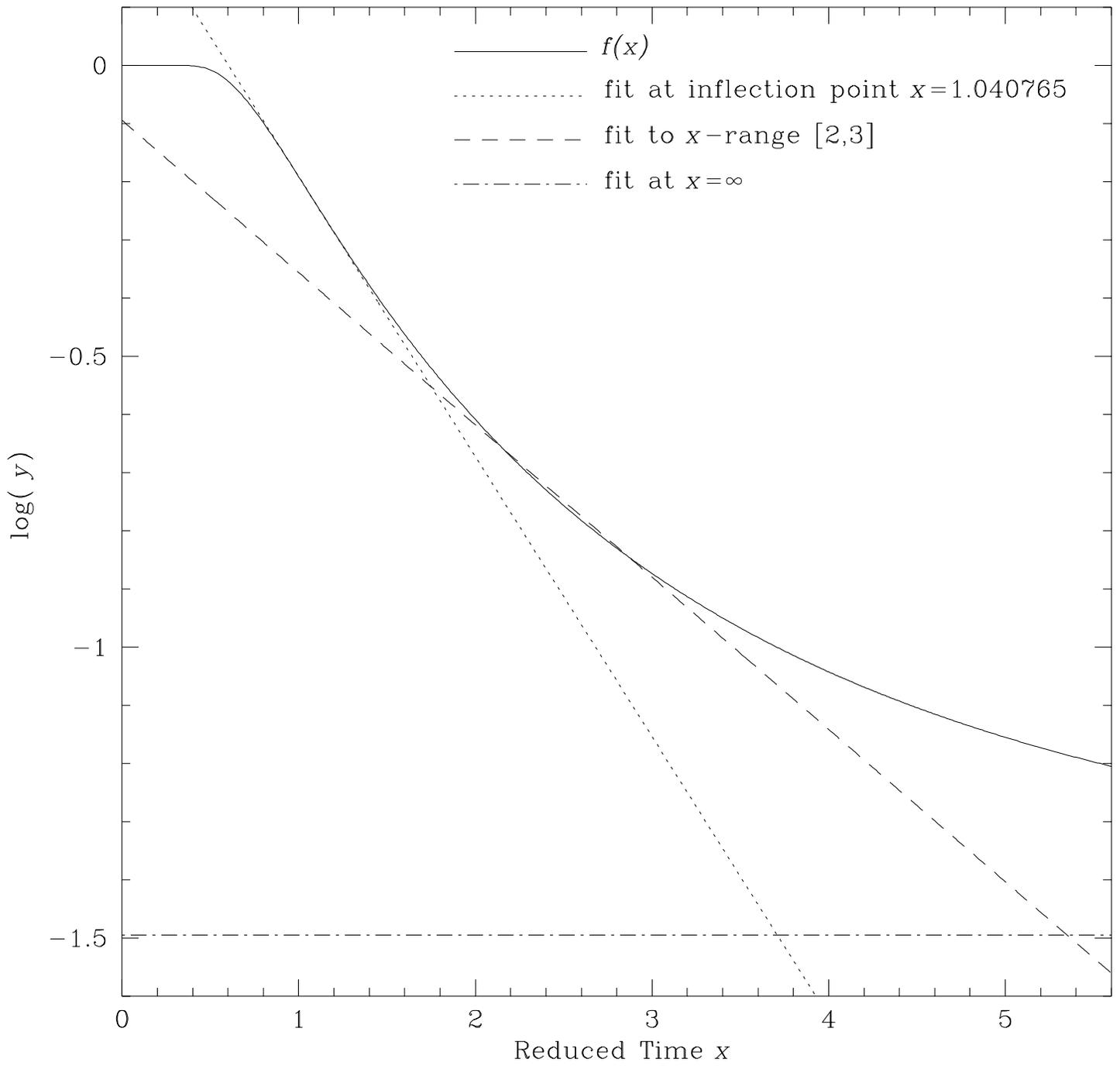
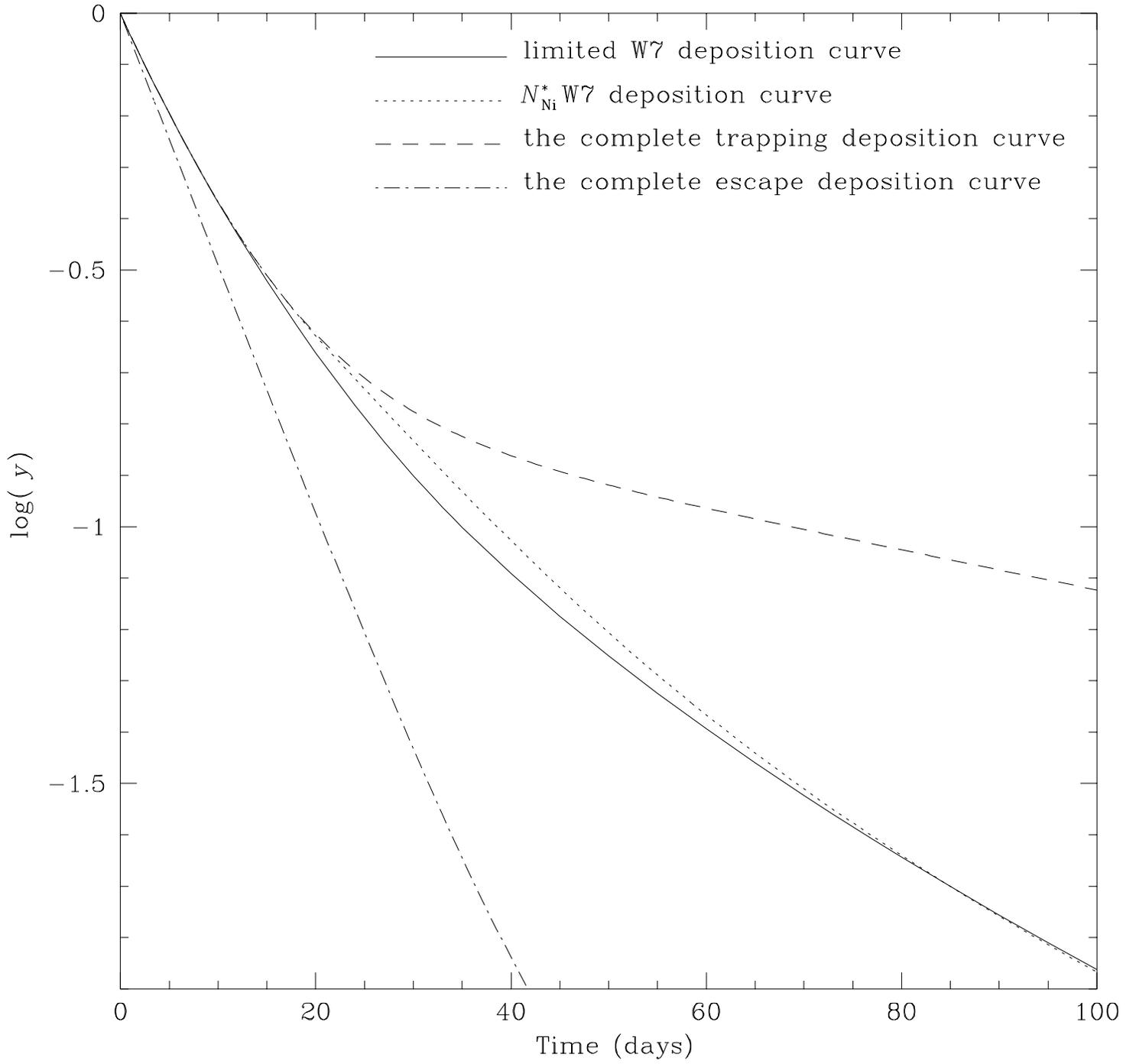
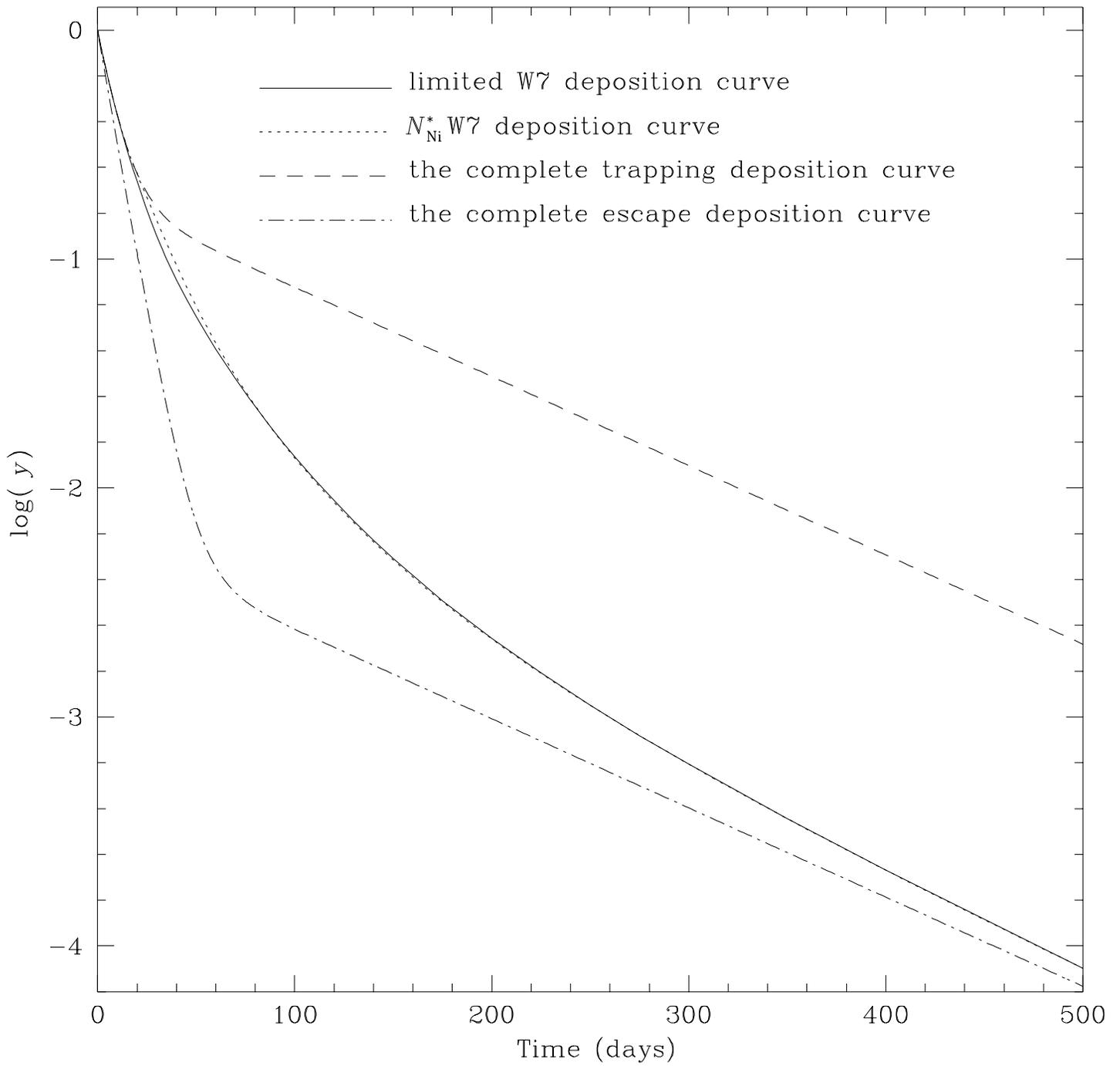
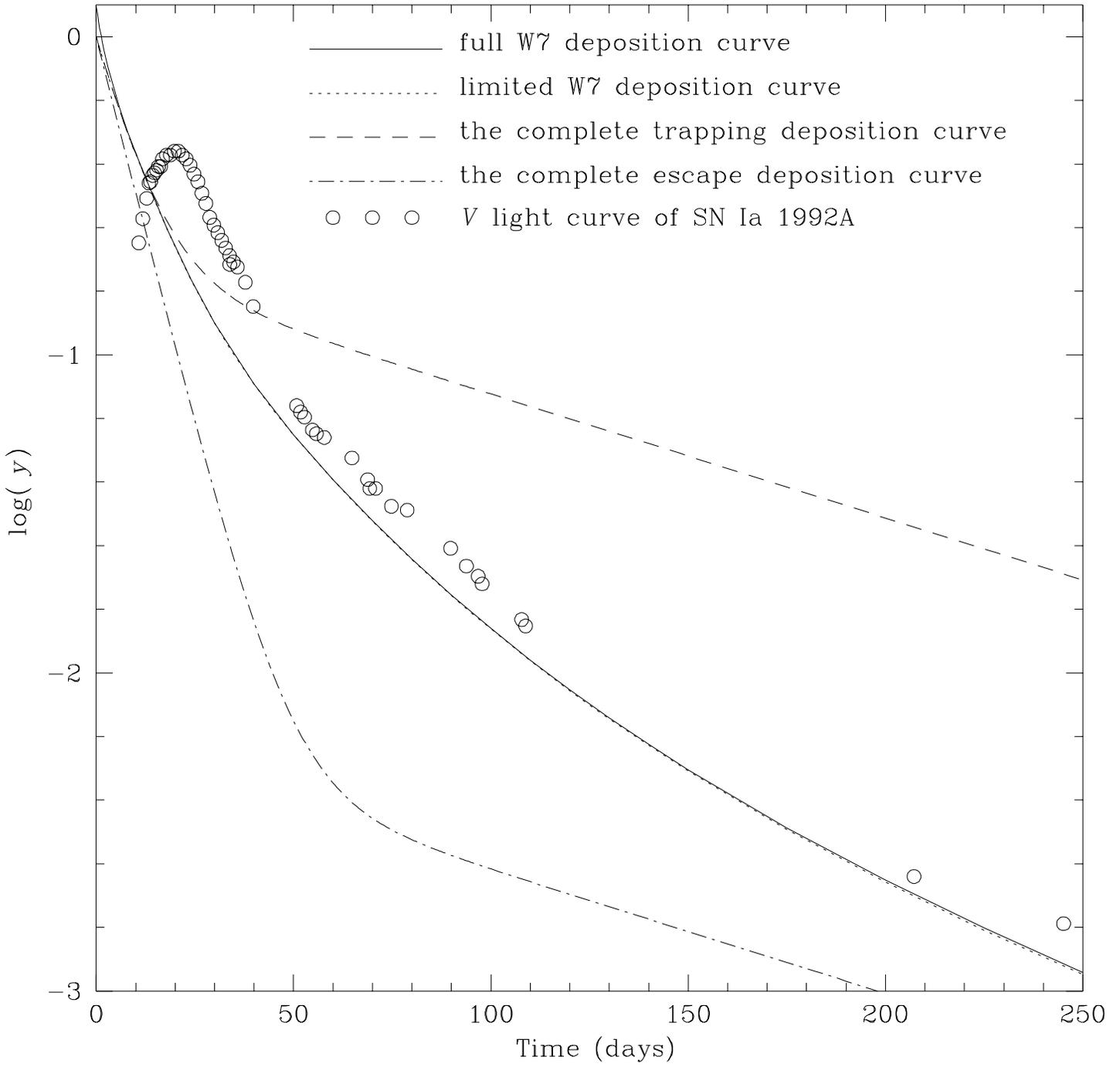
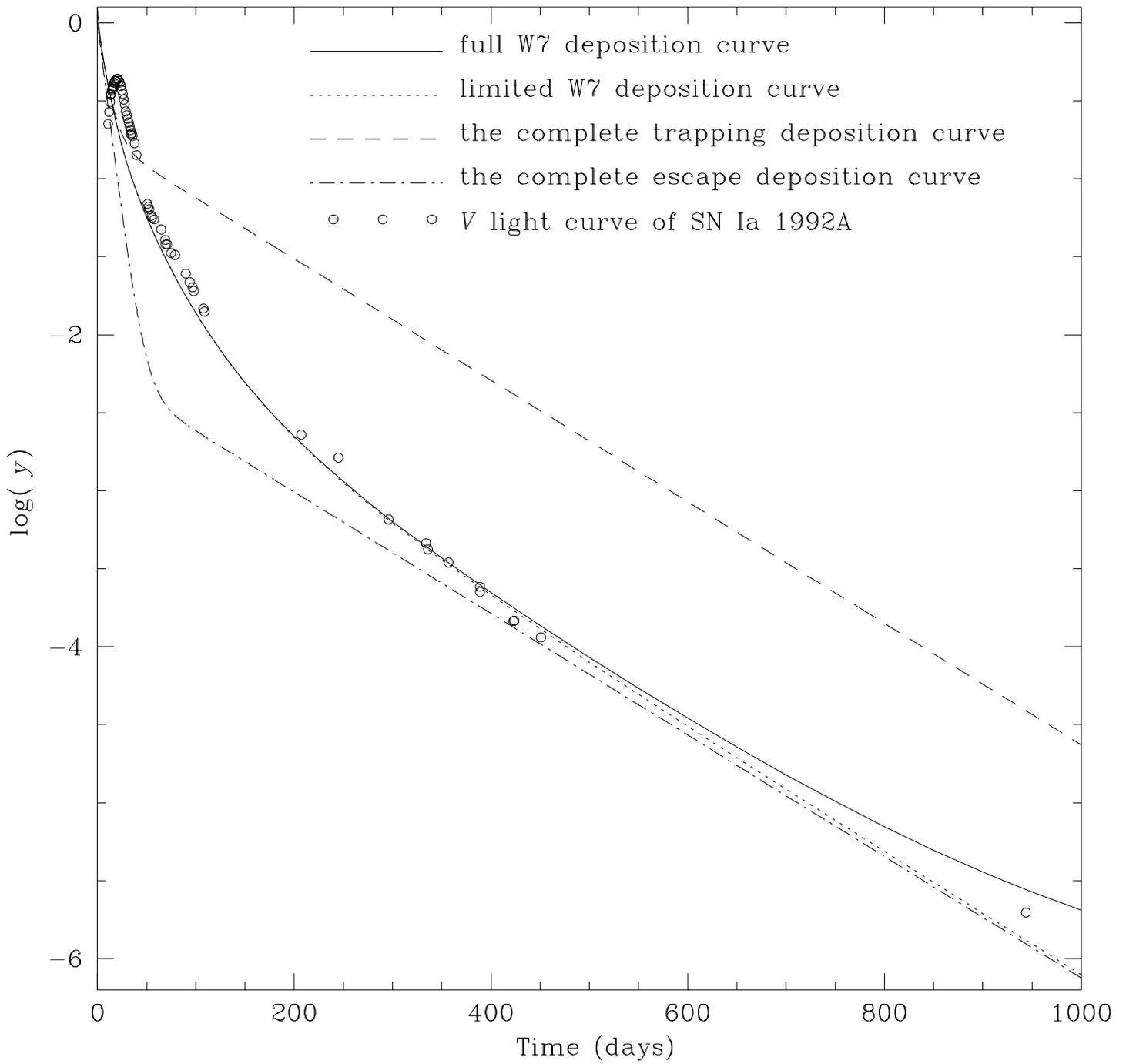
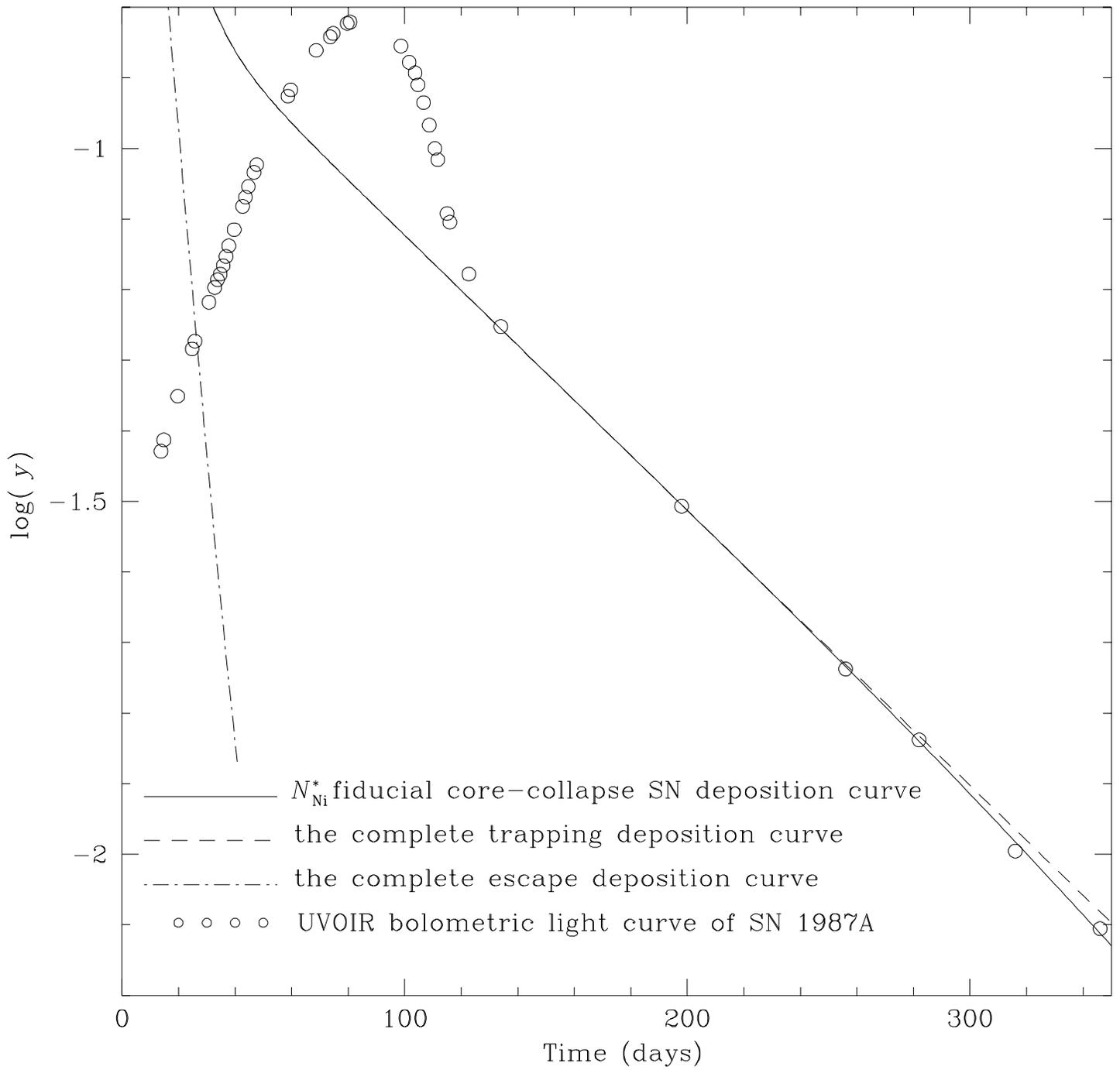
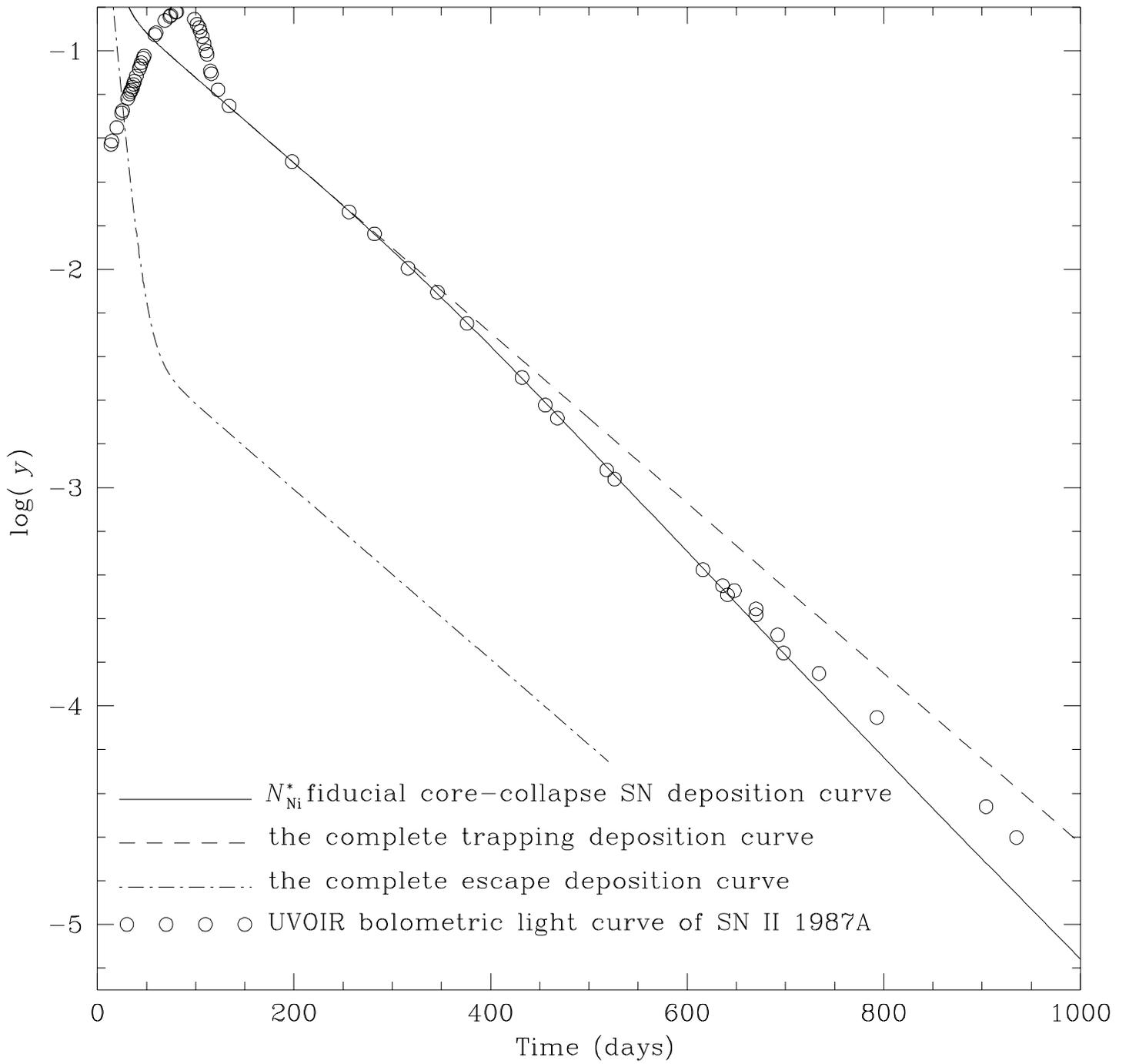
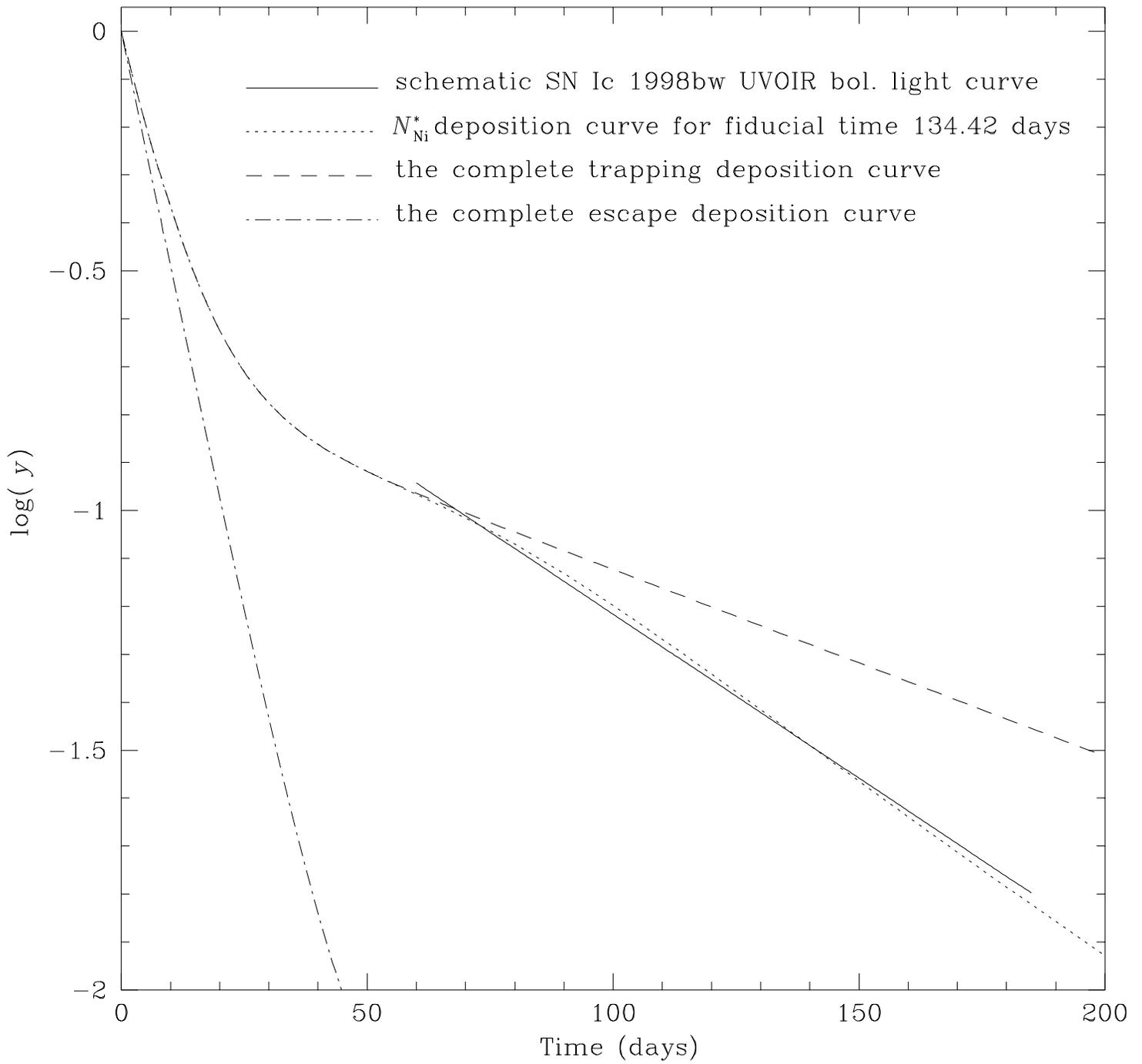
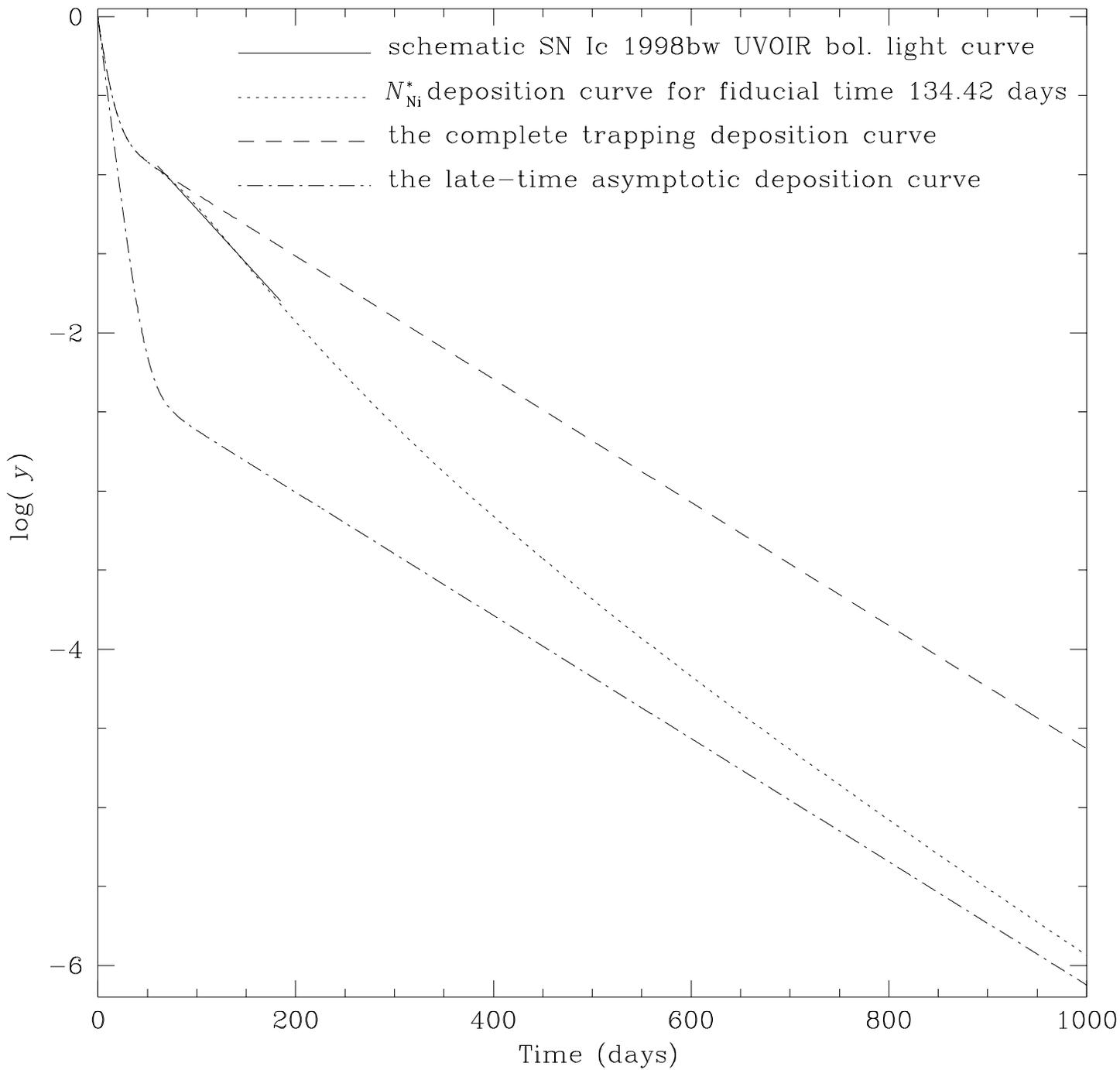
\bye